# RAMAN SPECTRA DATABASE OF THE GLASS BEADS EXCAVATED ON MAPUNGUBWE HILL AND K2, TWO ARCHAEOLOGICAL SITES IN SOUTH AFRICA

**Aurélie Tournié**
**Linda C. Prinsloo**
**Philippe Colomban**

*About two hundred coloured glass beads (red, yellow, green, blue, white, black, pink, plum) were selected among the thousands of beads excavated on Mapungubwe hill and at K2, archaeological sites in the Limpopo valley South Africa, and have been studied with Raman scattering. The glass matrix of the beads was classified according to its Raman signature into 3 main sub-groups and corroded glass could also be identified. At least seven different chromophores or pigments (lazurite, lead tin yellow type II, Ca/Pb arsenate, chromate, calcium antimonate, Fe-S "amber" and a spinel) have been identified. Many of the pigments were only manufactured after the 13th century that confirms the presence of modern beads in the archaeological record. This calls for further research to find a way to reconcile the carbon dating of the hill, which currently gives the last occupation date on the hill as 1280 AD with the physical evidence of the modern beads excavated on the hill.*

**2009-2010**

## 1. DATABASE

This work was conducted at the University of Pretoria in the Physics department during a post-doctorate contract (2009 - 2010).

The aim of this investigation was to compile a database of the beads excavated at the Mapungubwe and K2. This Raman spectral database compiled incorporates the results from previous studies (Prinsloo and Colomban, 2008; Tournié *et al*., JRS, 2010) and can be used as a guide to identify beads with a fast and non-destructive technique.

An assemblage of 175 beads that appeared to be different in shape, size and colour were selected from the formal Mapungubwe collection under the stewardship of the Department of UP Arts. The accessioned bead collection is extensive and has accumulated over a 75 year period of archaeological excavation. A majority of the beads are provenanced material, whilst many others have identifying labels and containers providing information that can be linked to archival catalogue entries and field notebooks. Only beads excavated on Mapungubwe hill and K2 sites were selected and most of the beads selected are from the older excavations (1934-1938) and include the M3 beads series (Gardner, 1963) not analysed by Davison, Saitowitz and Wood.

The database is classified by colour: blue (Figure 1), red (Figure 2), yellow (Figure 3), green (Figure 4), black (Figure 5), white (Figure 6) and other beads like pink, striped, quartz and tiny (Figure 7). All the beads were analysed in different spots to make sure that the spectra are reproductible. This catalogue compiles: photographs of each of the beads analysed, museum reference number, site (Mabungubwe hill or K2), baseline corrected Raman spectrum, type of glass matrix ($Na_2O$ or $Na_2O/CaO$) and the pigment identified. The approximate quantity for some of the more unusual beads found in the collection is also indicated.

After these figures we're going to present the Mapungubwe and K2 sites, then the experimental part will be described and the glass matrix and pigments will be classified and compared to those established using morphological and technological parameters. The pigments will be classified according to the period of manufacture, which can be related to the date the beads were exported to Africa and therefore provide an approximate date of the site where it was excavated.

## Figure 1: BLUE BEADS

**BEADS** (scale= 1mm) | **REF. MUSEUM** (quantity) | **RAMAN SPECTRA** (after baseline subtraction, a.u. = arbitrary unit) | **GLASS MATRIX PIGMENT**

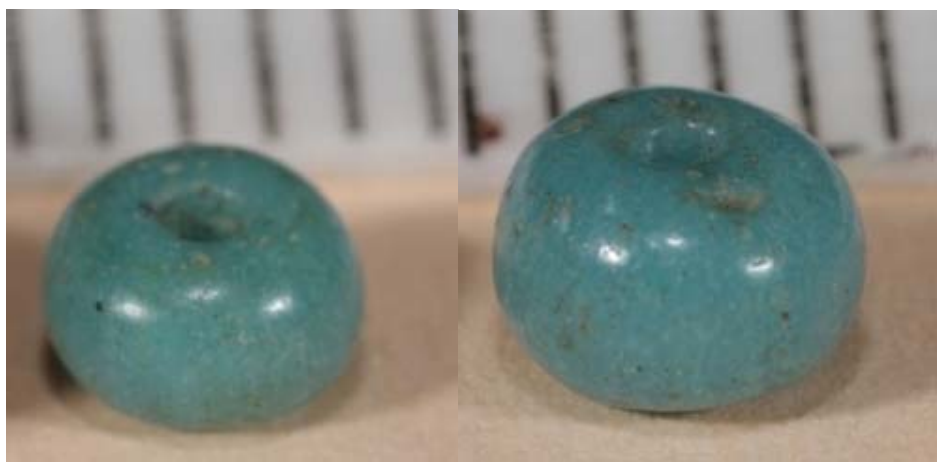

Accession: **C/4354**
Box C0241

Map. Hill

(~30)

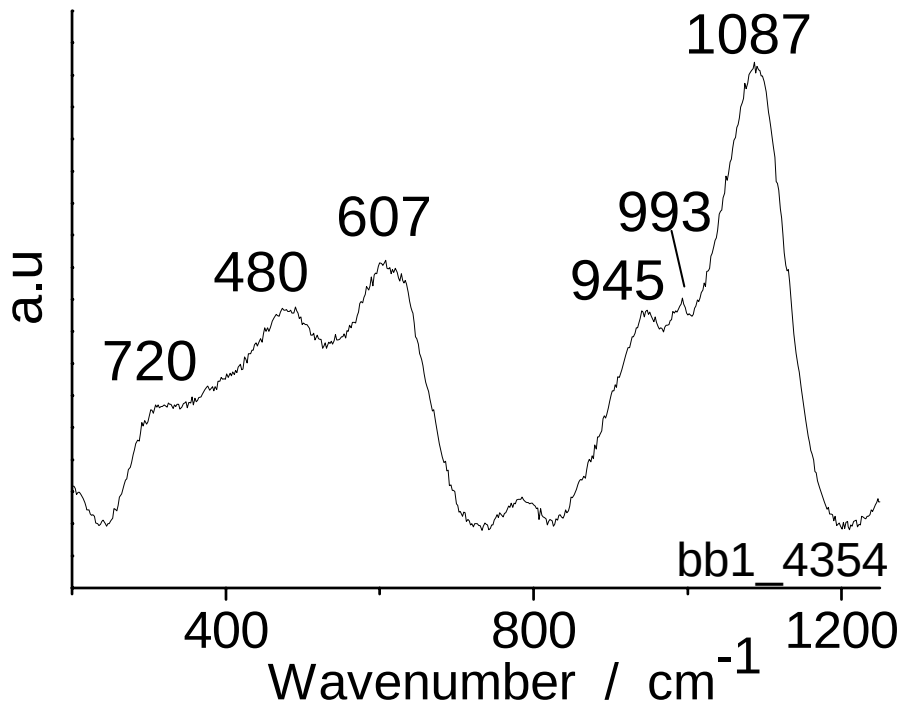

$Na_2O/CaO$

Copper

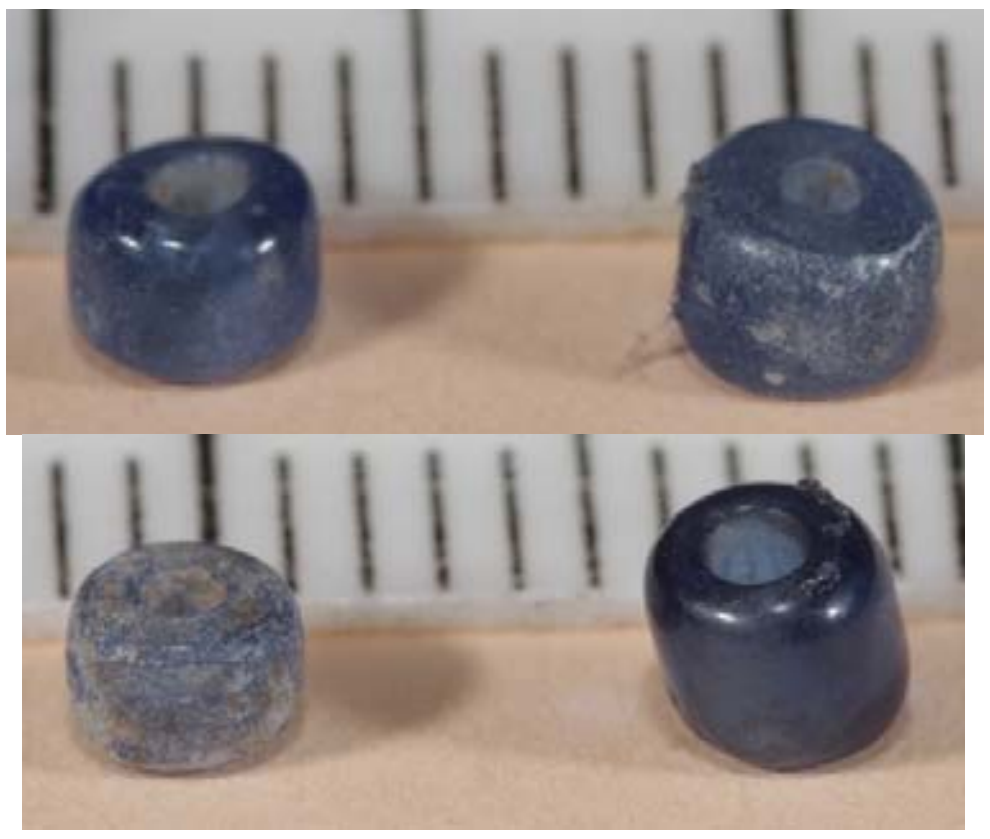

Accession: **C/952**
Box C0046

Map. Hill

(~100)

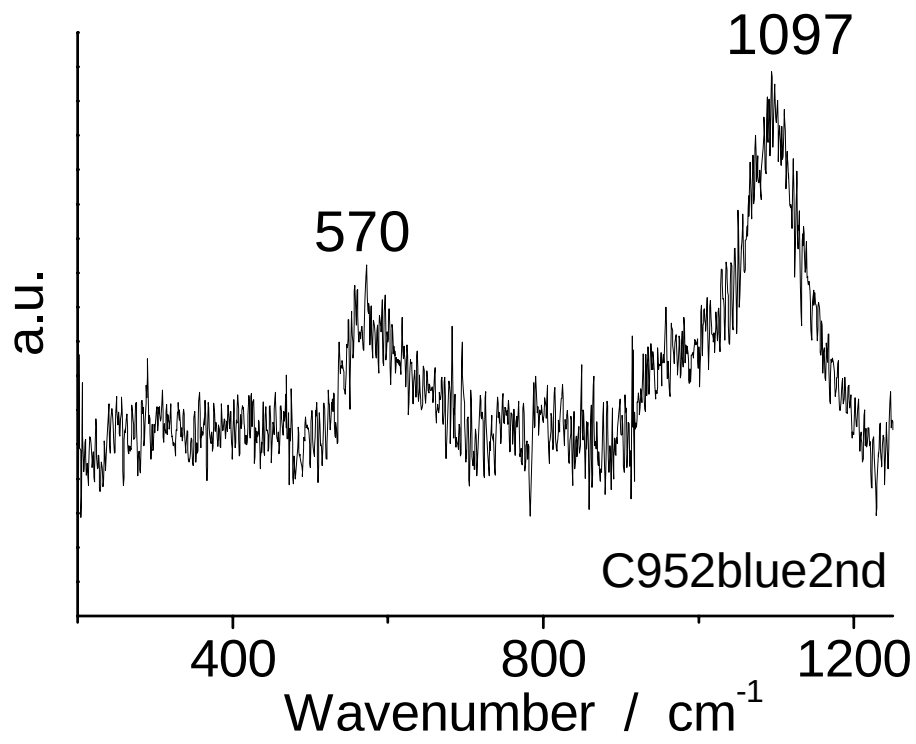

$Na_2O/CaO$

Cobalt

* T64000

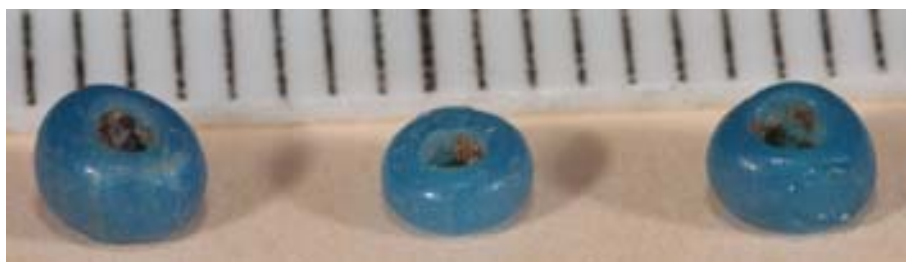

Accession: **C/740**
Box C0038

Map. Hill

(~300-400)

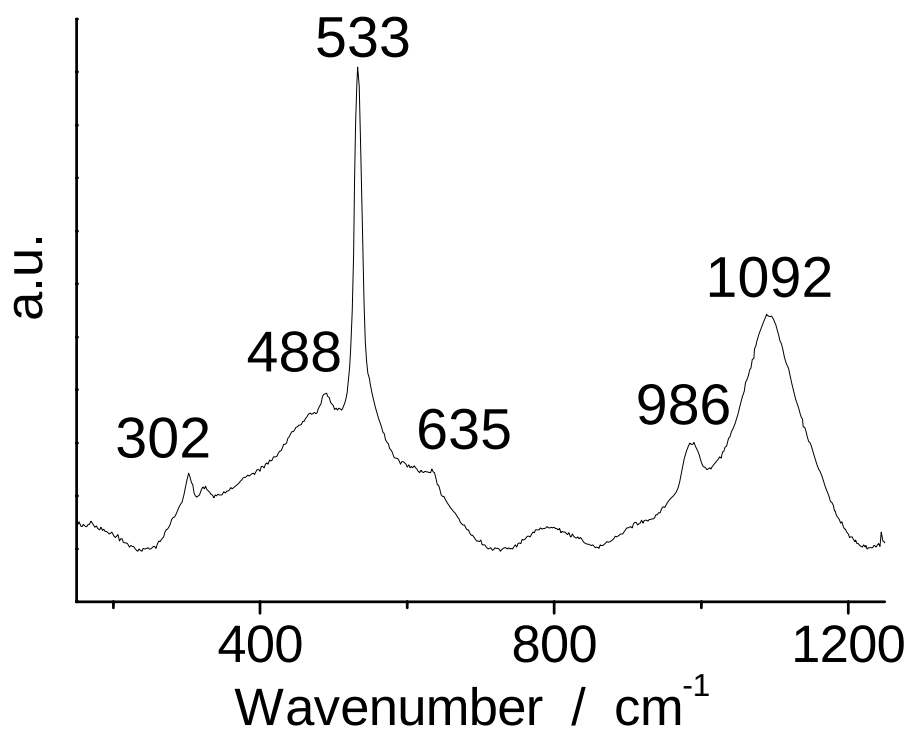

Lazurite
19th c.

* T64000

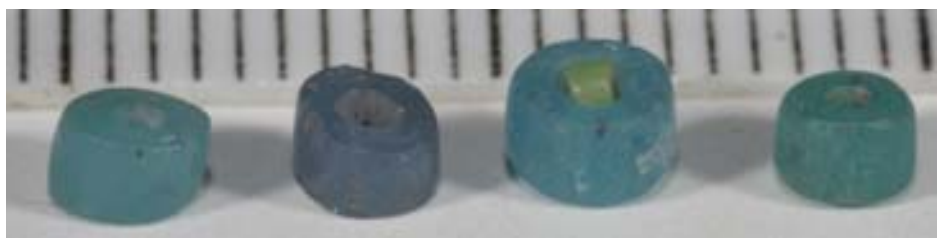

Accession:
**C/952**
Box C0046

Map. Hill

(~800)

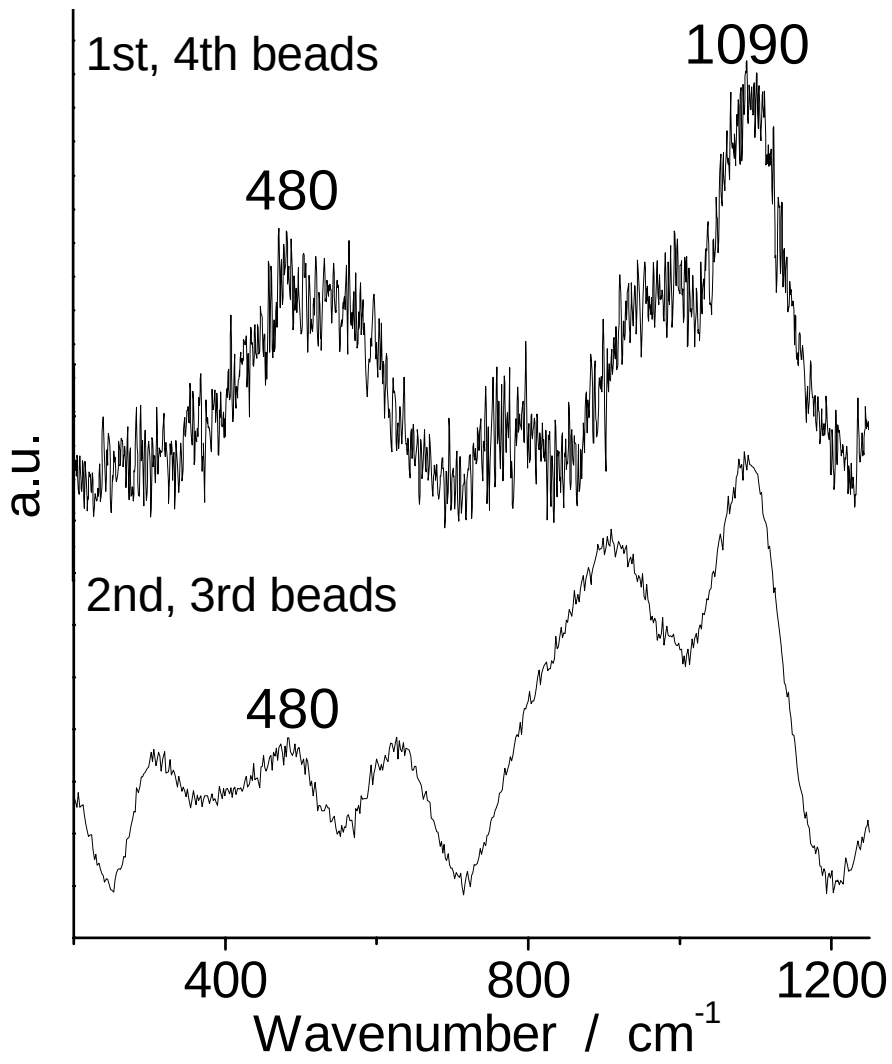

$Na_2O$

Cobalt

Beads 1,3 and 4 looks like copper

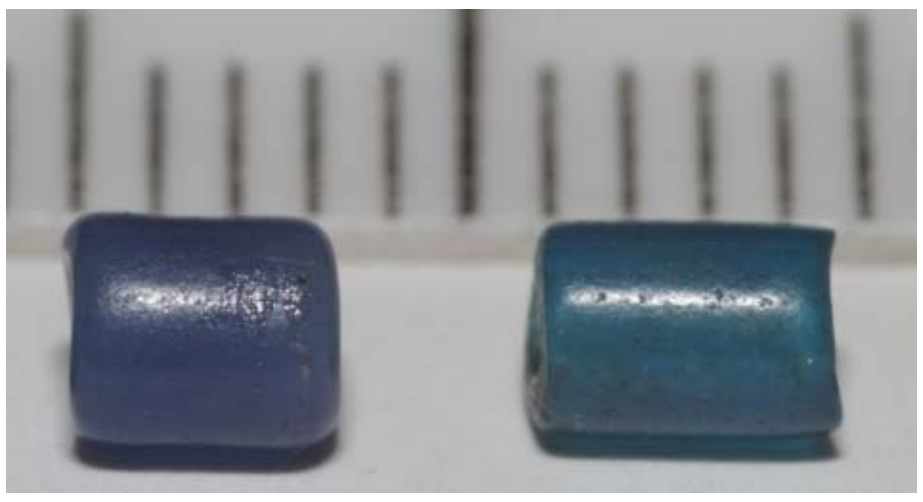

Accession:
**C/4333**
Box 0241

Map. Hill

(~10)

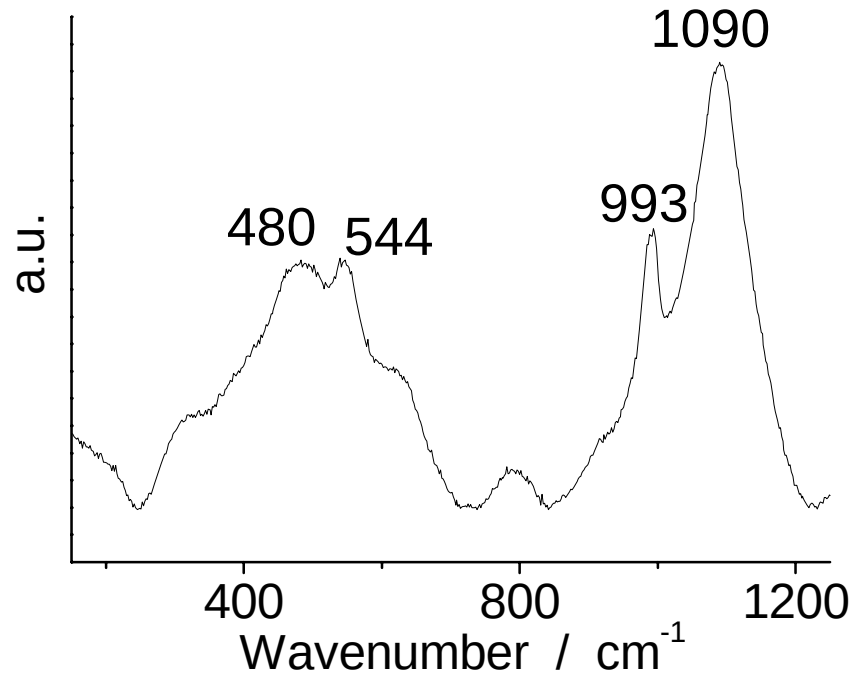

$Na_2O$

Lazurite
19th c.

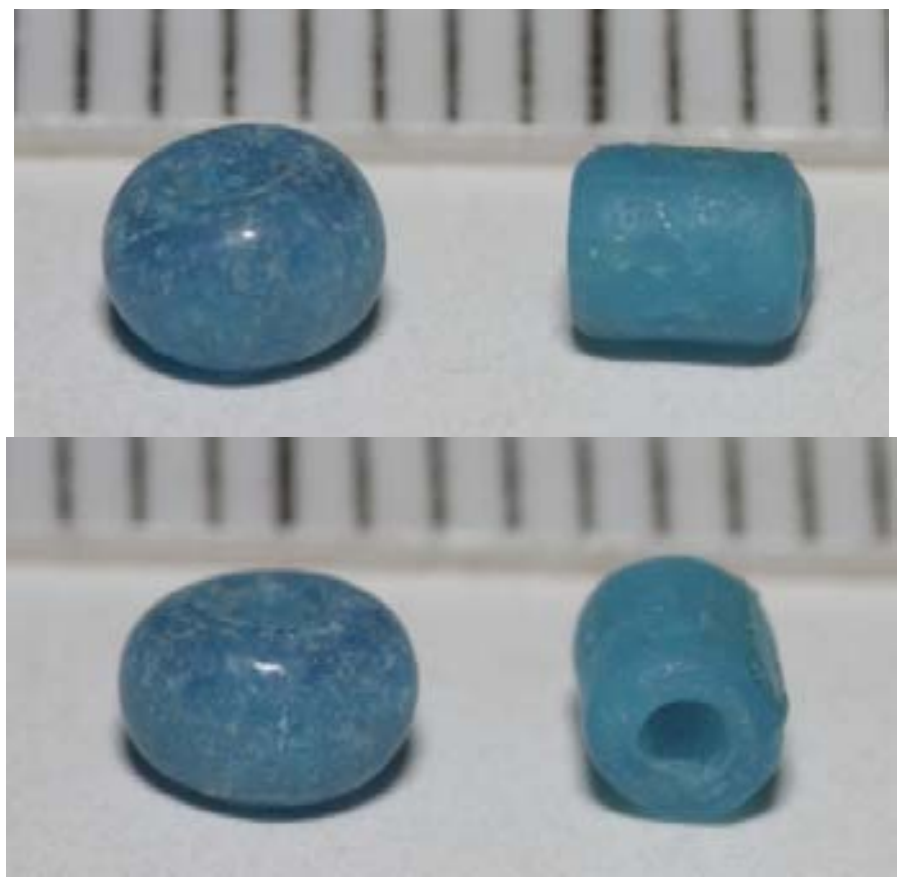

Accession:
**C/762**
Box 0041

Map. Hill

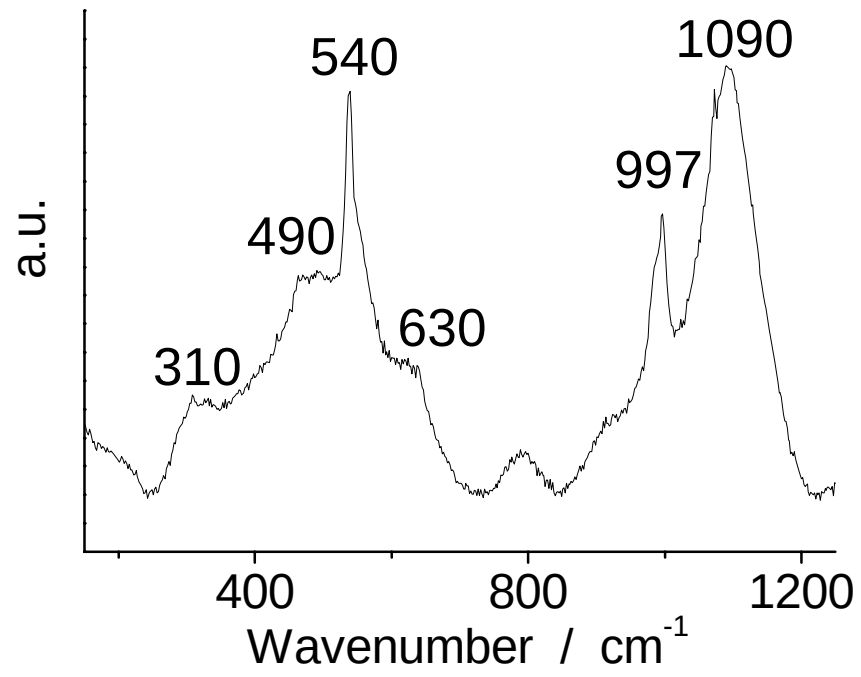

Lazurite
19th c.

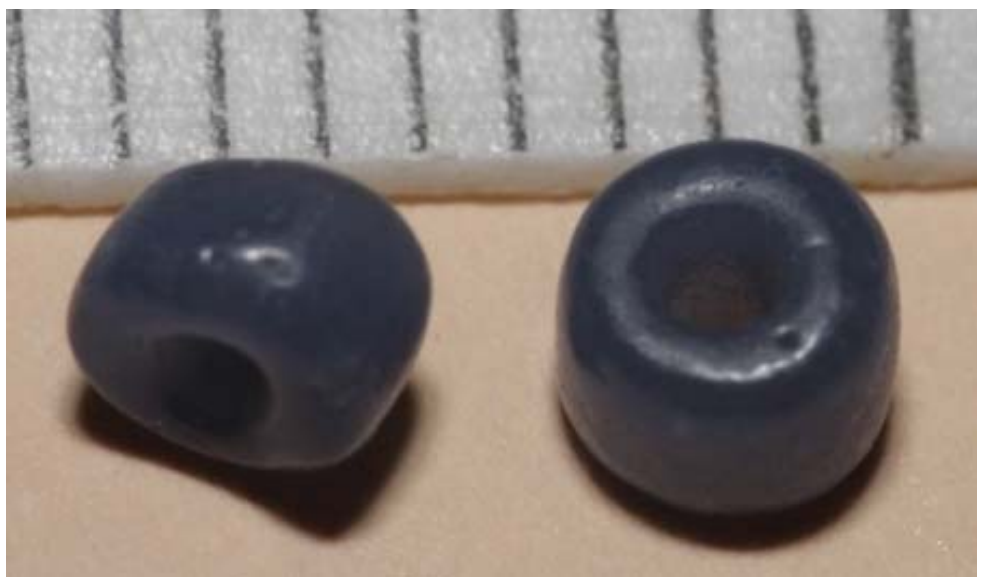

Accession:
**C1846**
Box C0055

Map. Hill, Southern Terrace

(~153)

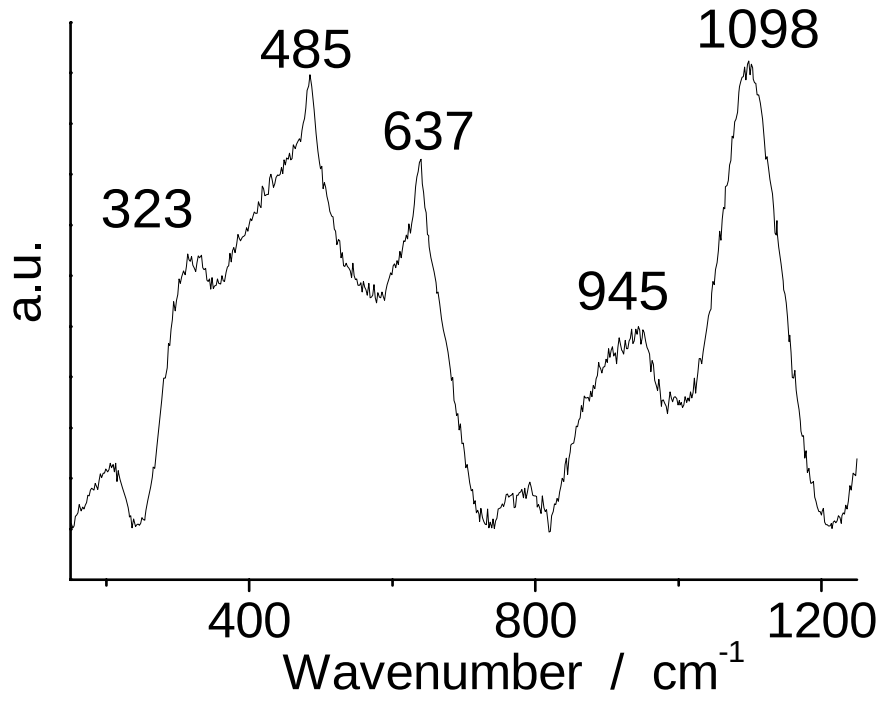

$Na_2O$

$Ca_2Sb_2O_7$
European
17-18th c.

Cu
Cobalt

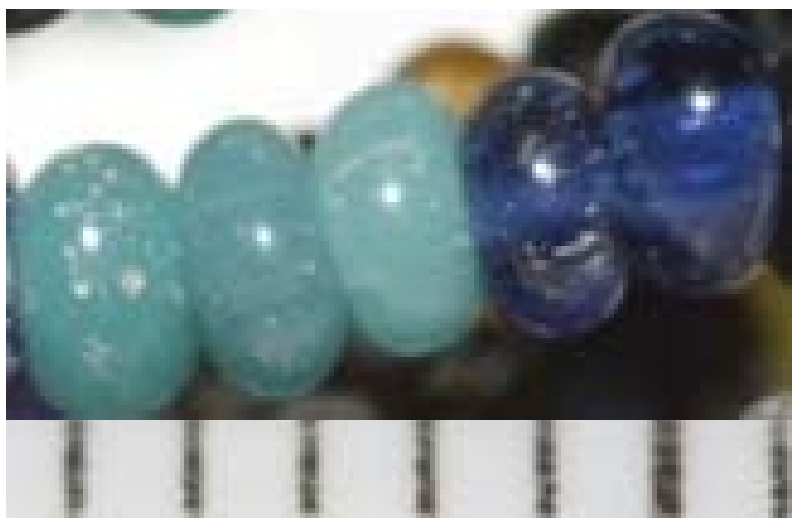

Accession:
**C934**
Box C0047

K2

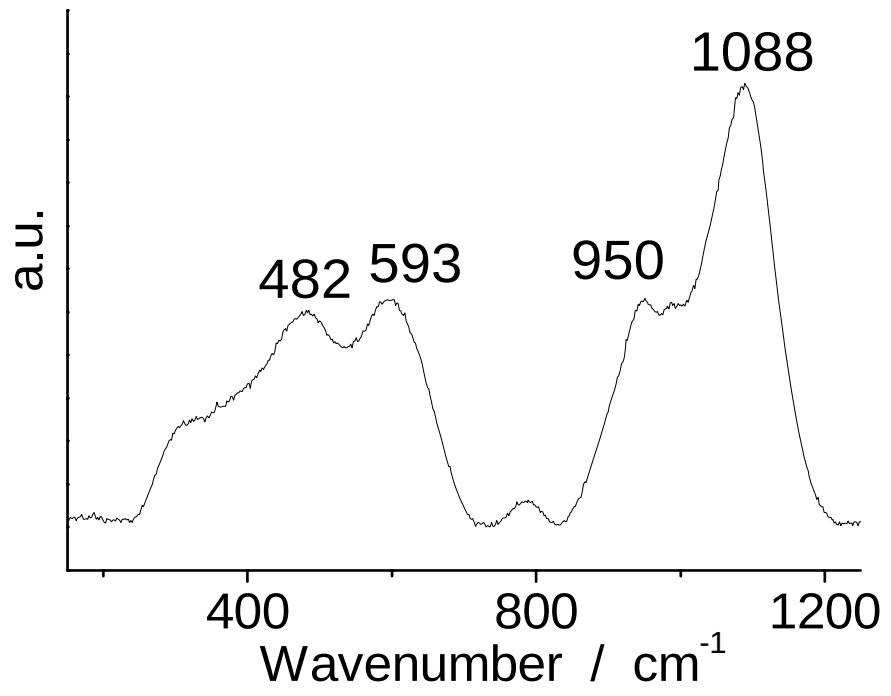

$Na_2O/CaO$

Cobalt for dark blue, Copper for turquoise (these are Mapungubwe oblates)

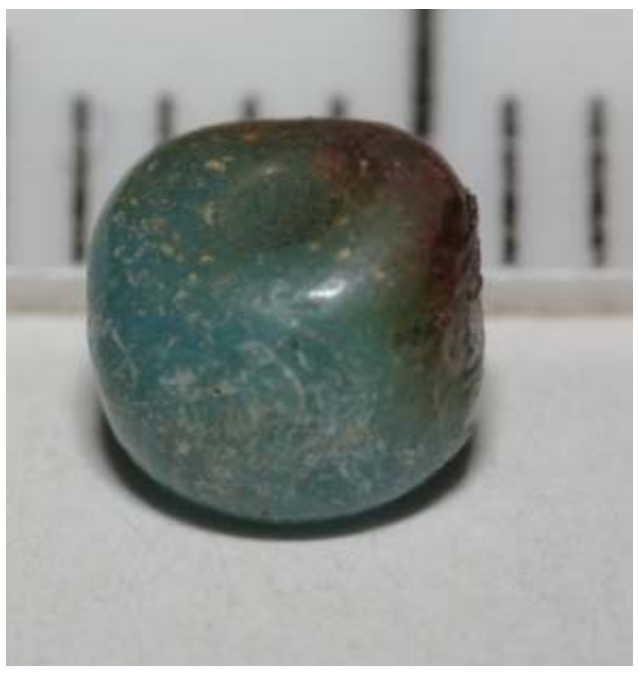

Accession:
**C/725**
Box 0036

Map. Hill

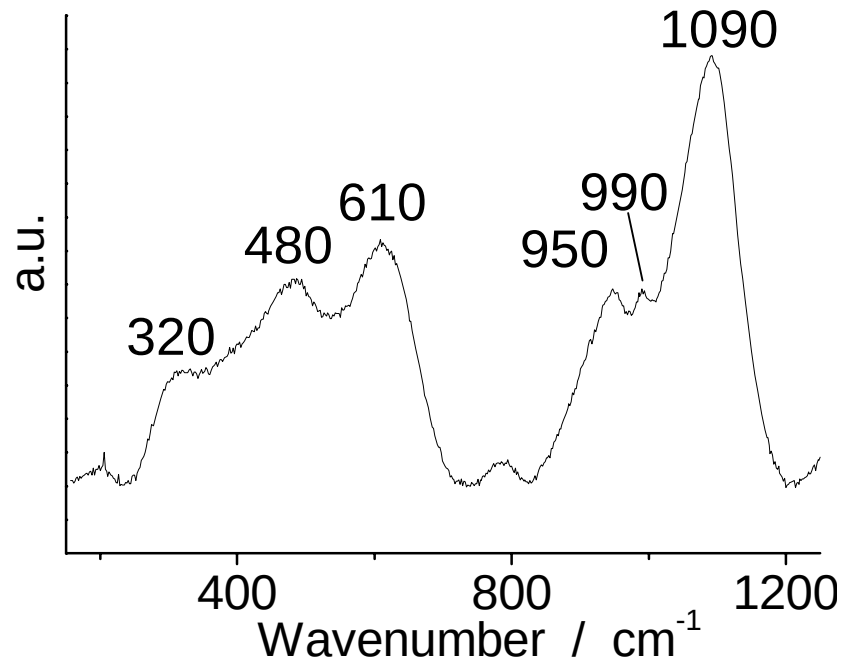

$Na_2O/CaO$

Copper

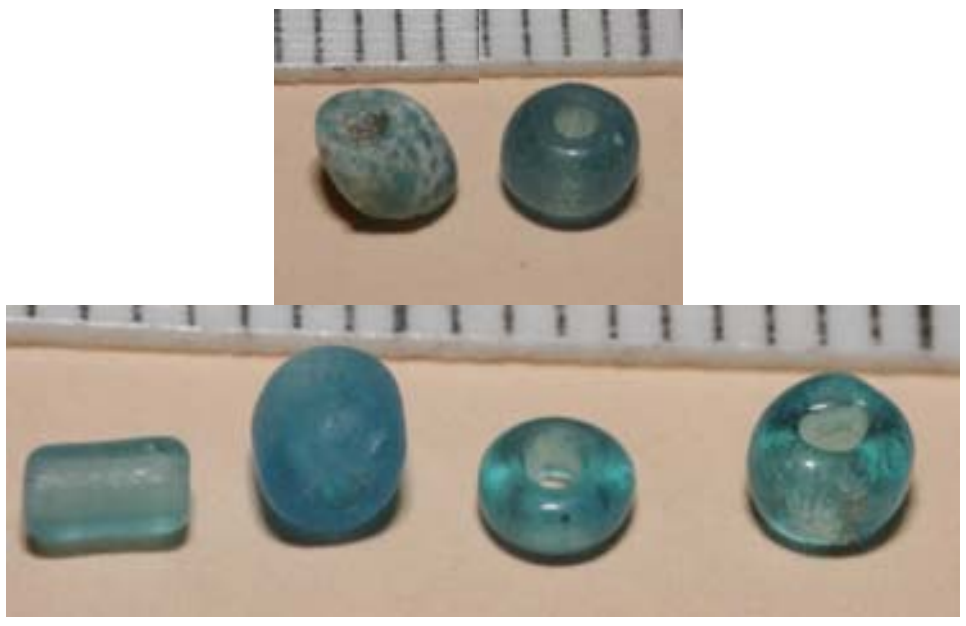

Accession:
**C1855**
Box C0055

K2

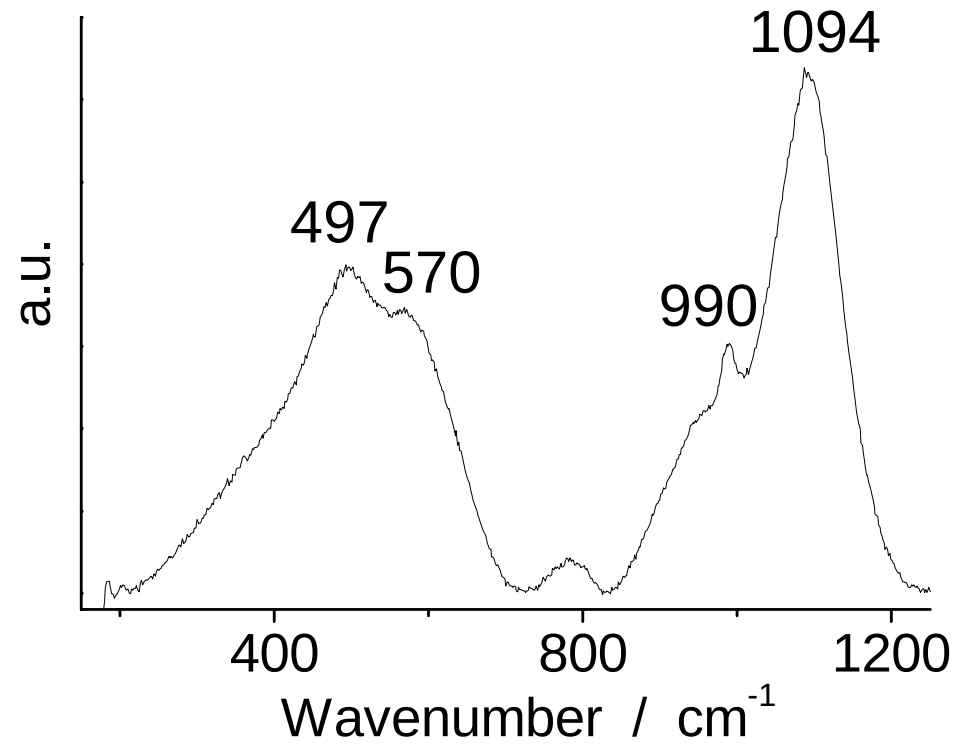

$Na_2O$

Cu

* Infinity

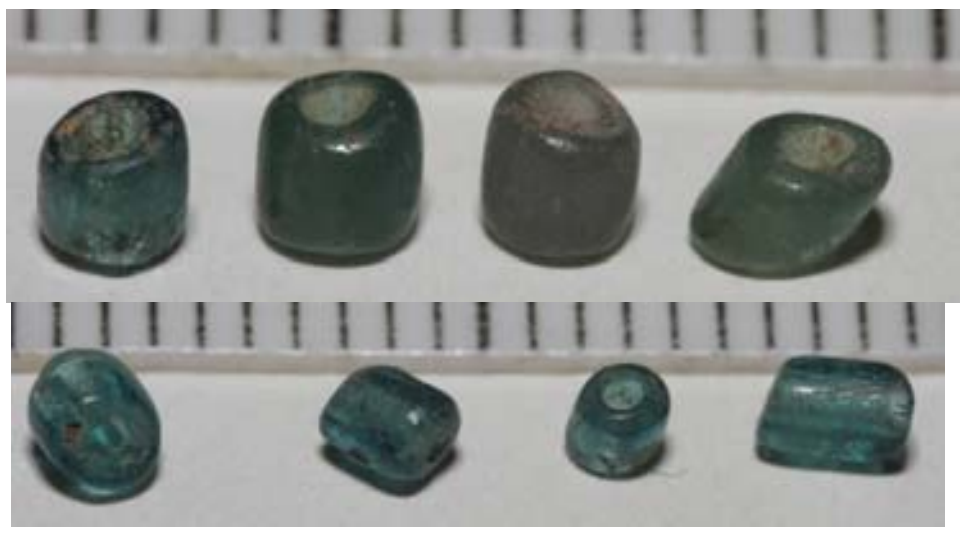

Accession:
**C198**
Box C0035

K2

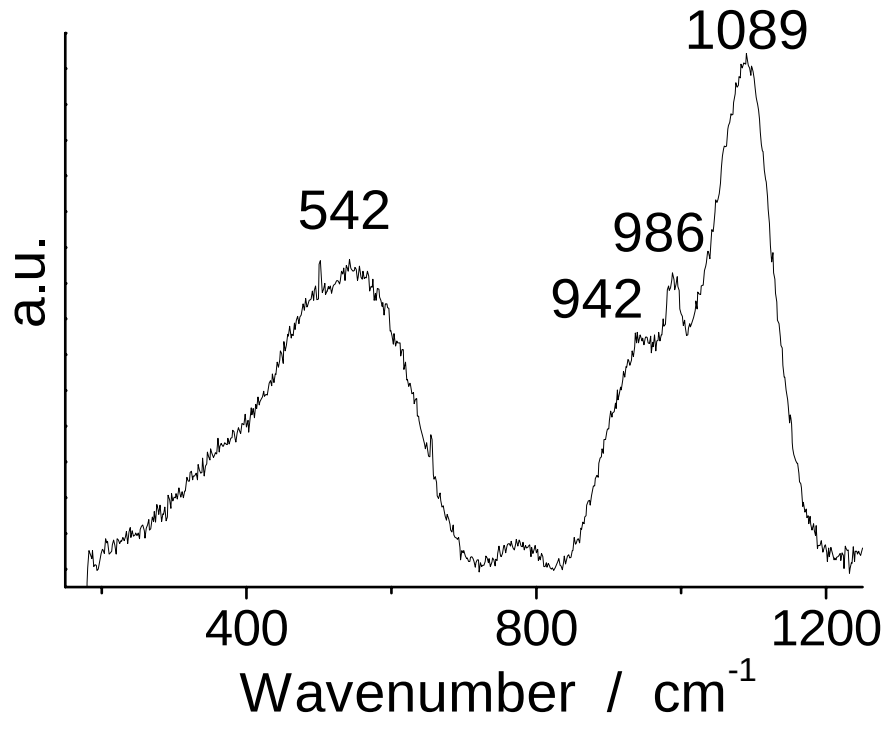

$Na_2O$

Cu

* Infinity

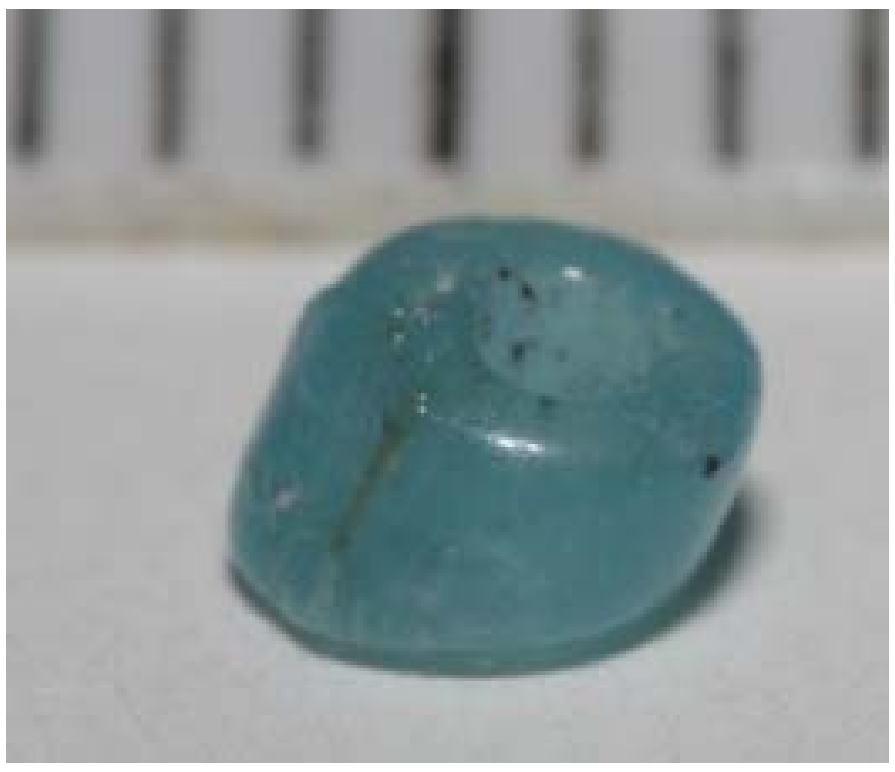

Accession:
**C1873**
Box C0050

K2

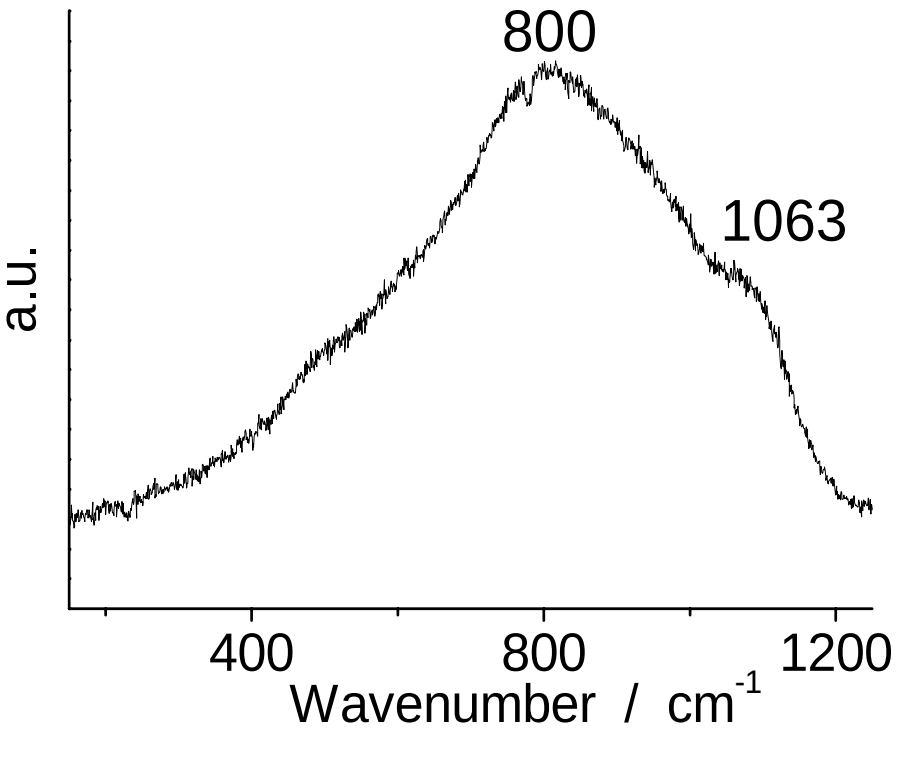

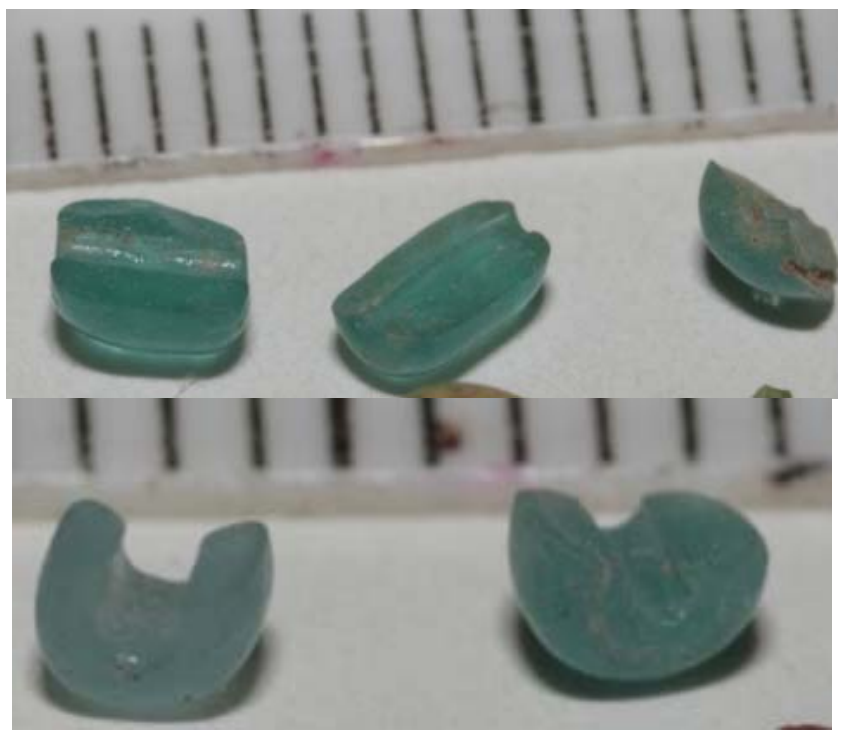

Accession:
**C1515**
Box C0054

Southern
Terrace

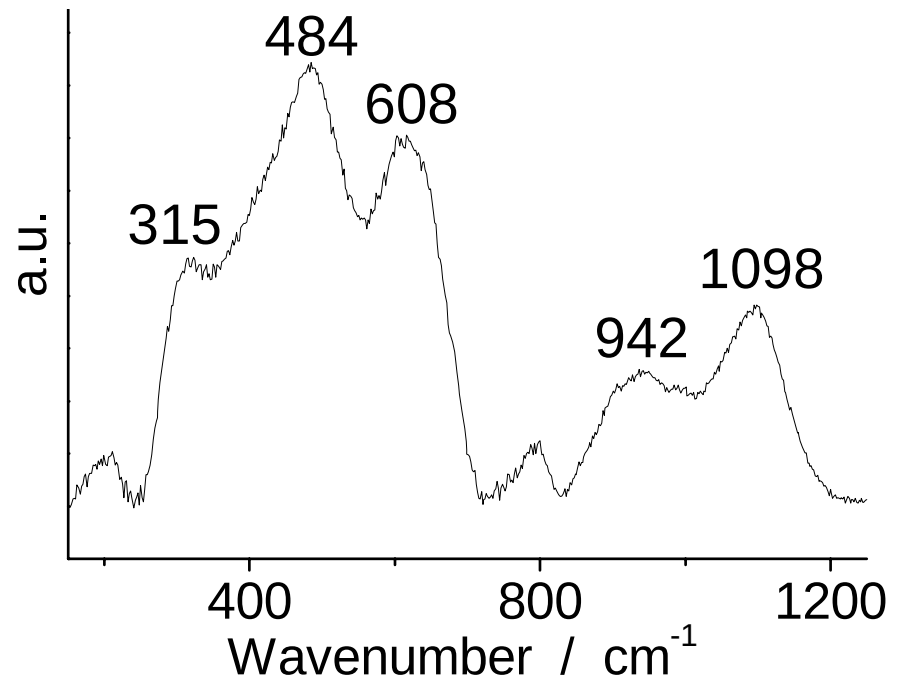

$Na_2O$

Cu

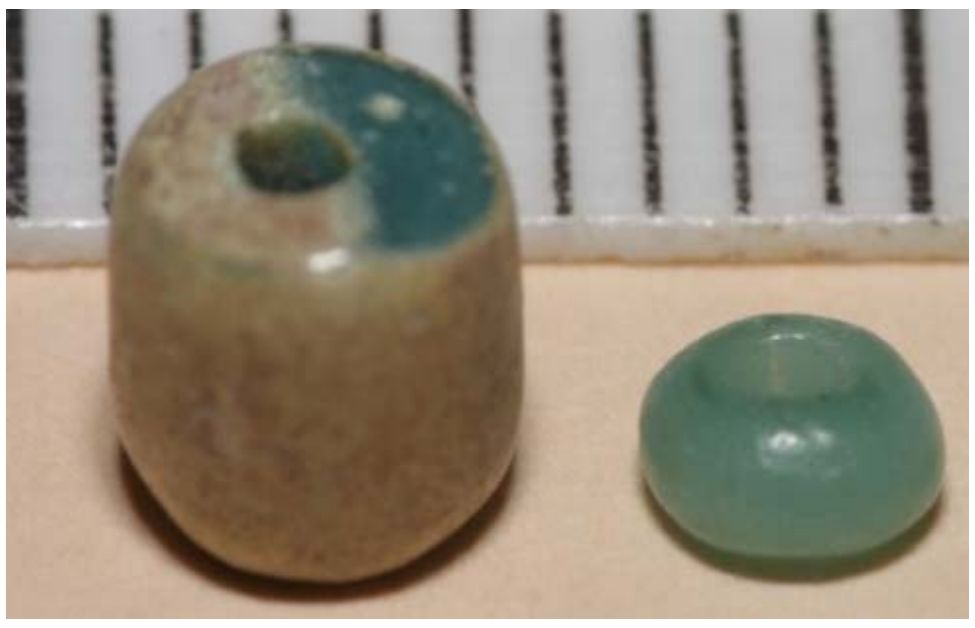

Accession:
**C1497**
Box C0053

Southern
Terrace

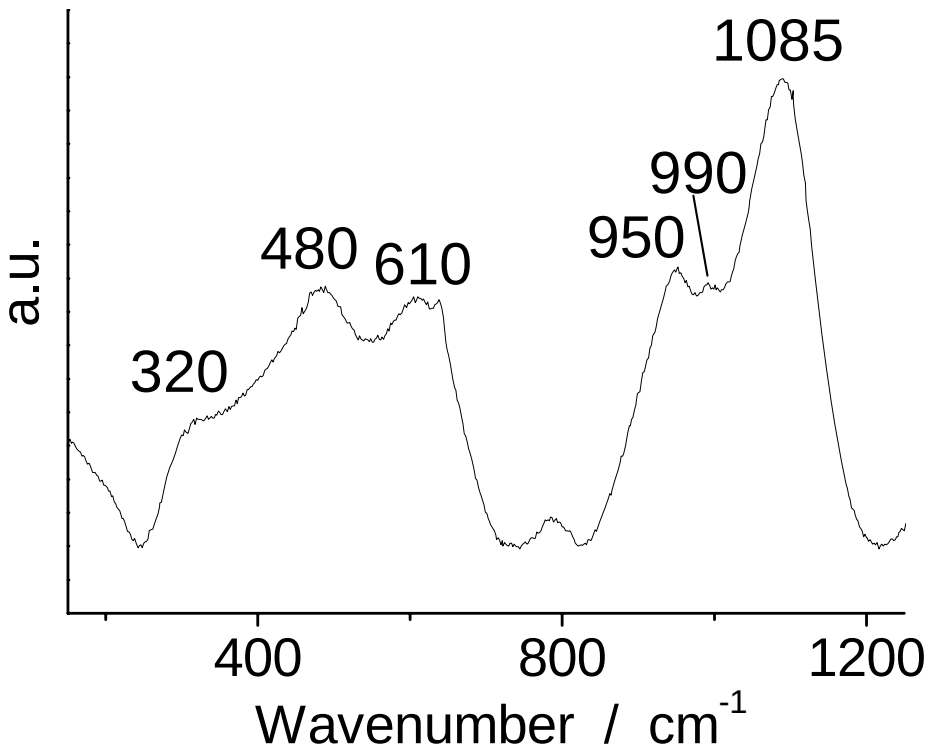

$Na_2O$

Cu

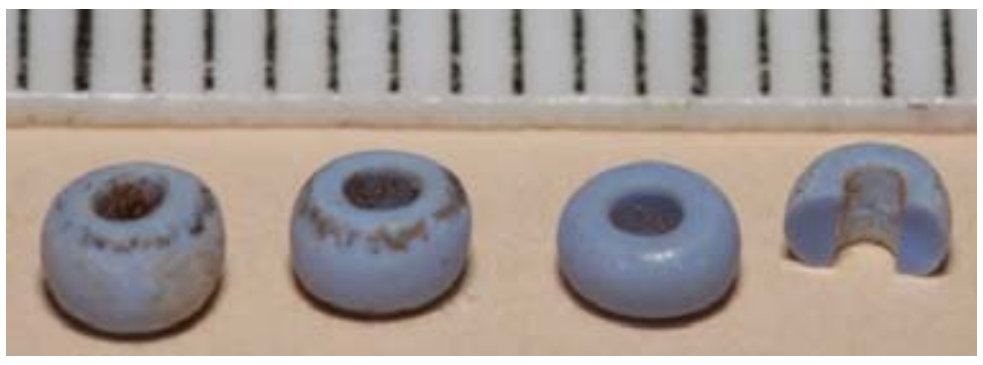

Accession:
**C743**
Box C0038

K2

(~400)

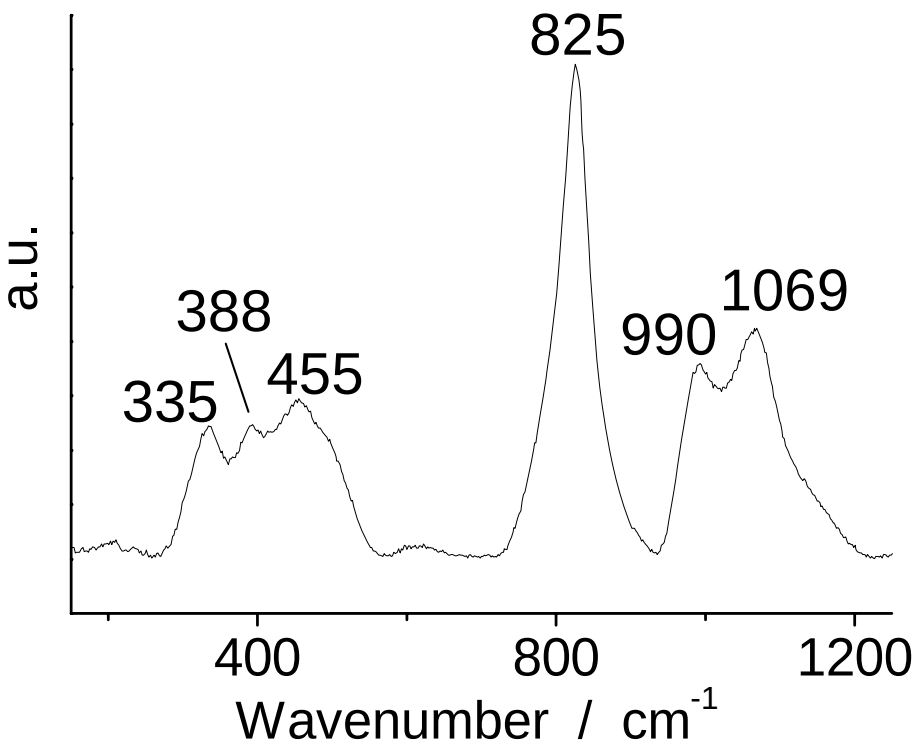

$Na_2O$

Ca and Pb
Arsenate
16th c.

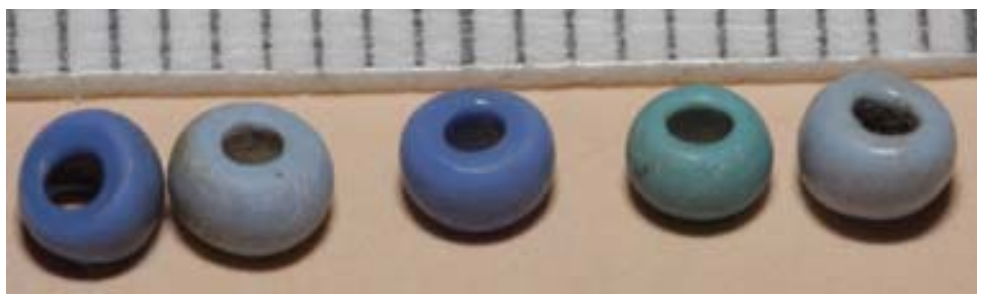

Accession:
**C1550**
Box C0054

K2

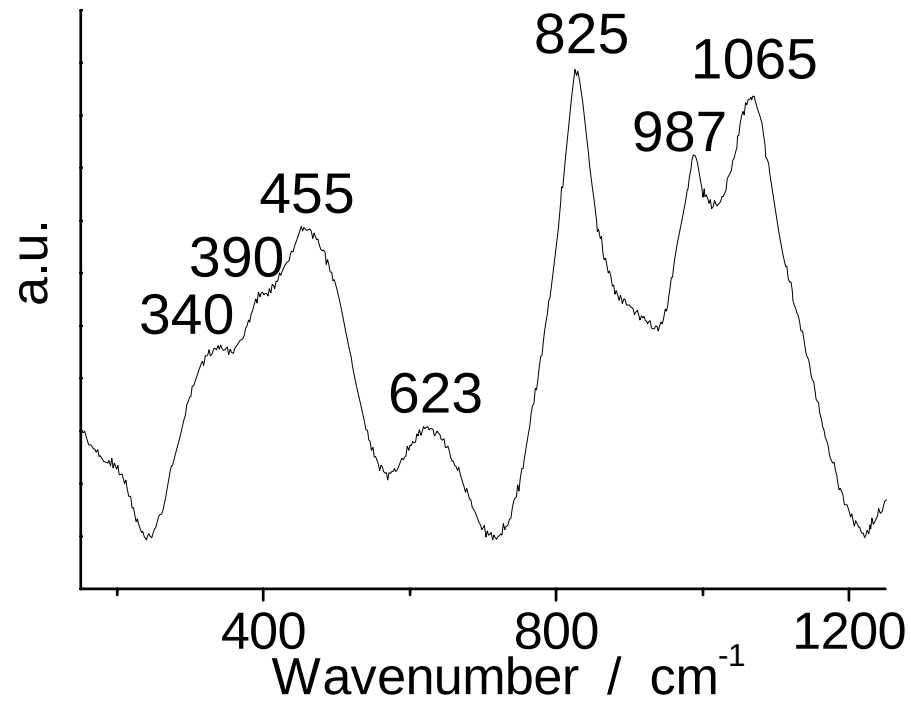

$Na_2O$

Ca and Pb
Arsenate
16th c.

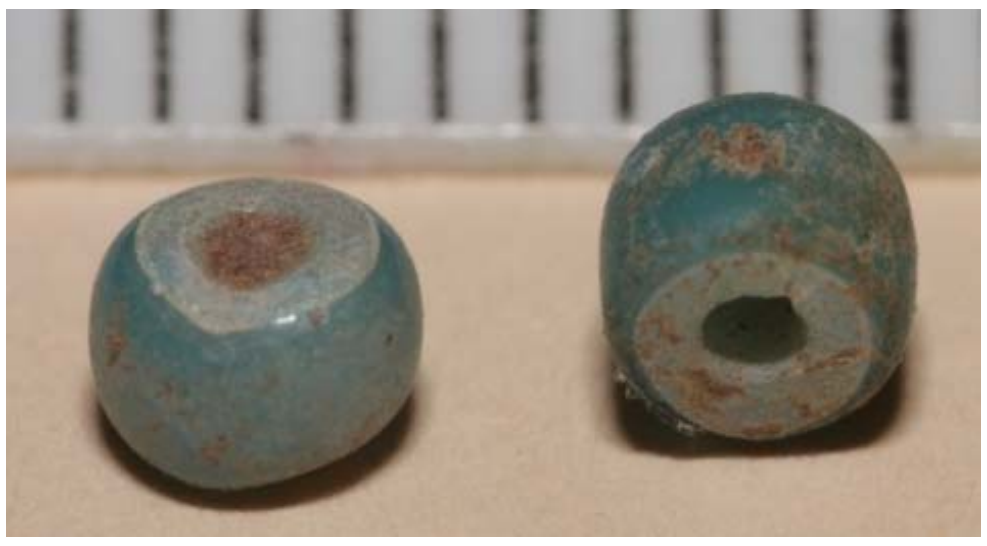

Accession:
**C723**
Box C0036

K2

(~10)

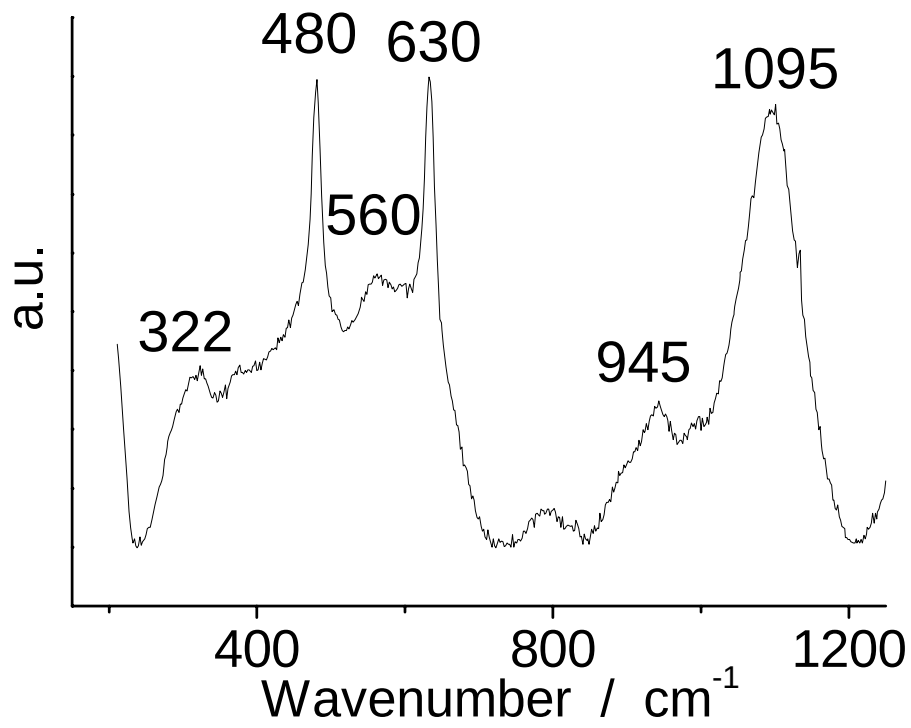

$Na_2O/CaO$

$Ca_2Sb_2O_7$
European
17-18th c

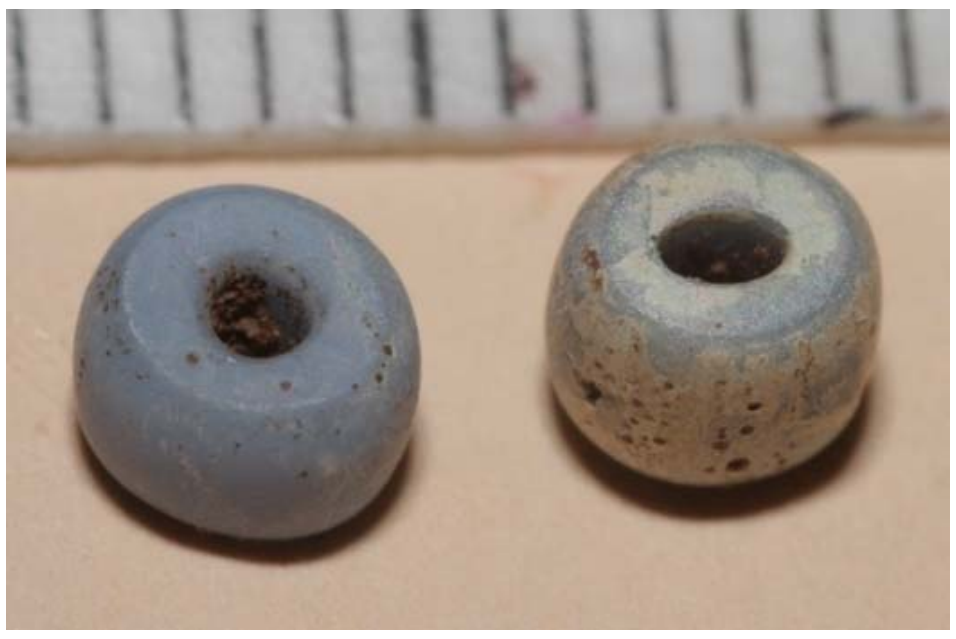

Accession:
**C745**
Box C0038

K2

(~250)

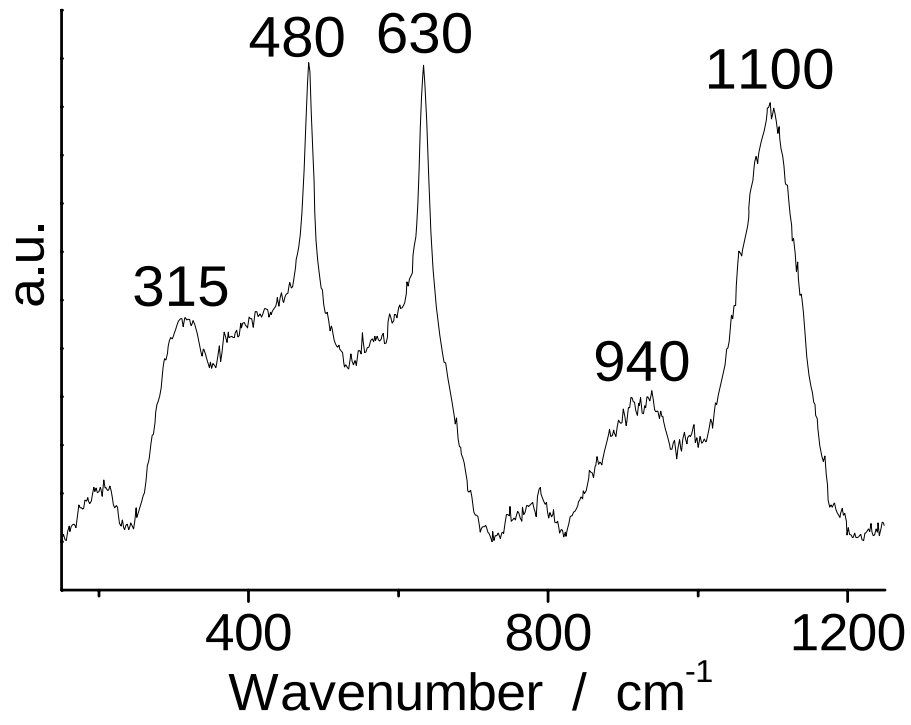

$Ca_2Sb_2O_7$
European
17-18th c.

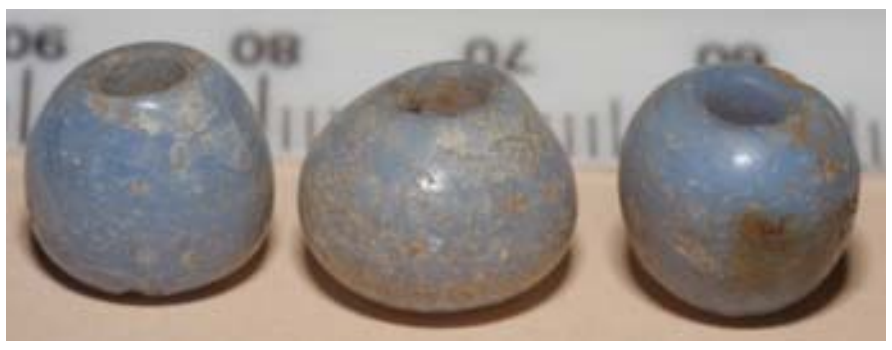

Accession:
**C731**
Box C0037

Map. Hill

(~50)

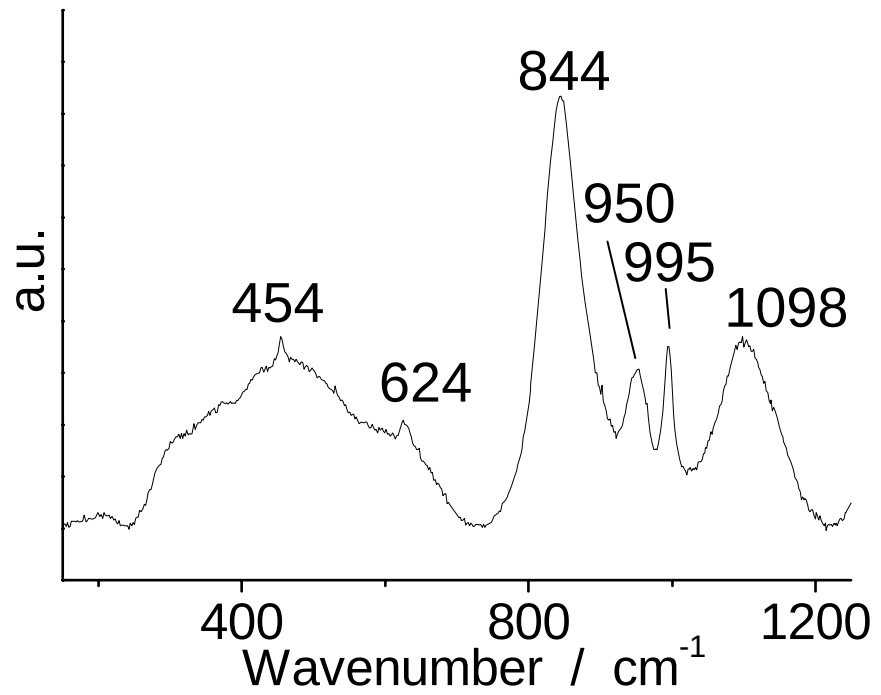

$CrO_4$ ions

$Ca_2Sb_2O_7$
European
17-18e c.

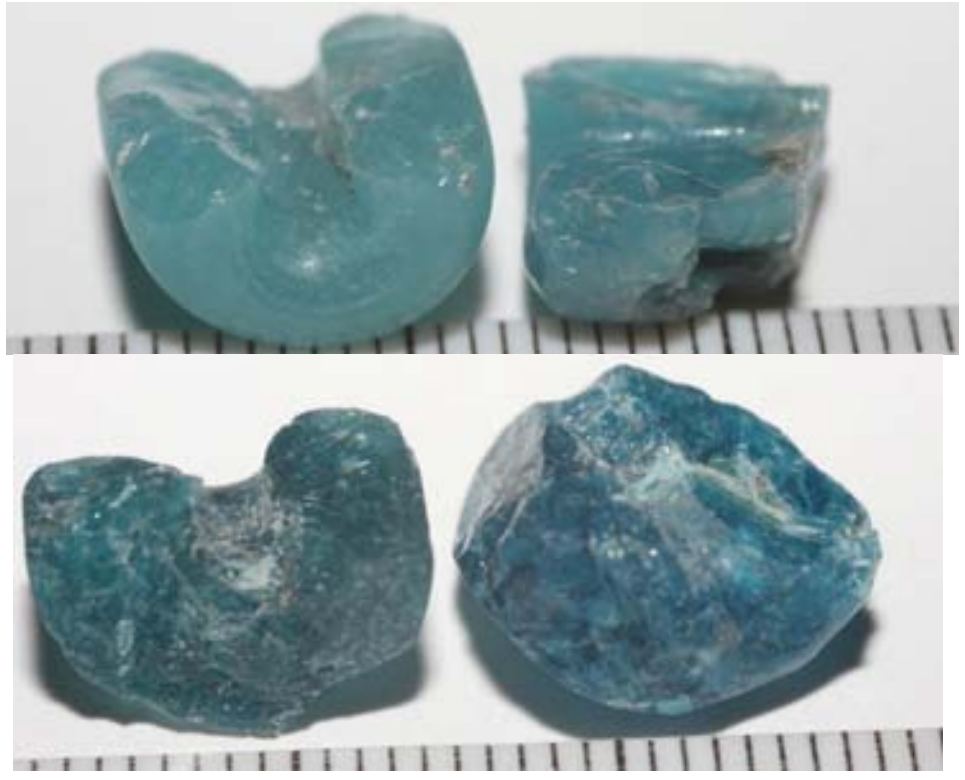

Accession:
**C765**
Box C0039

Map. Hill

*Garden roller*

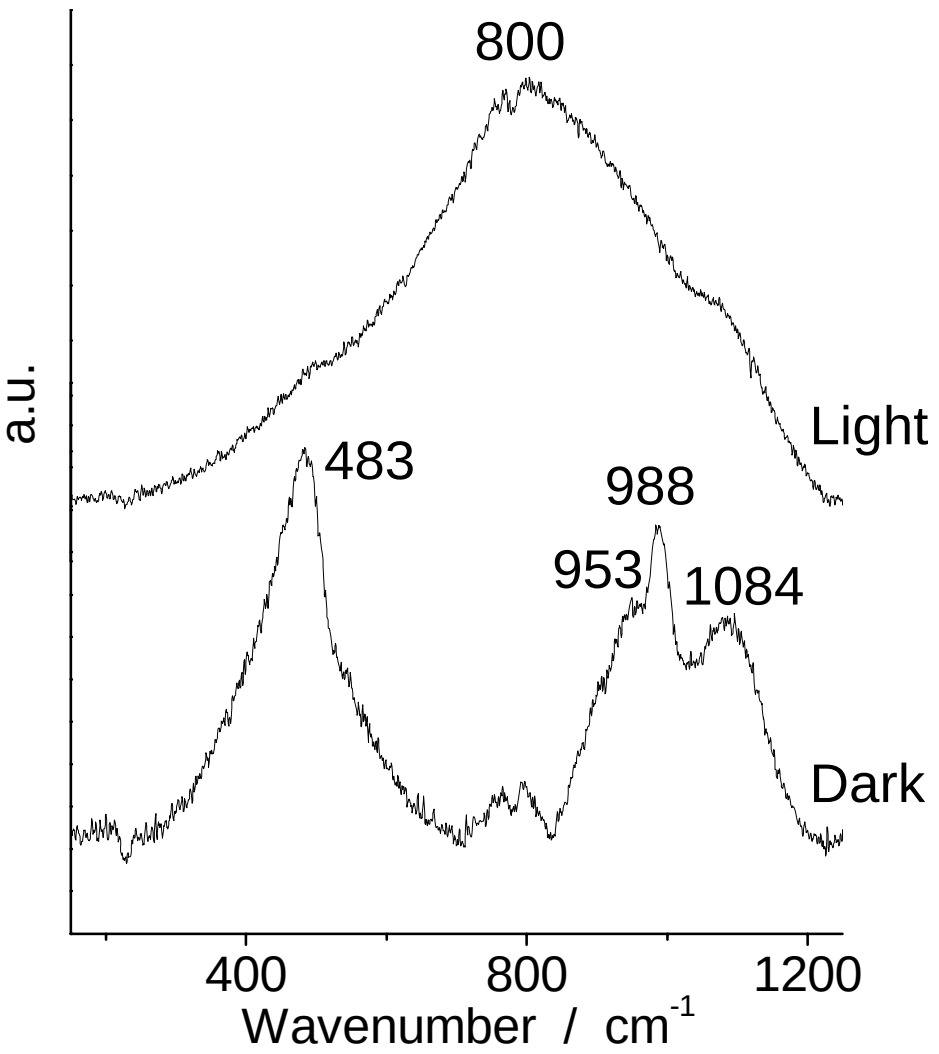

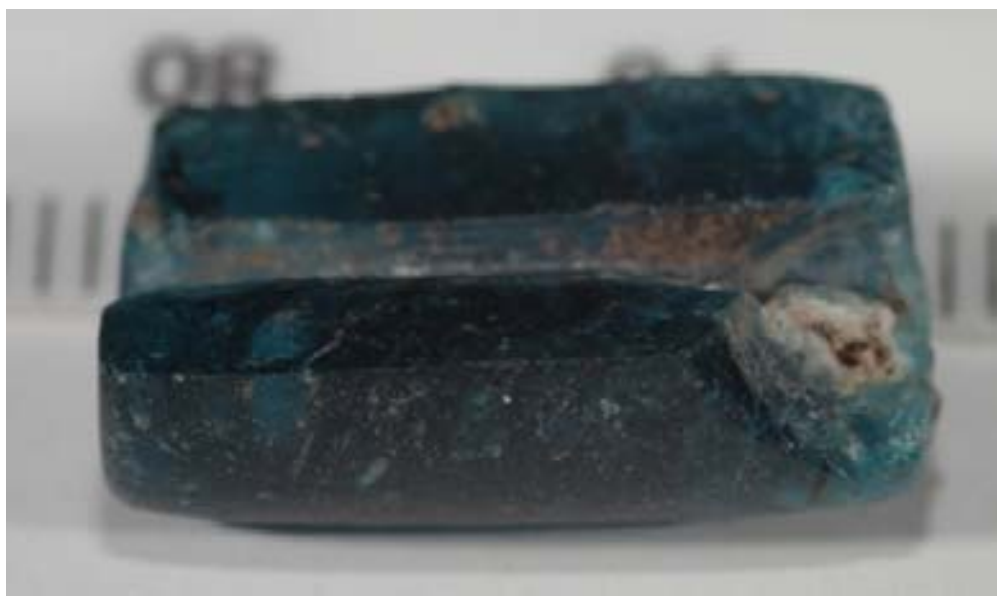

Accession:
**C/335**
Box 0035

Map. Hill

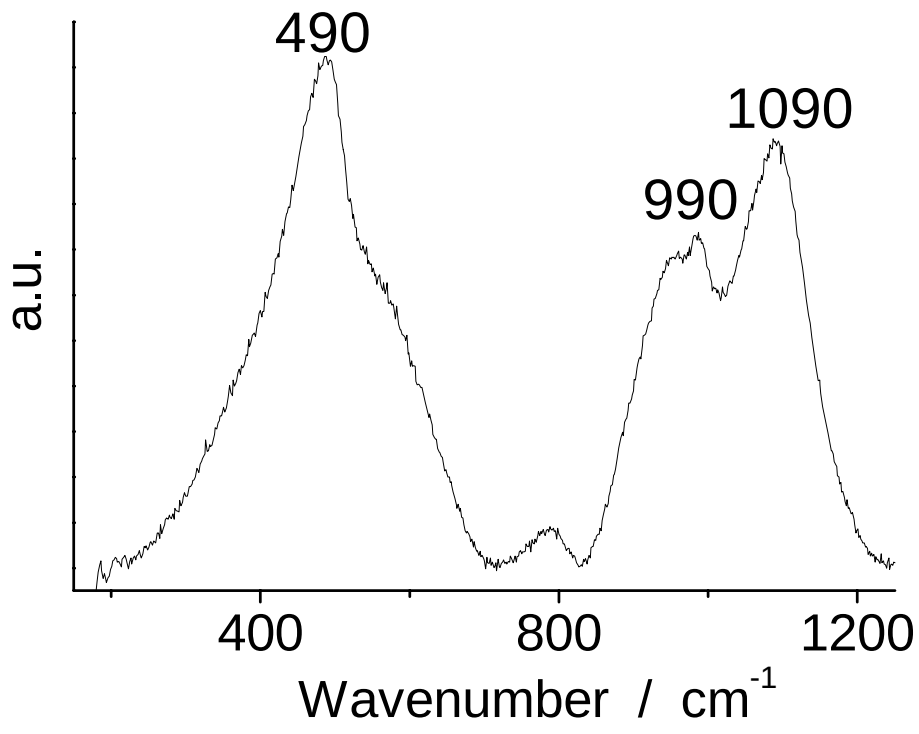

$Na_2O/CaO$

Cu

* Infinity

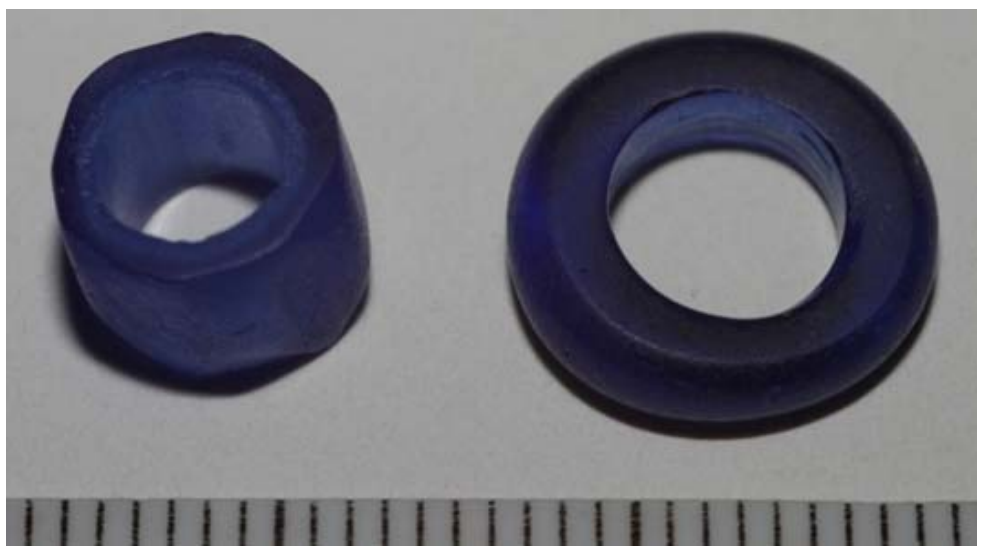

A B

Accession:
**C1581**
Box C0054

K2

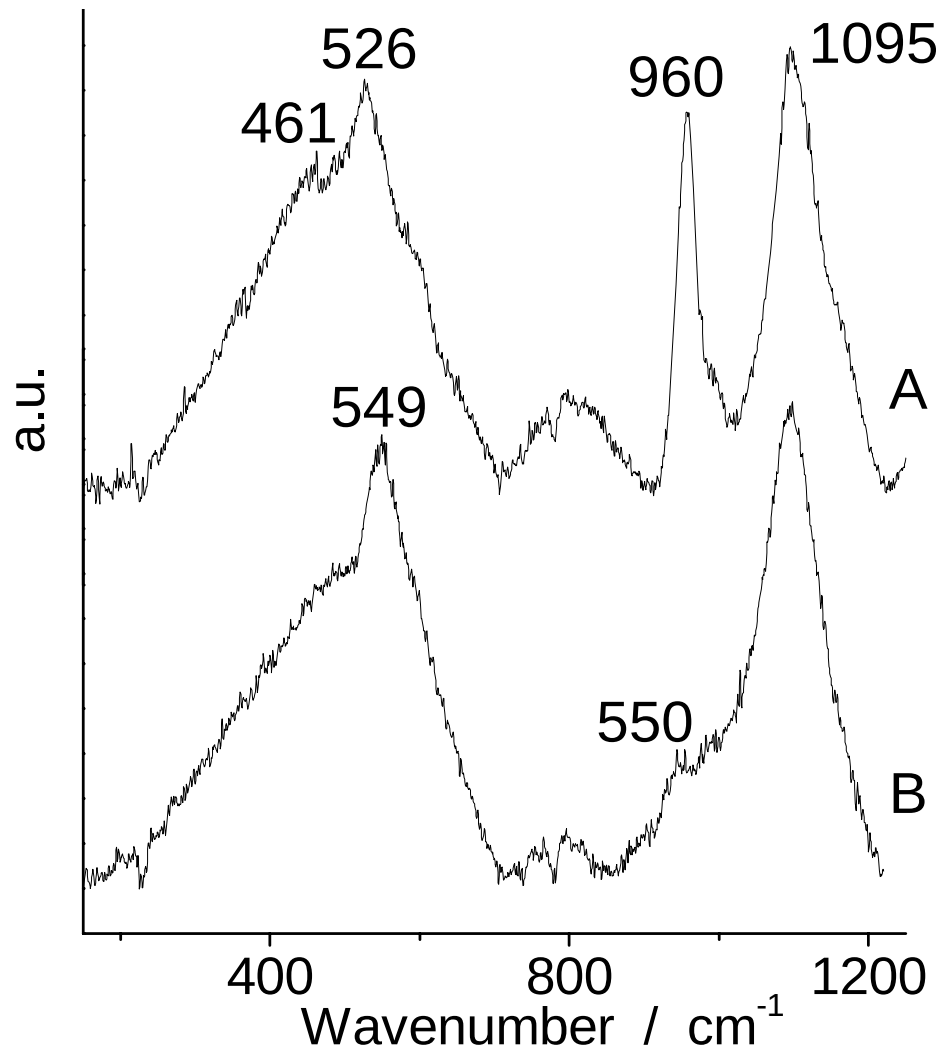

Calcium phosphate

* T64000

## Figure 2: RED BEADS

| BEADS (scale= 1mm) | REF. MUSEUM (quantity) | RAMAN SPECTRA (after baseline subtraction, a.u. = arbitrary unit) | GLASS MATRIX PIGMENT |
|---|---|---|---|

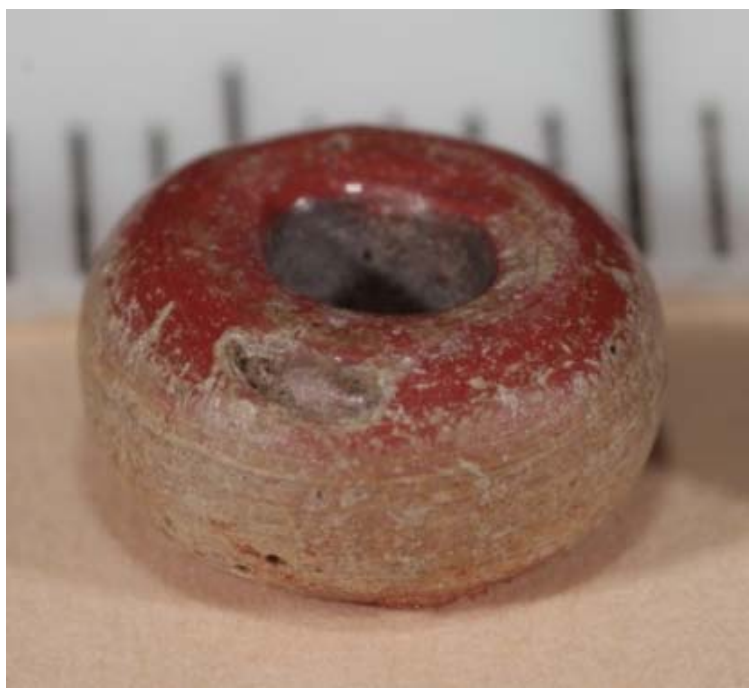

Accession: **C/4354** Box C0241

Map. Hill

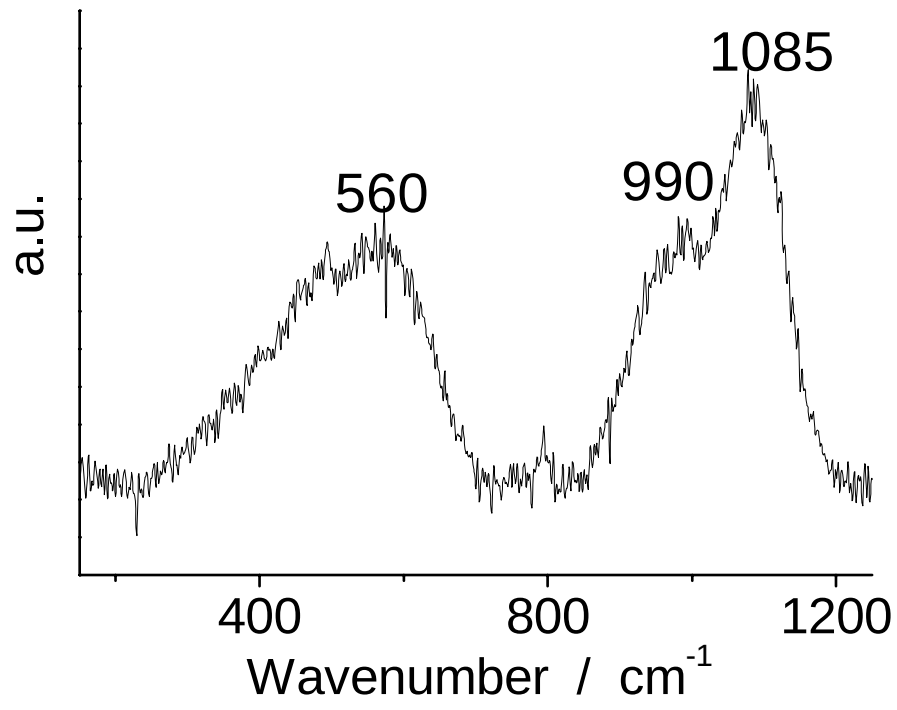

$Na_2O/CaO$

* T64000

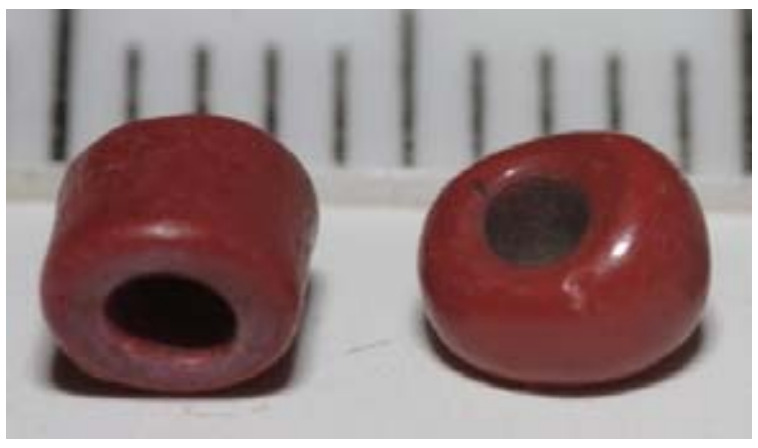

Accession: **C/762** Box 0041

Map. Hill

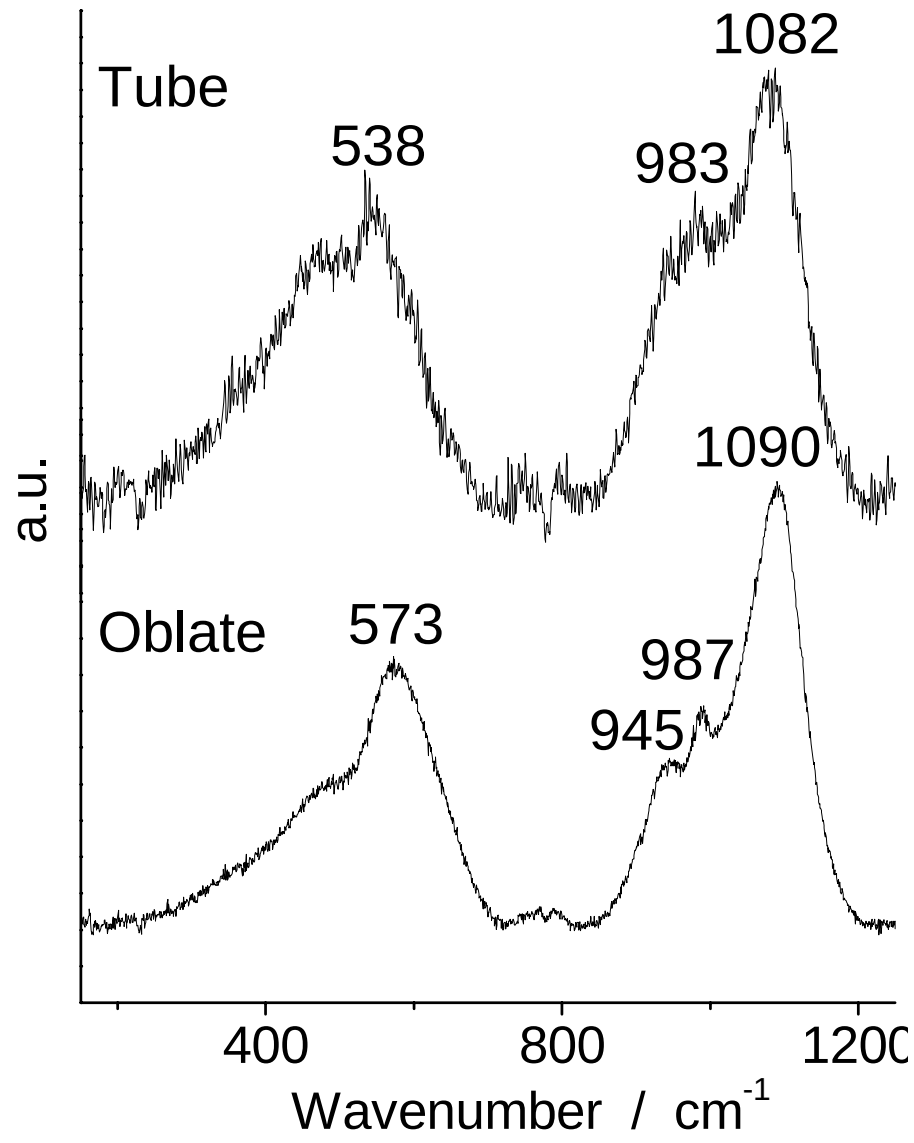

$Na_2O/CaO$

* T64000

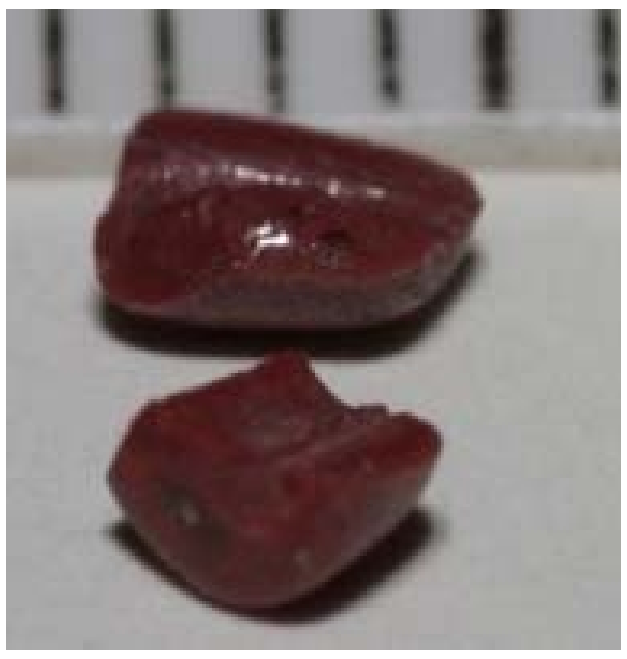

Accession: **C1515** Box C0054

Southern Terrace

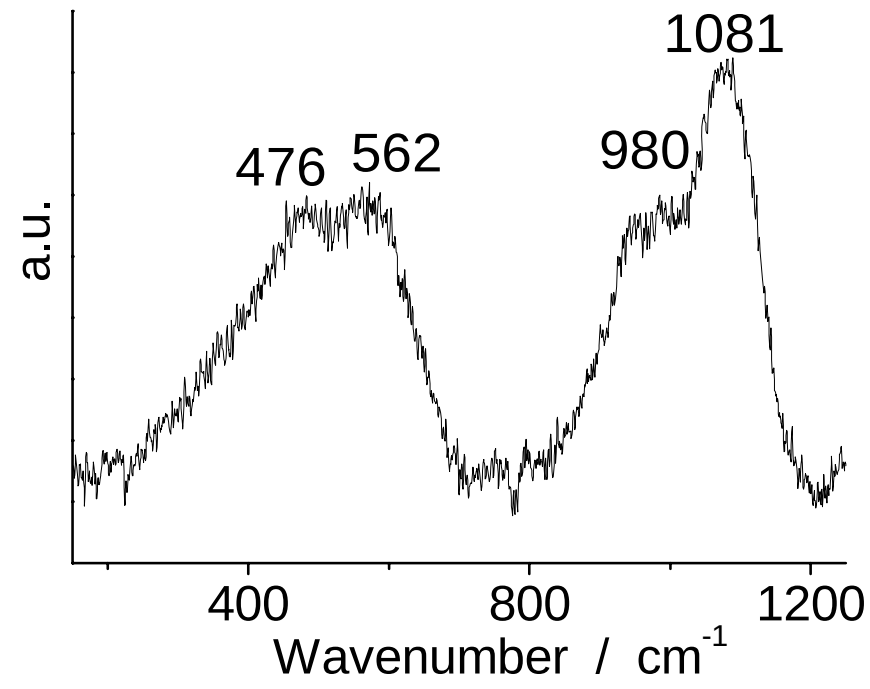

$Na_2O/CaO$

* T64000

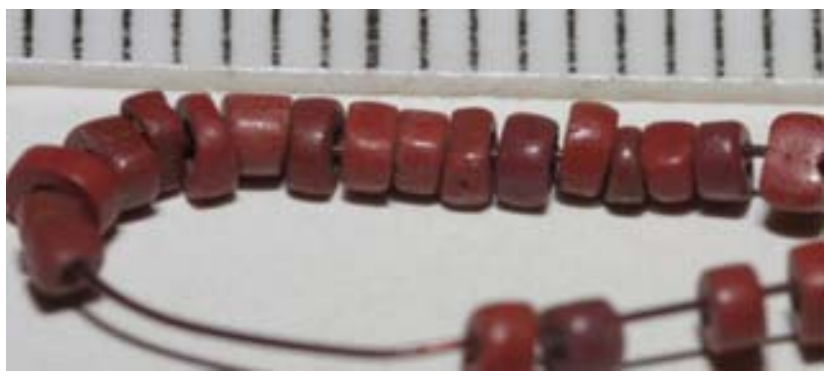

Accession: **C952** Box C0046

Map. Hill

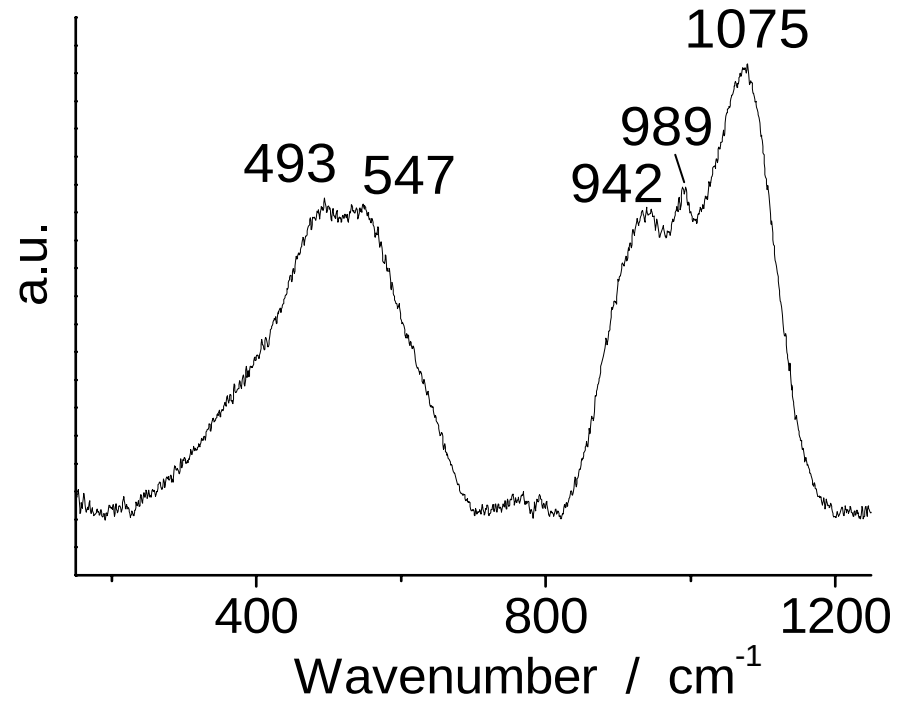

* T64000

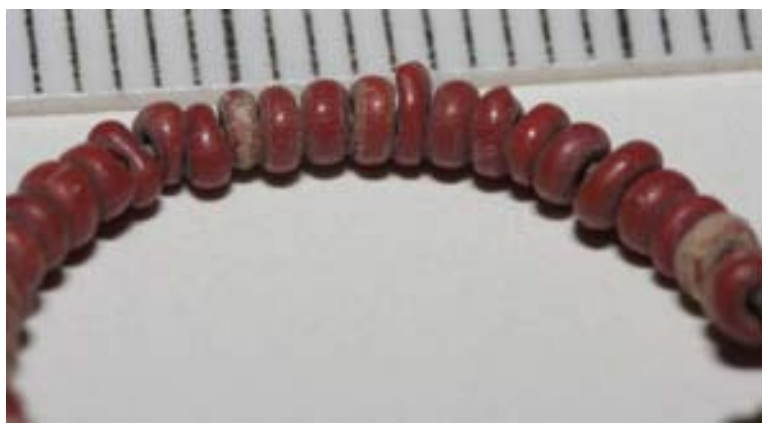

Accession:
**C952**
Box C0045

Map. Hill

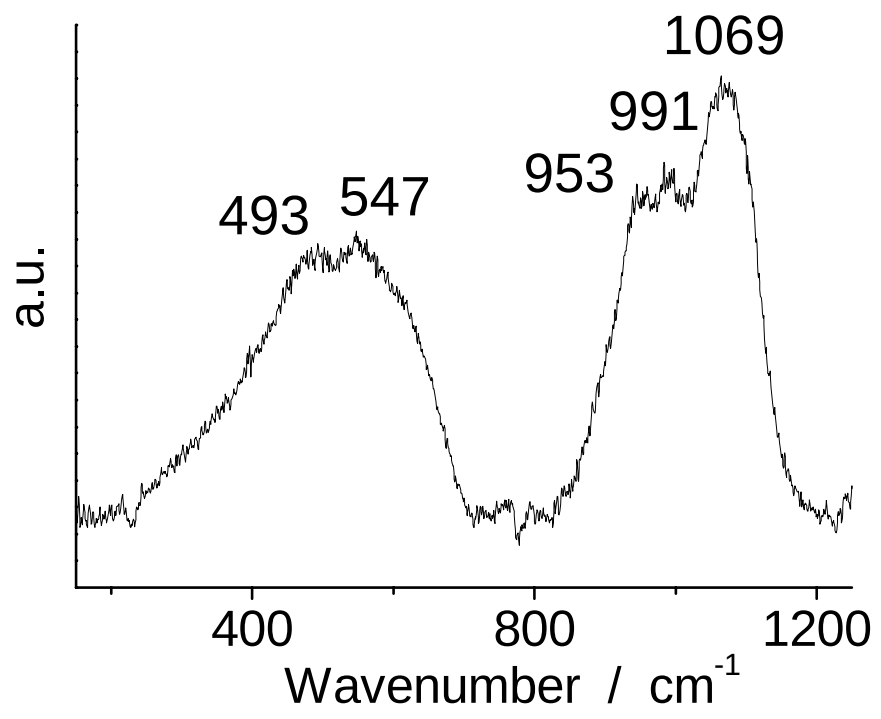

$Na_2O/CaO$

* T64000

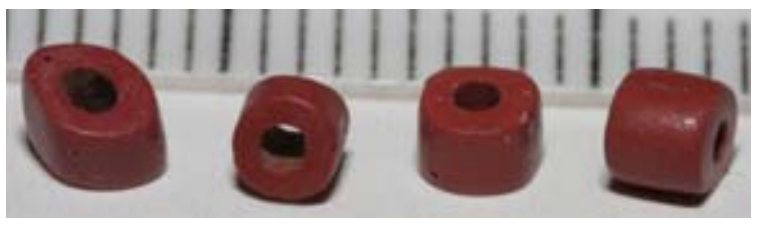

Accession:
**C952**
Box C0046

Map. Hill

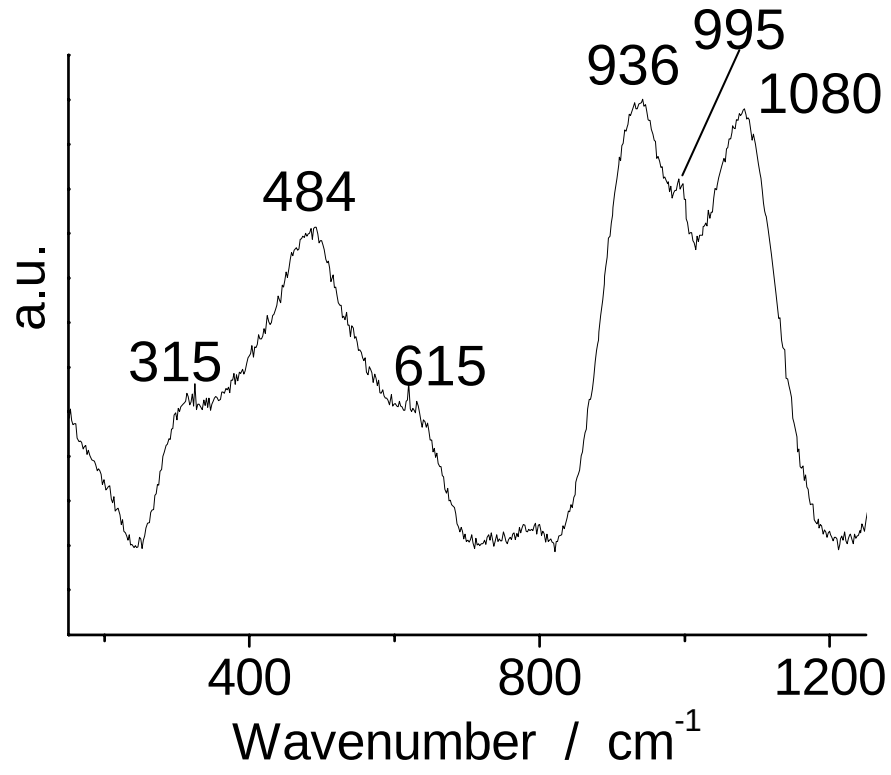

Cu

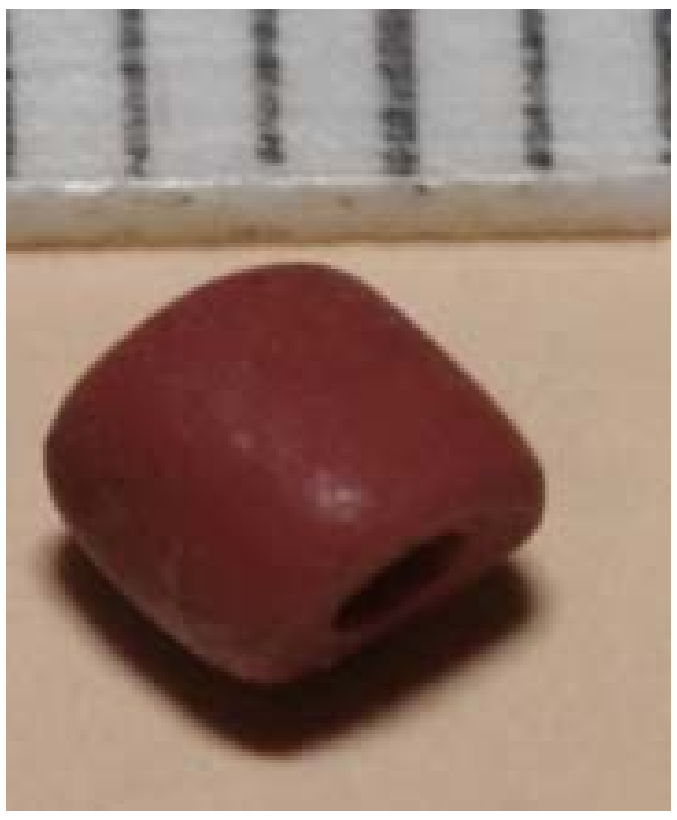

Accession:
**C1855**
Box C0055

K2

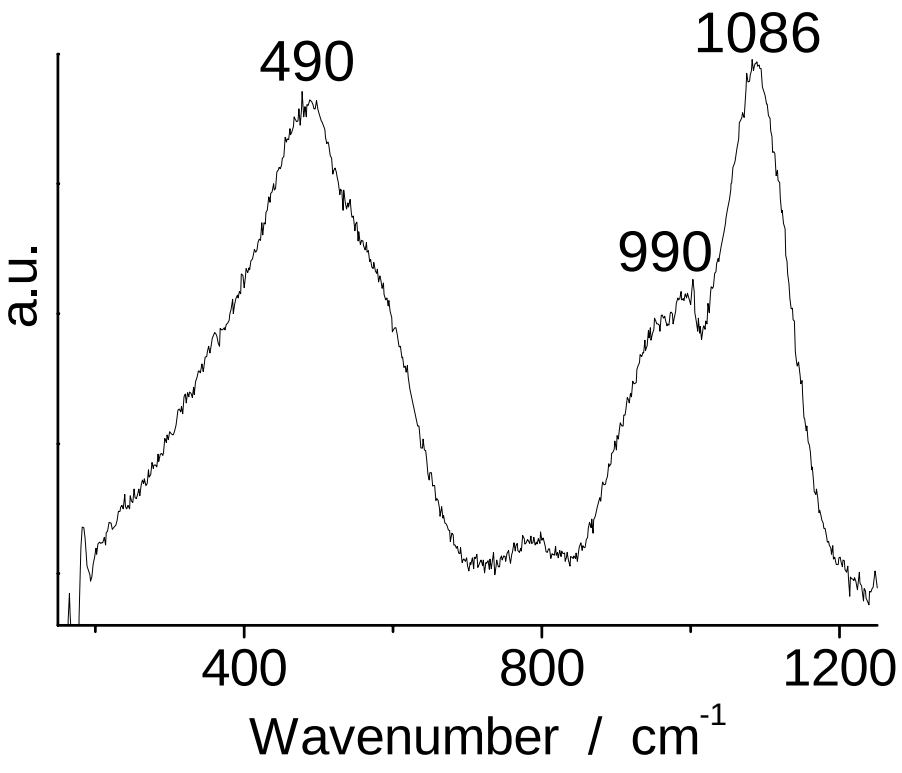

$Na_2O$

Cu

* Infinity

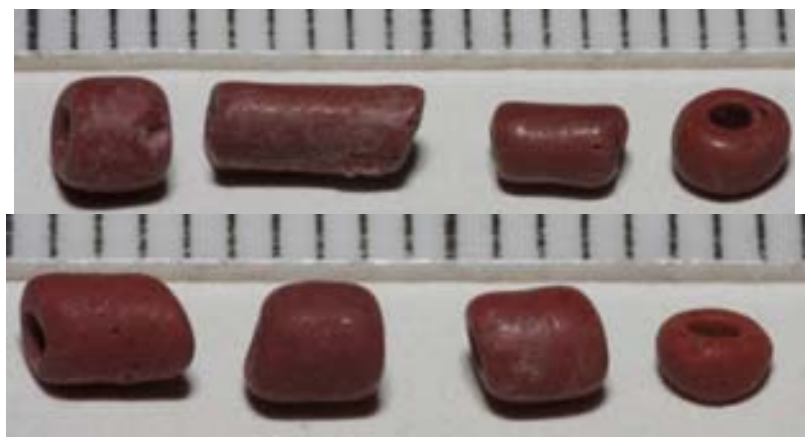

Accession:
**C1636**
Box C0055

K2

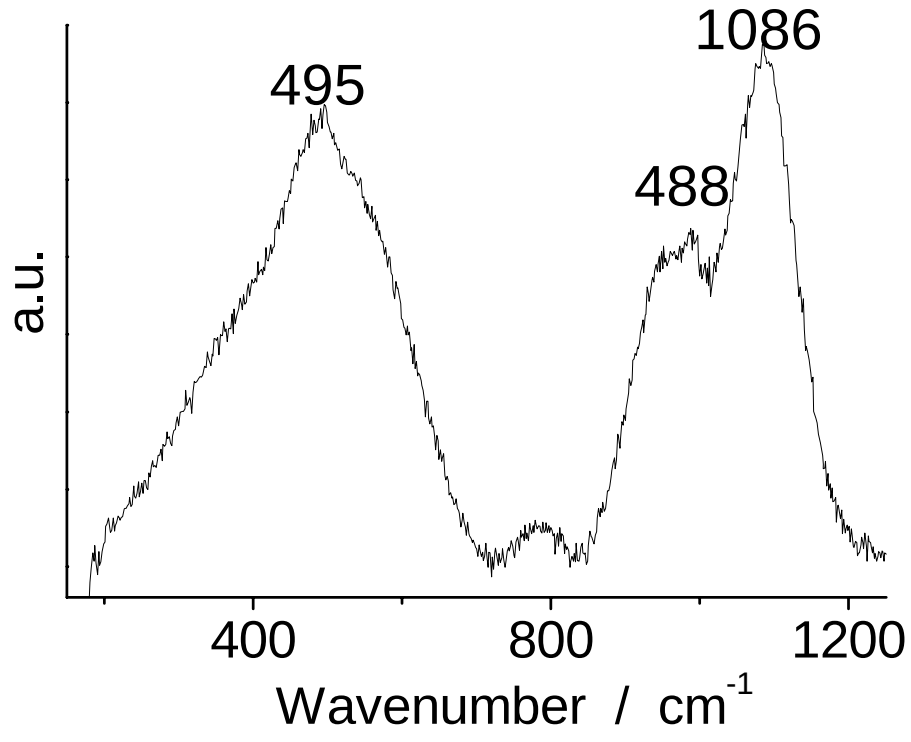

$Na_2O$

Cu

* Infinity

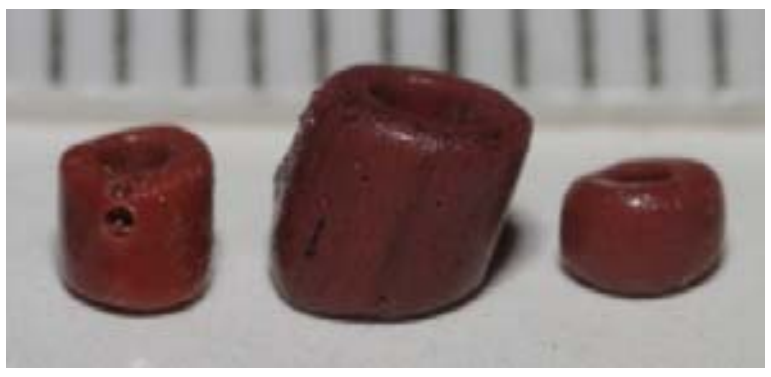

Accession:
**C1873**
Box C0050

K2

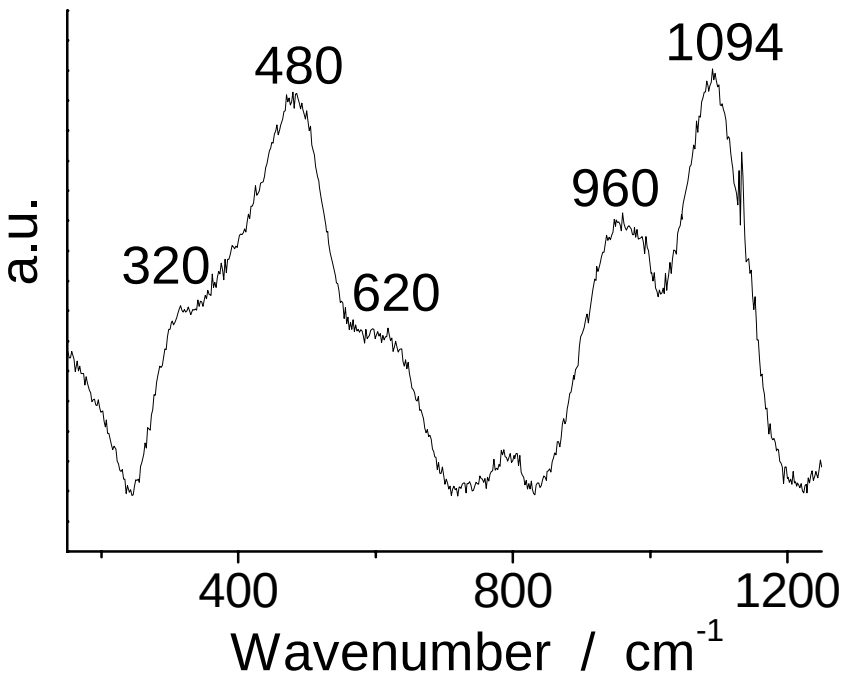

$Na_2O$

Cu

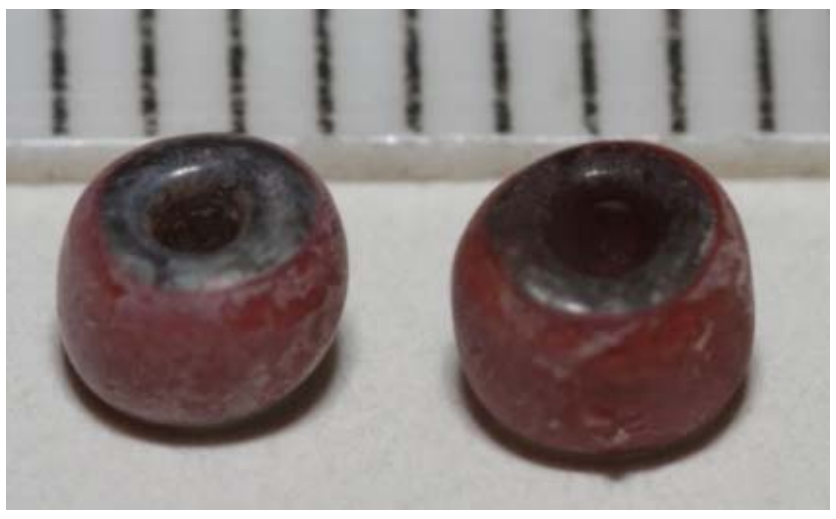

Accession:
**C952**
Box C0046

(~500-600)

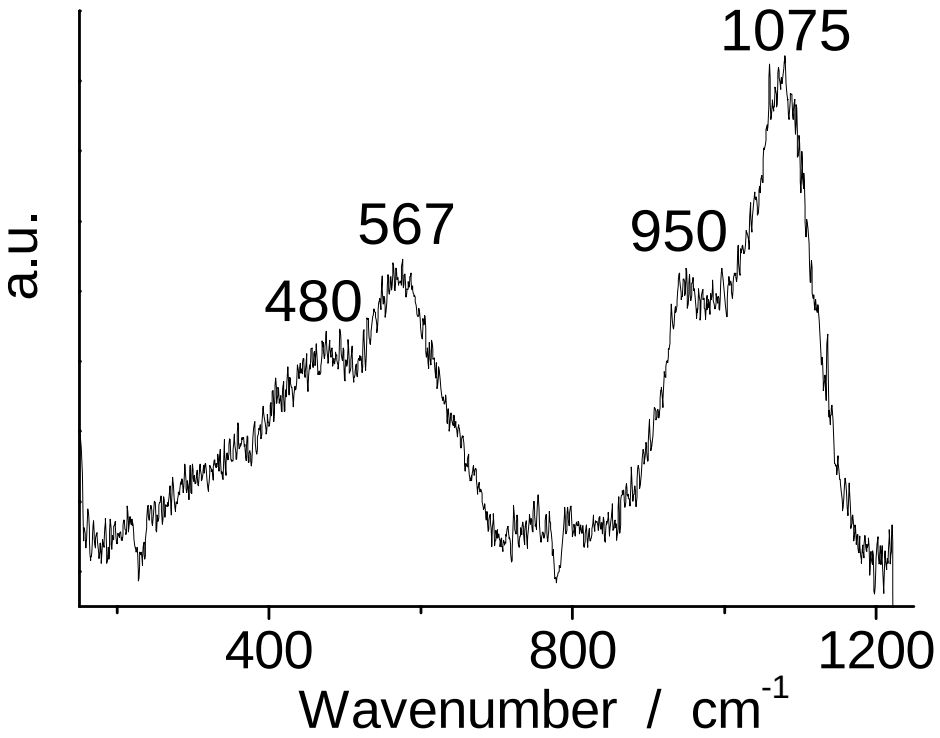

$Na_2O/CaO$

*Ext part*

* T64000

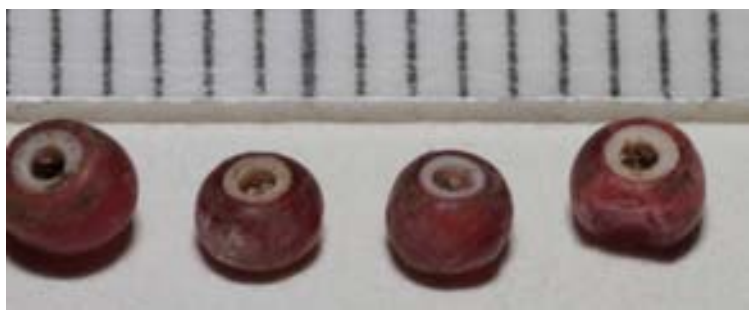

Accession:
**C732**
Box C0036

K2

(~50)

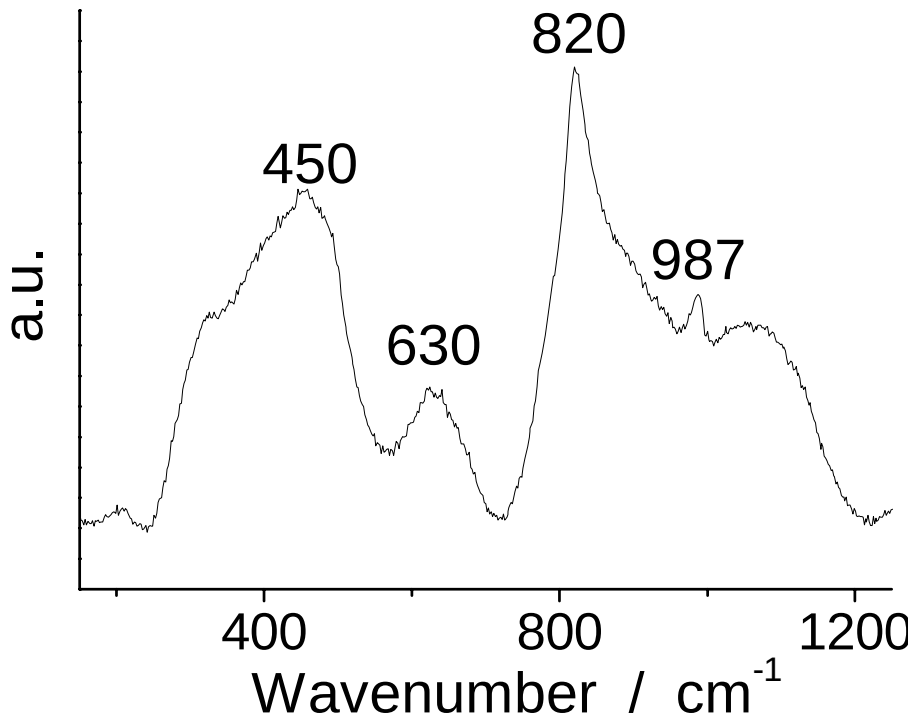

Ca and Pb
Arsenate
16th c.

*White part*

Red with
white base
= 1836

## Figure 3: YELLOW / ORANGE

| BEADS (scale= 1mm) | REF. MUSEUM (quantity) | RAMAN SPECTRA (after baseline subtraction, a.u. = arbitrary unit) | GLASS MATRIX PIGMENT |
|---|---|---|---|
| 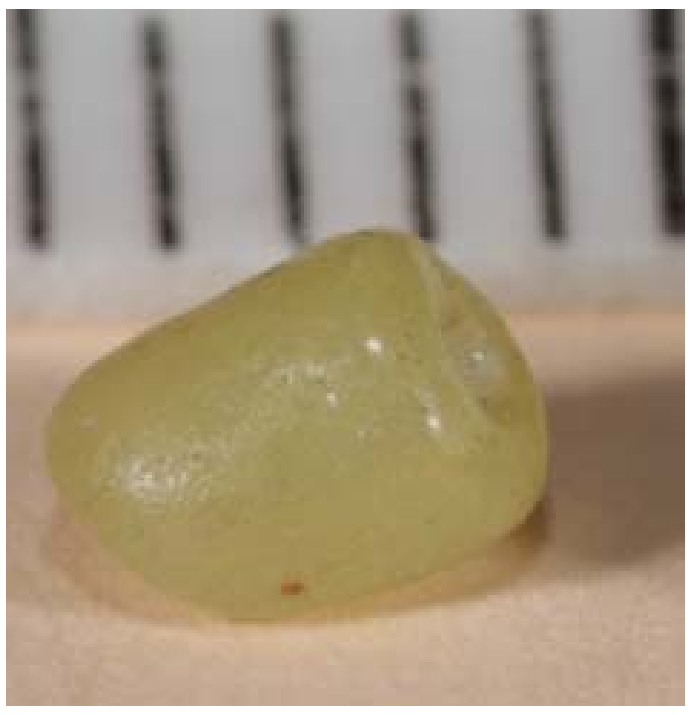 | Accession: **C/4354** Box C0241<br>Map. Hill<br>(~10) | 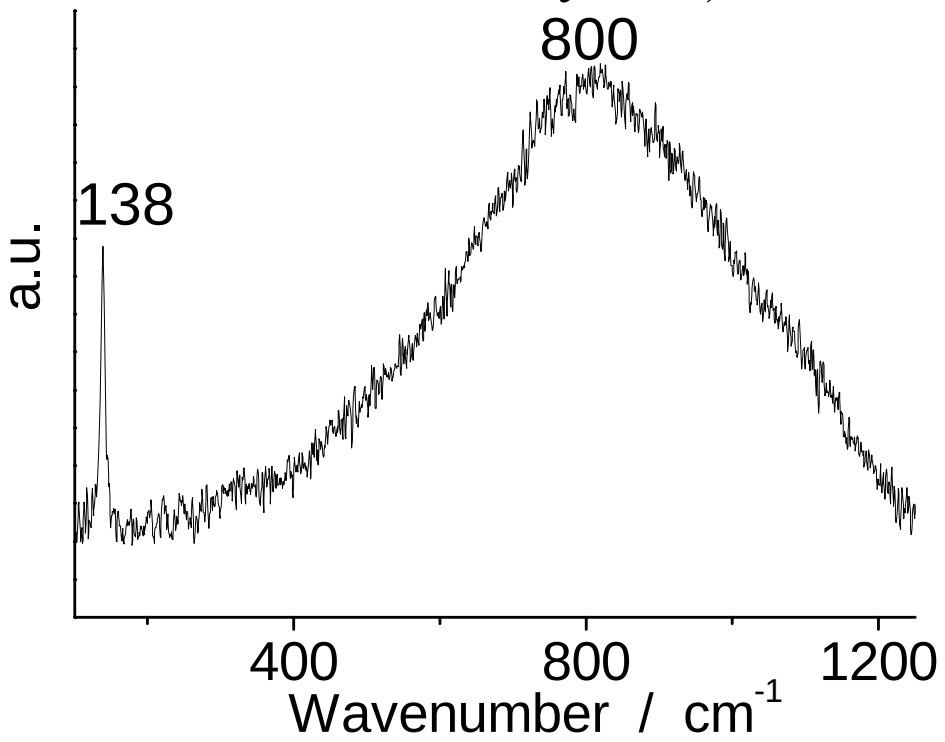 | ??<br>$Pb_2Sb_2O_7$ Naples Yellow Renaissance |
| 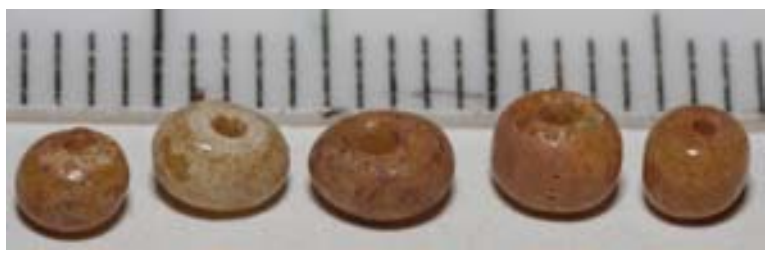 | Accession: Box<br>Map. Hill<br>Large orange after XRF | 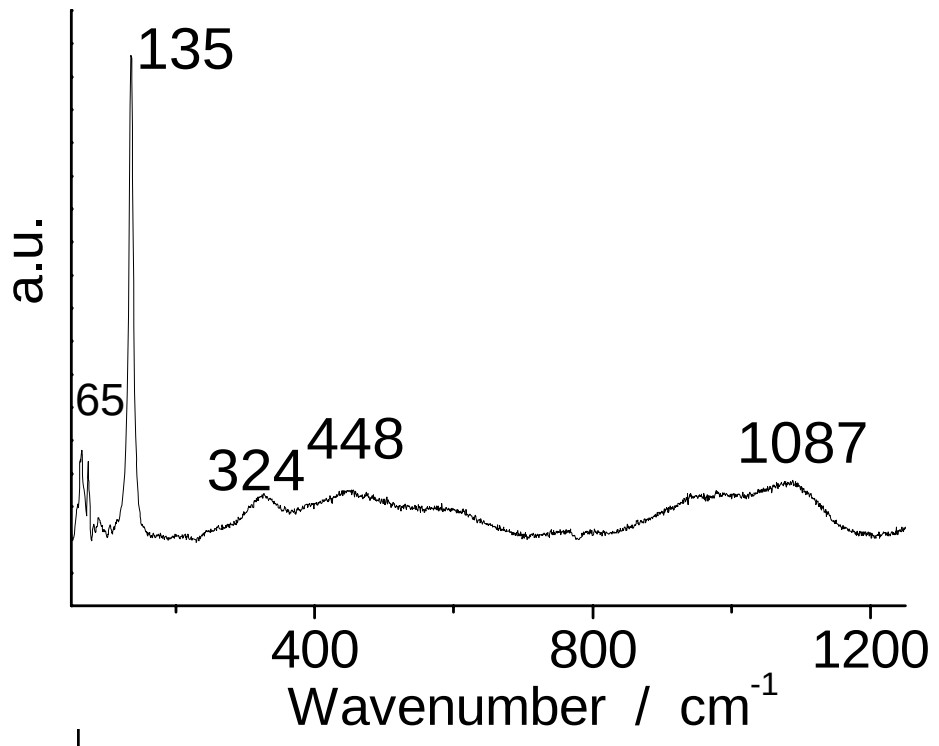 | $Na_2O$<br>$Pb_2Sb_2O_7$ Naples Yellow Renaissance<br>*T64000 |
| 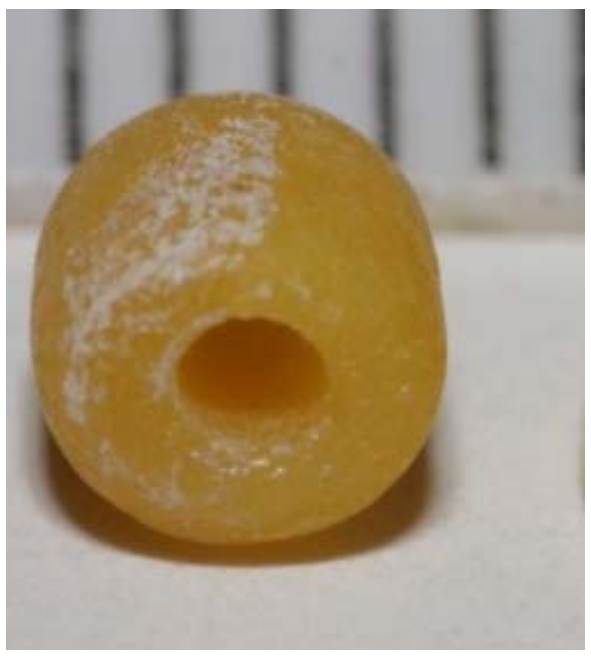 | Accession: **C952** Box C0046<br>Map. Hill<br>(~50-100) | 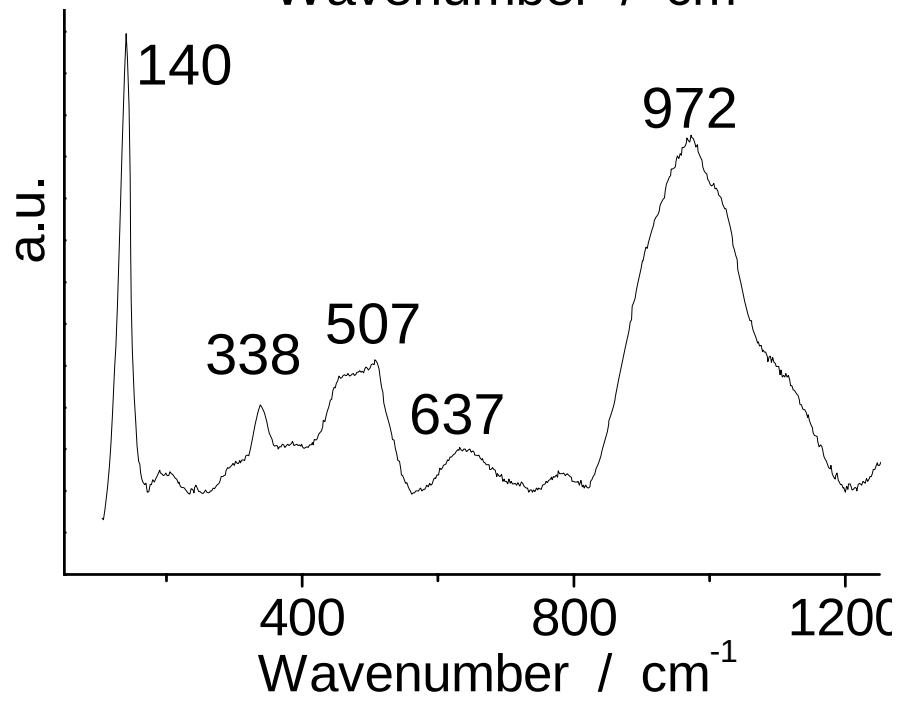 | $Pb_2Sb_2O_7$ Naples Yellow Renaissance<br>Stannate $CaSb_2O_6$/ $PbSnO_4$ |
| 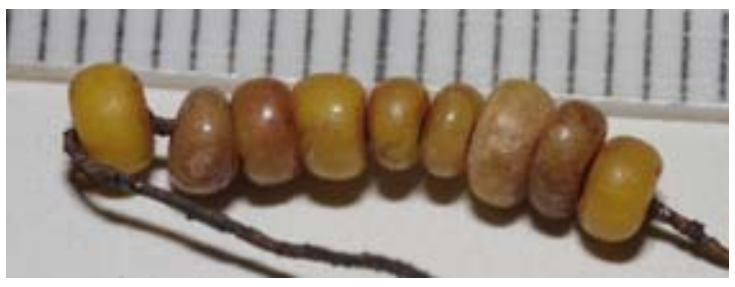 | Accession: **C2378** Box C0054<br>K2 | 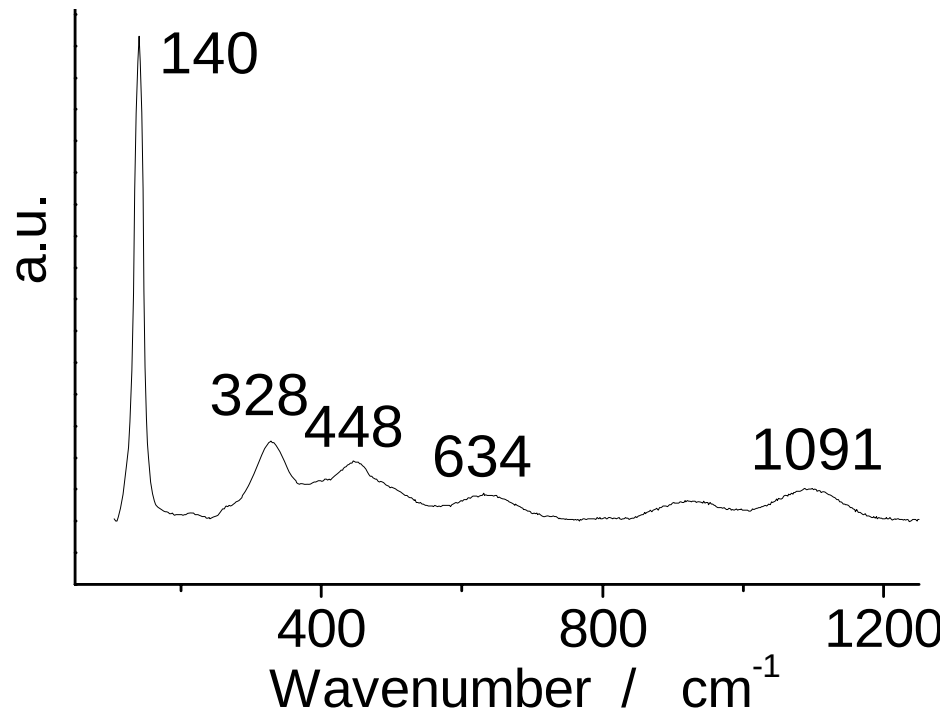 | $Pb_2Sb_2O_7$ Naples Yellow Renaissance |

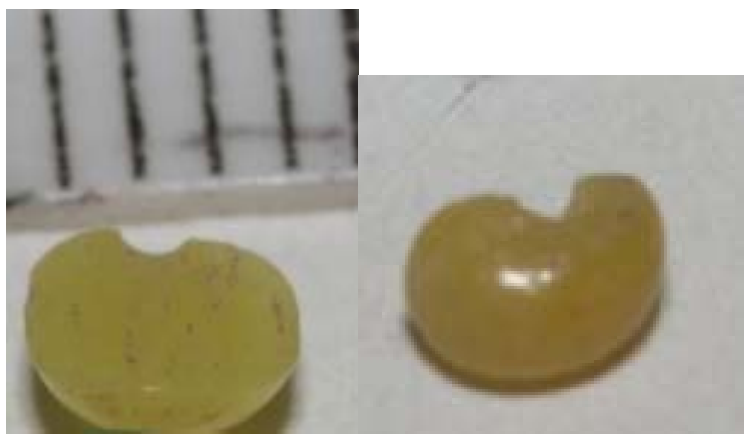

Accession:
**C1515**
Box C0054

Southern
Terrace

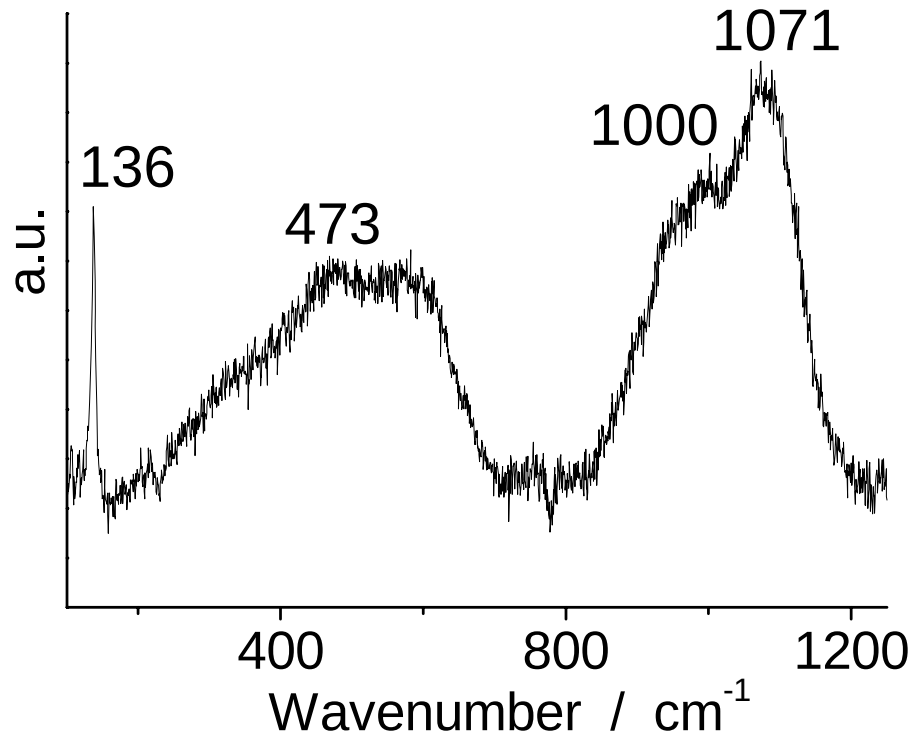

$Na_2O$

$Pb_2Sb_2O_7$
Naples Yellow
Renaissance

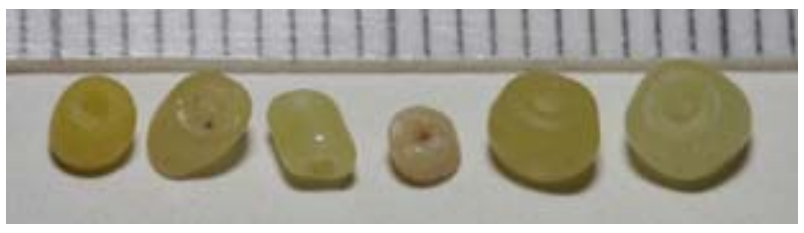

Accession:
**C1855**
Box C0055

K2

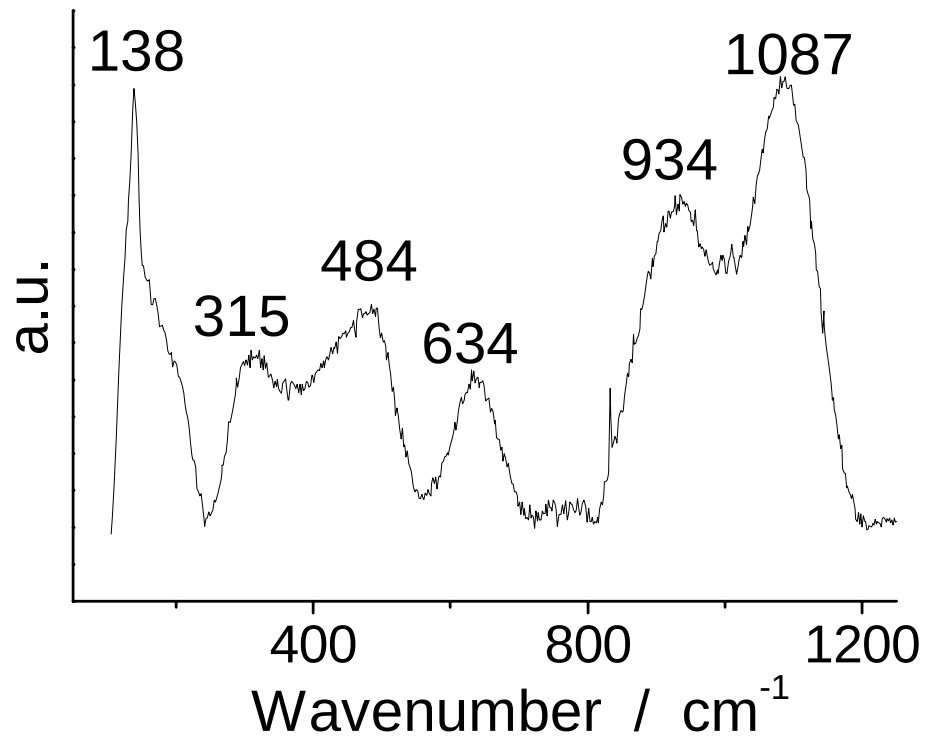

$Na_2O$

$Pb_2Sb_2O_7$
Naples Yellow
Renaissance

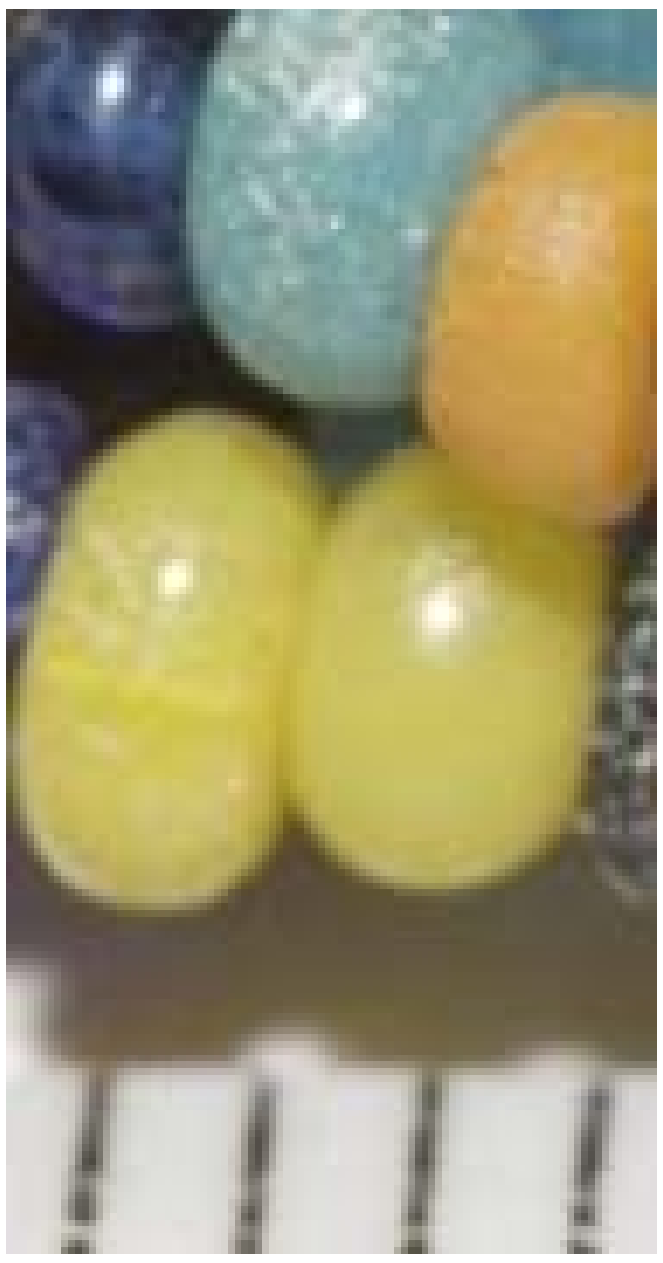

Accession:
**C934**
Box C0047

K2

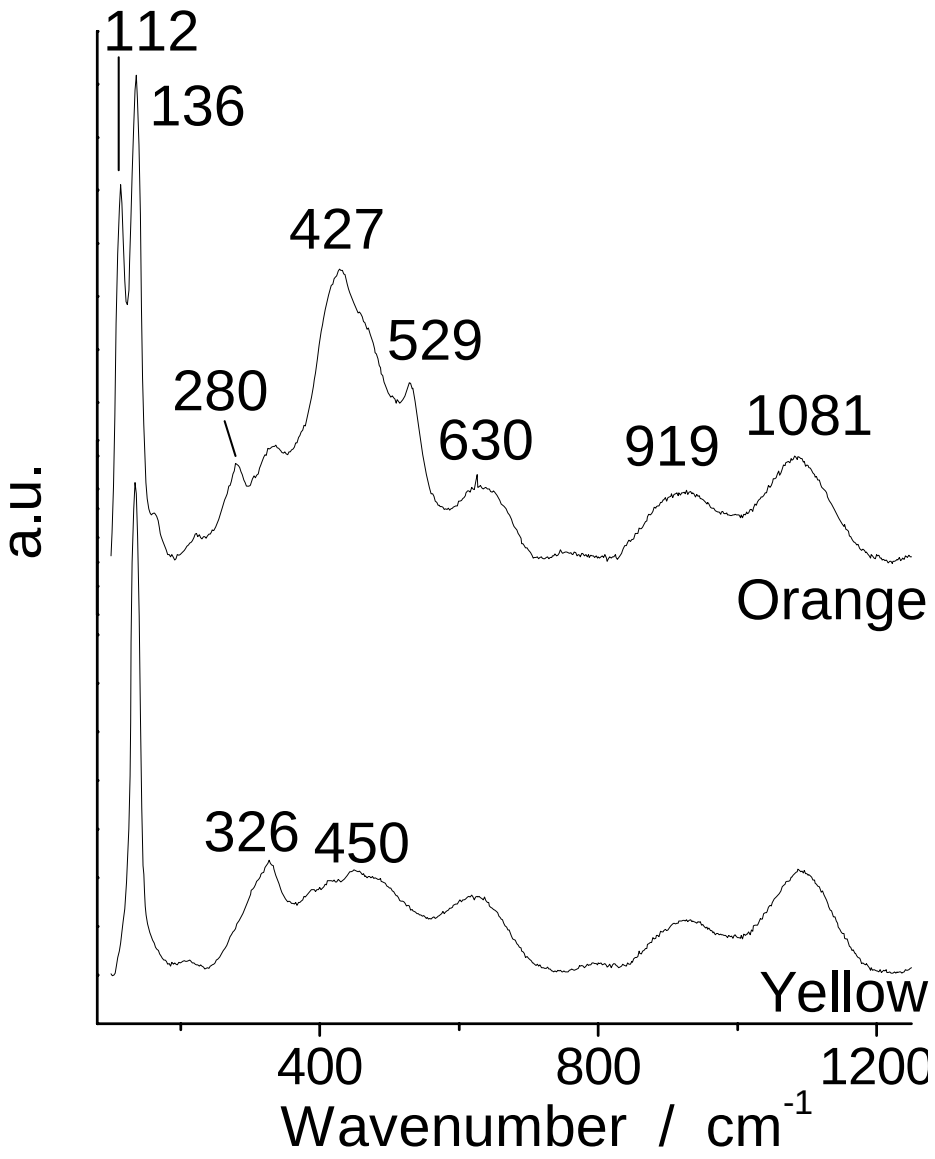

Stannate
$CaSb_2O_6$/

$PbSnO_4$
All of these beads are Mapungubwe oblates and according to XRF do not contain Sb

$Pb_2SnO_4$
Lead tin
Yellow
Antiquity

## Figure 4: GREEN BEADS

| BEADS (scale= 1mm) | REF. MUSEUM (quantity) | RAMAN SPECTRA (after baseline subtraction, a.u. = arbitrary unit) | GLASS MATRIX PIGMENT |
|---|---|---|---|

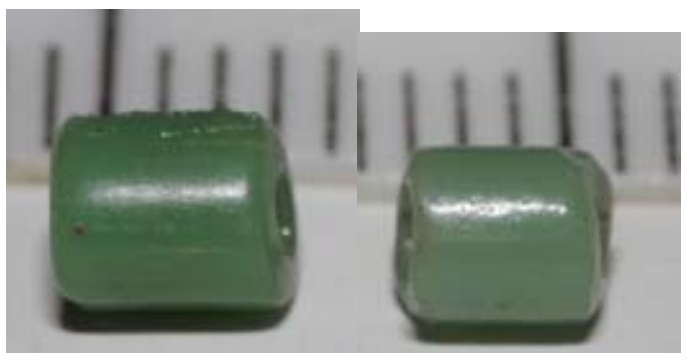

A D

Accession: **C/4333** Box 0241

Map. Hill

(~10)

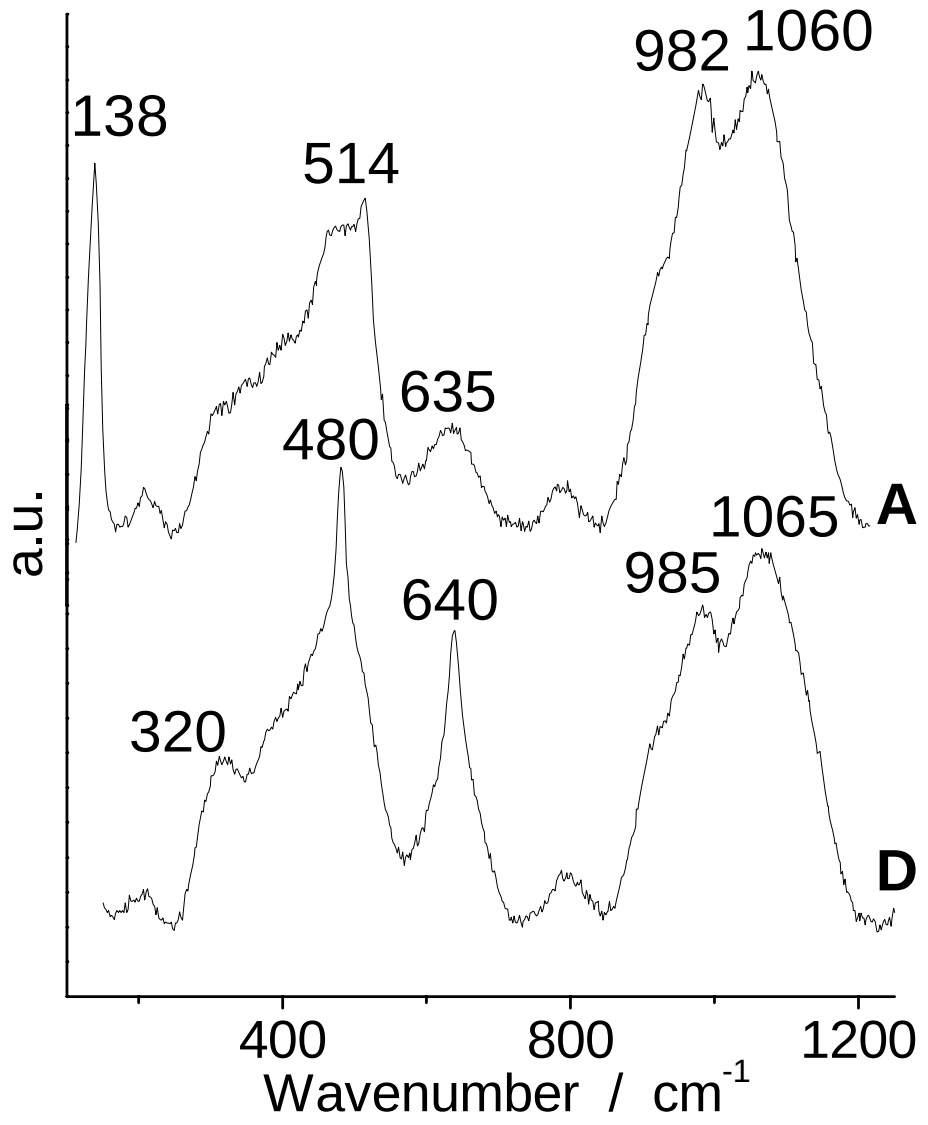

Stannate $CaSb_2O_6$/ $PbSnO_4$

$Na_2O$

$Ca_2Sb_2O_7$ European 17-18th c.

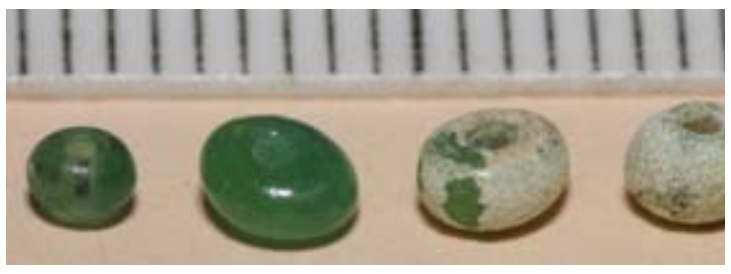

Accession: **C/144** Box 0049

Map. Hill

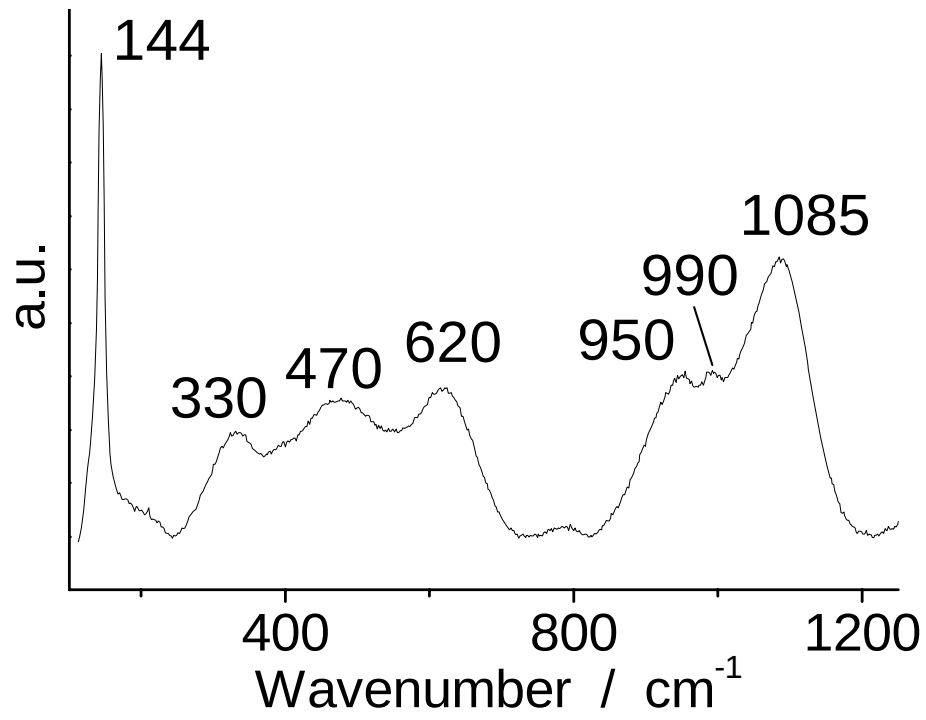

$Na_2O$/CaO

$Pb_2Sb_2O_7$ Naples Yellow Renaissance

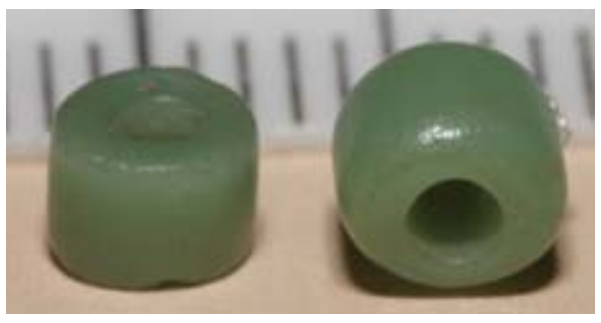

Accession: **C/1546** Box 0054

Map. Hill

(~10)

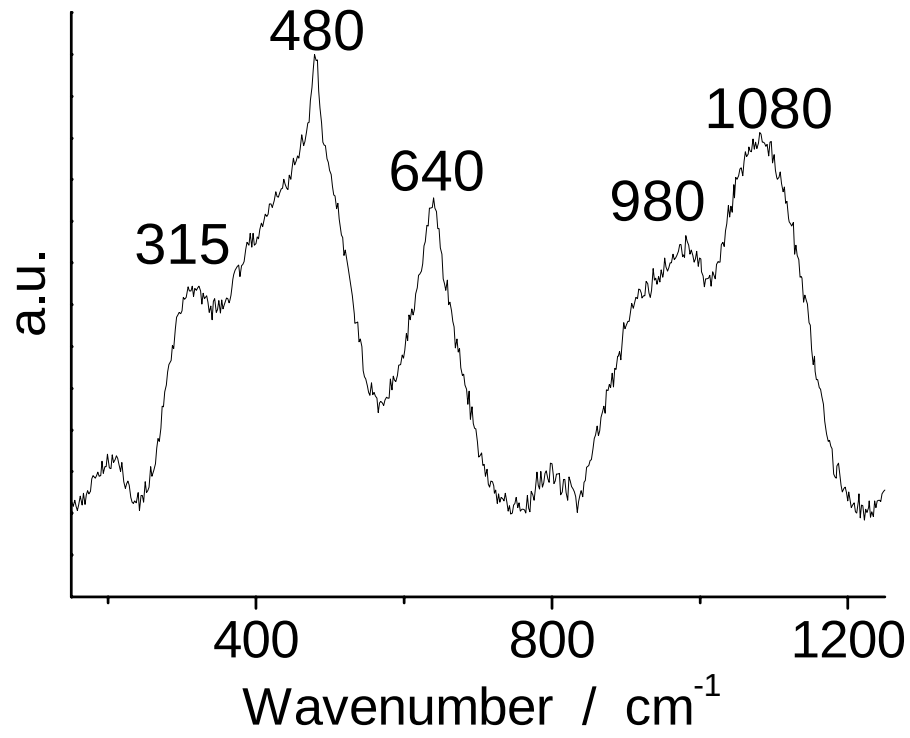

$Ca_2Sb_2O_7$ European 17-18th c.

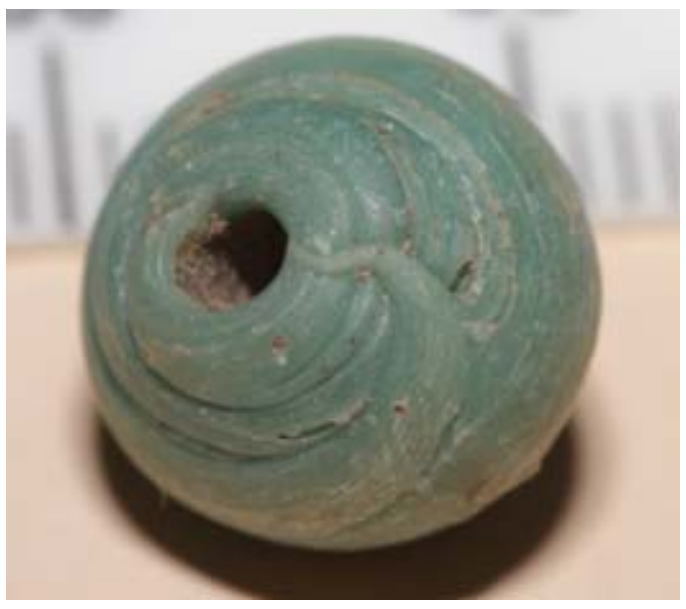

Accession: **C878** Box C0045

Map. Hill

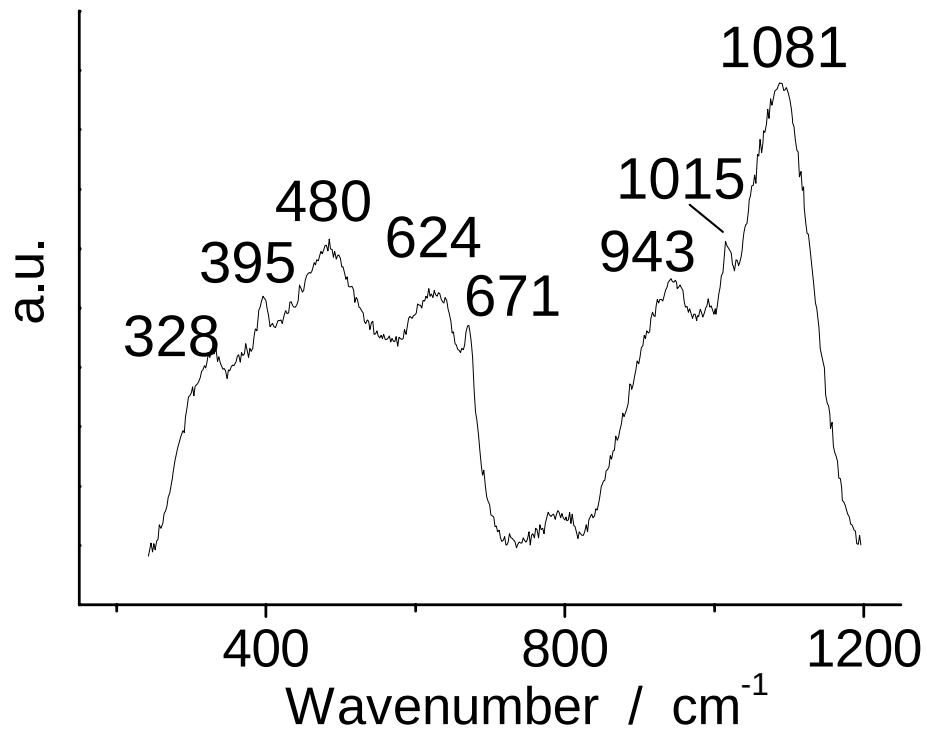

$Na_2O$

Clinopyroxene diopside

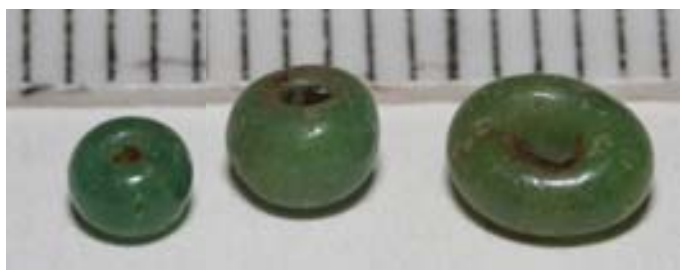

Accession:
**C2378**
Box C0054

K2

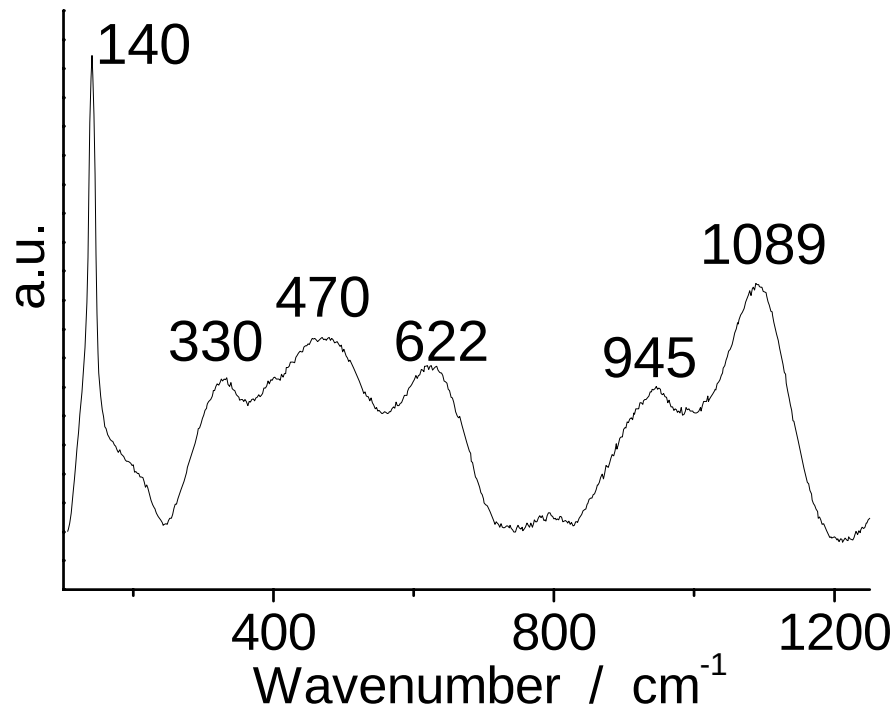

$Na_2O$

$Pb_2Sb_2O_7$
Naples Yellow
Renaissance

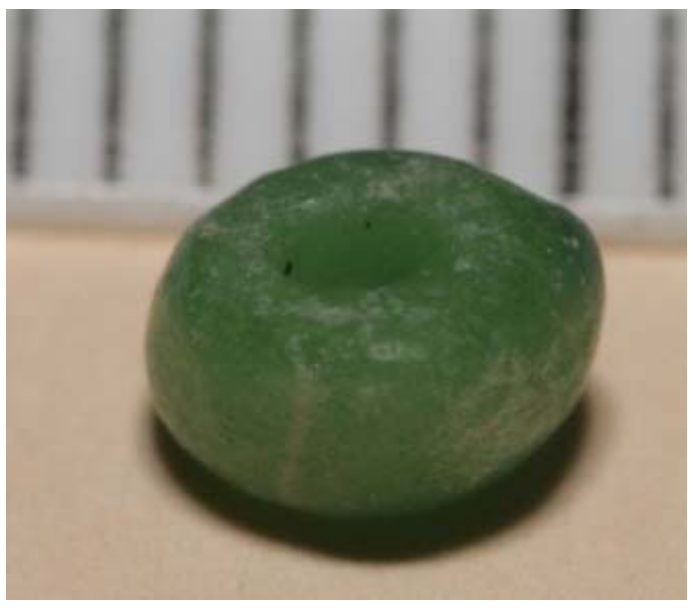

Accession:
**C1474**
Box C0053

Southern
Terrace

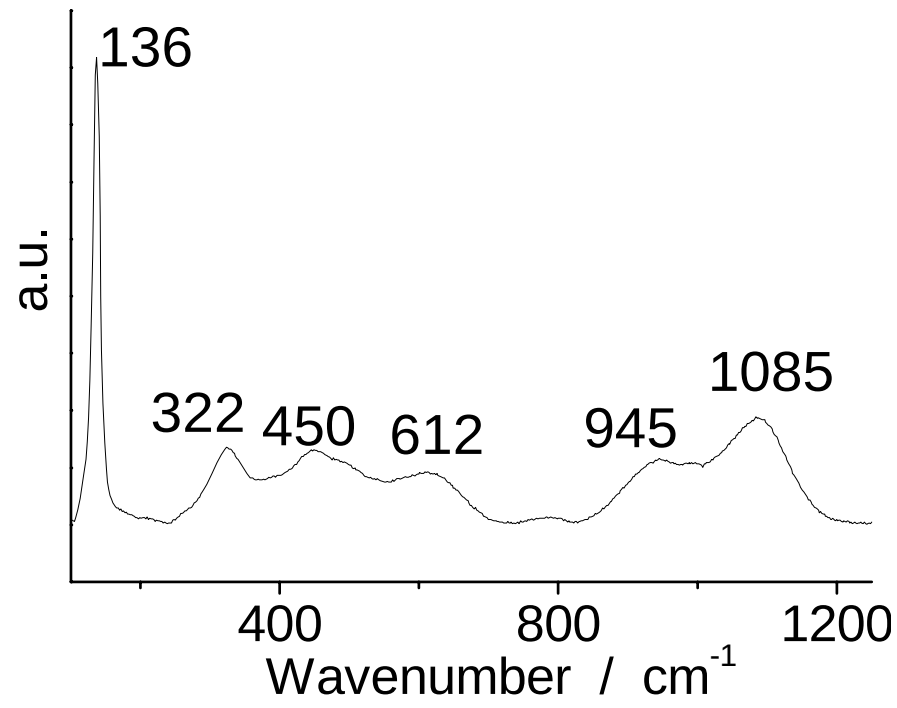

$Na_2O$

$Pb_2Sb_2O_7$
Naples Yellow
Renaissance

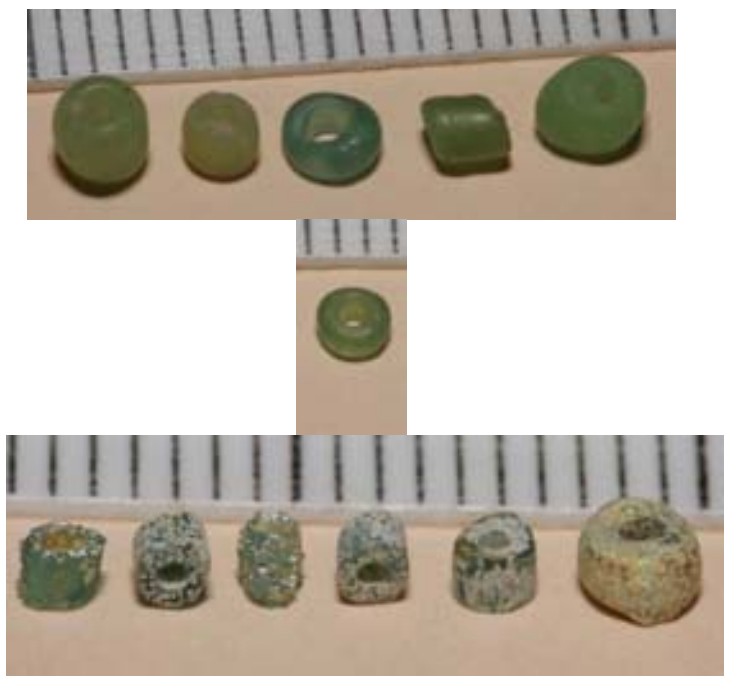

Accession:
**C1855**
Box C0055

K2

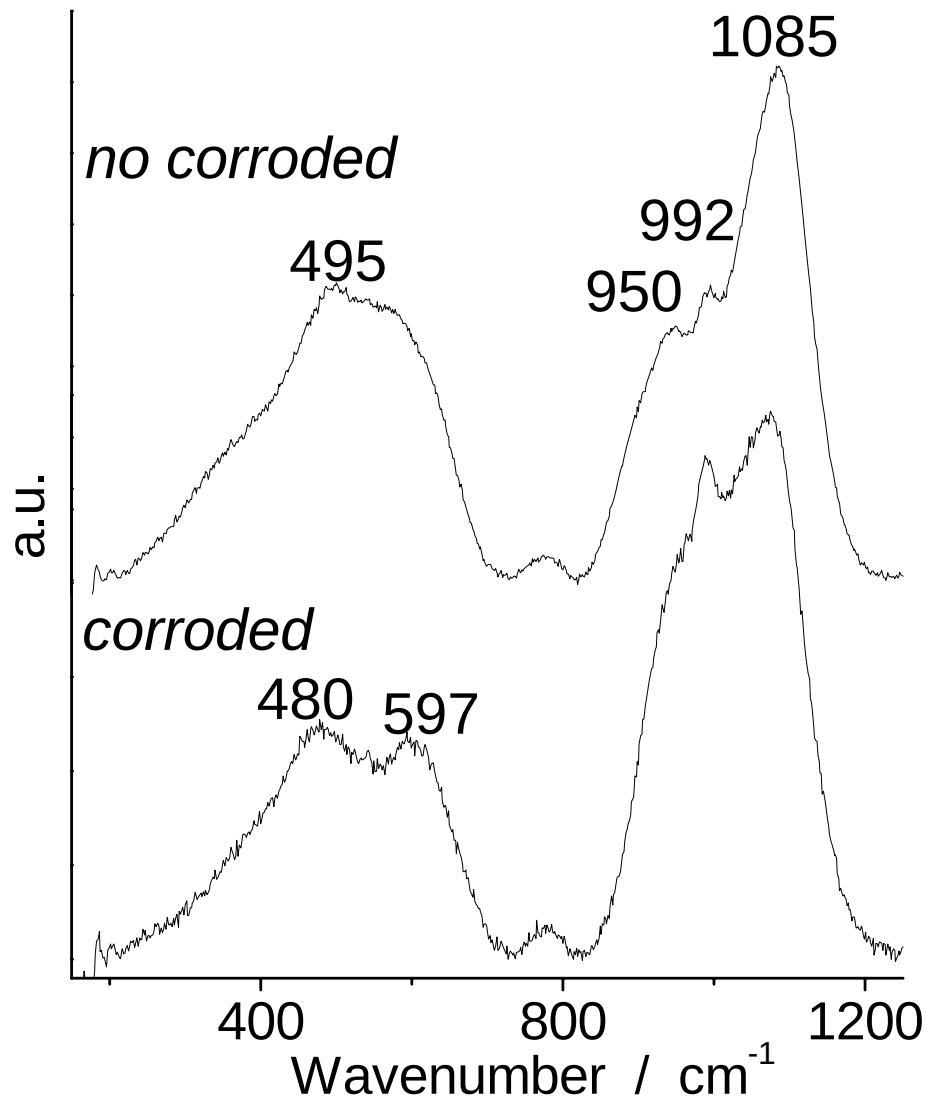

$Na_2O$

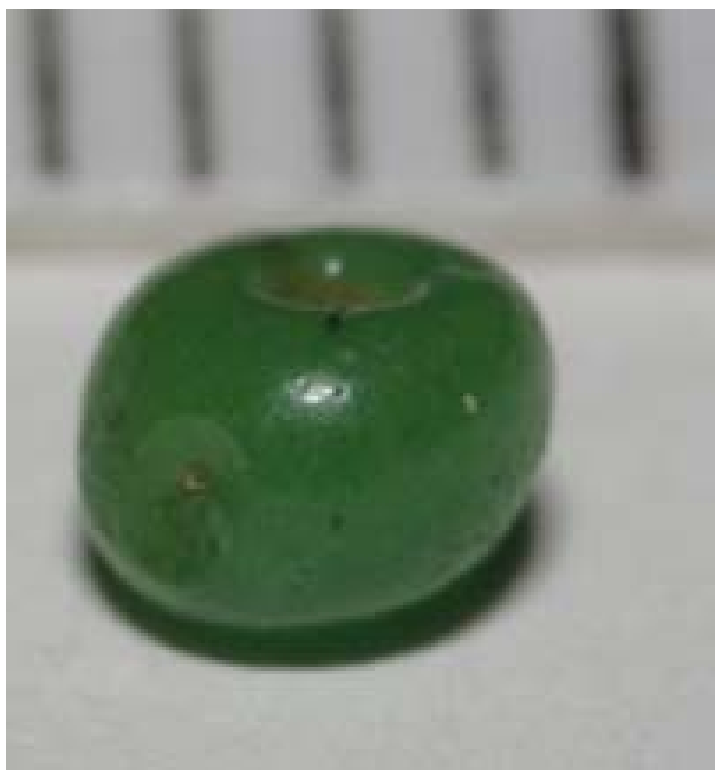

Accession:
**C1873**
Box C0050

K2

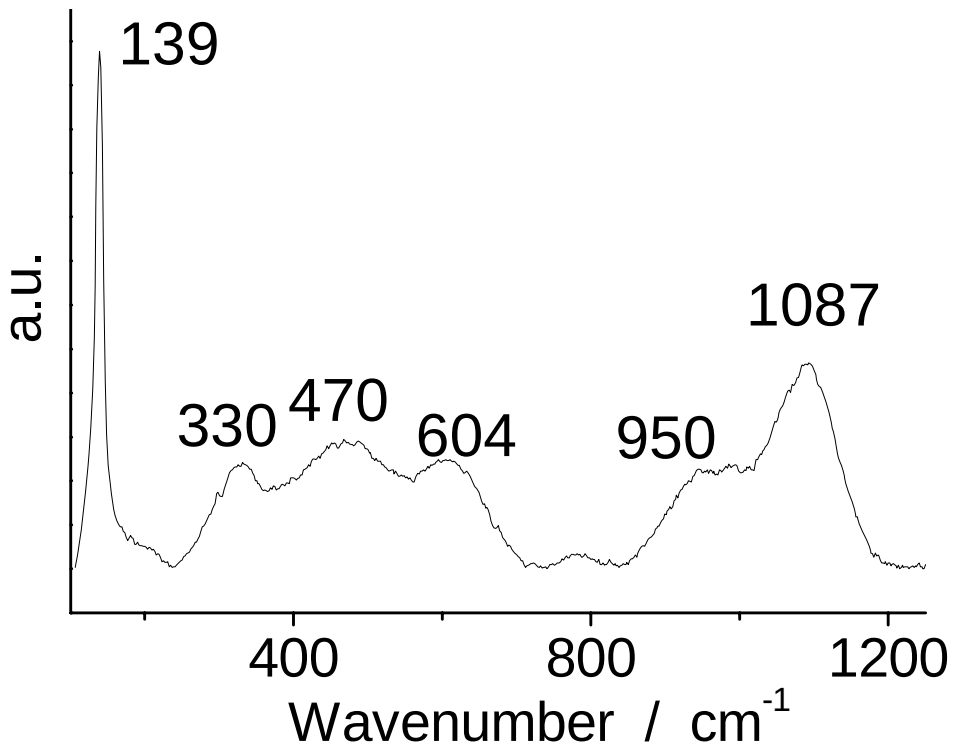

$Na_2O$

$Pb_2Sb_2O_7$
Naples Yellow
Renaissance

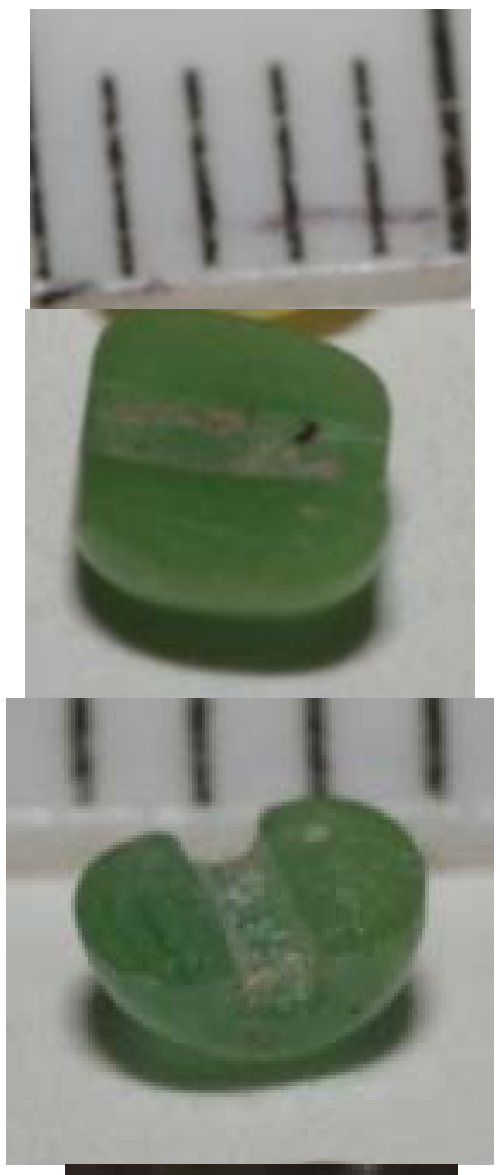

Accession:
**C1515**
Box C0054

Southern
Terrace

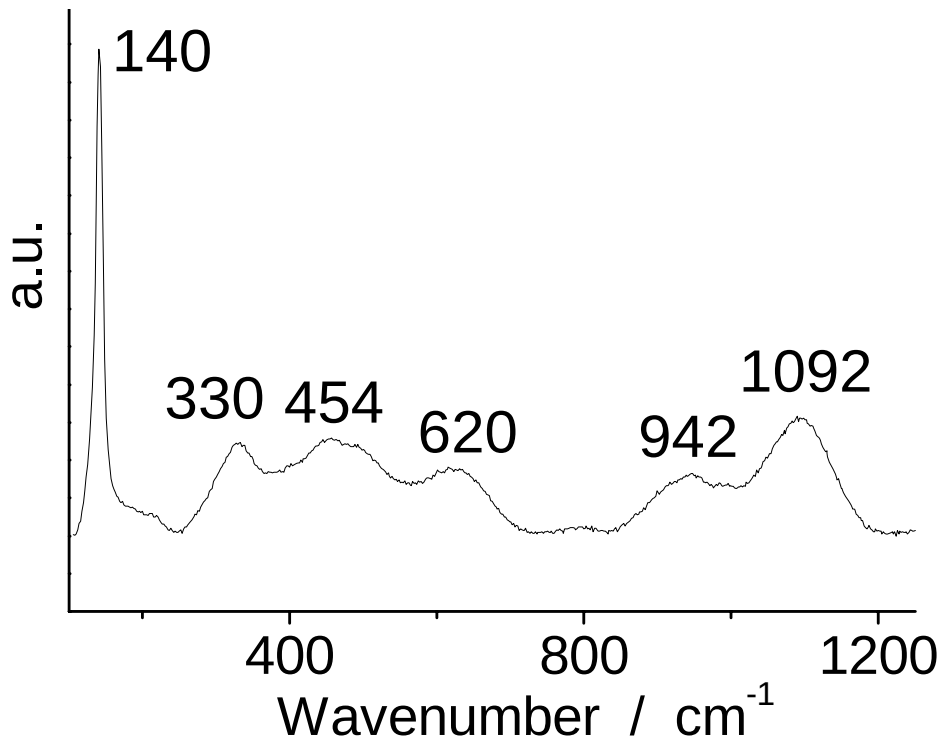

$Na_2O$

$Pb_2Sb_2O_7$
Naples Yellow
Renaissance

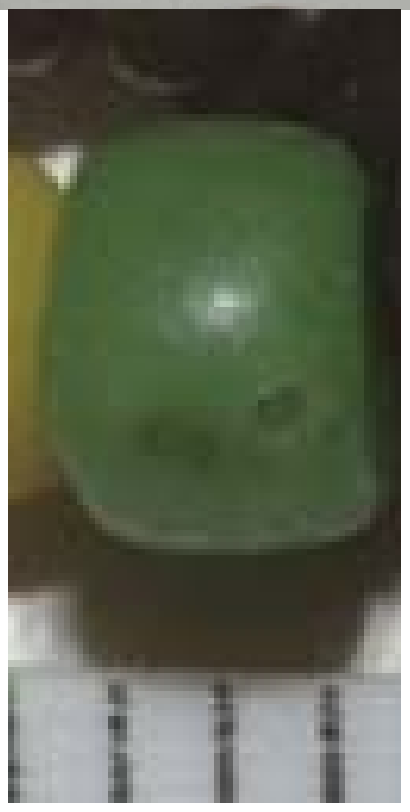

Accession:
**C783**
Box C0041

Greefswald

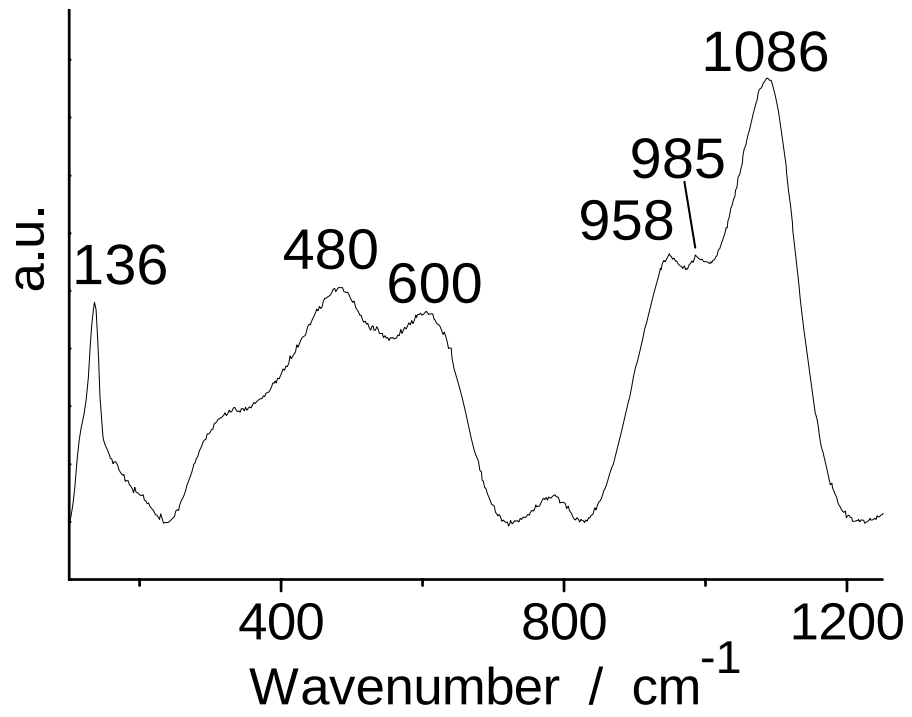

$Na_2O$

$Pb_2Sb_2O_7$
Naples Yellow
Renaissance

# Figure 5: BLACK BEADS

| BEADS (scale= 1mm) | REF. MUSEUM (quantity) | RAMAN SPECTRA (after baseline subtraction, a.u. = arbitrary unit) | GLASS MATRIX PIGMENT |
|---|---|---|---|
| 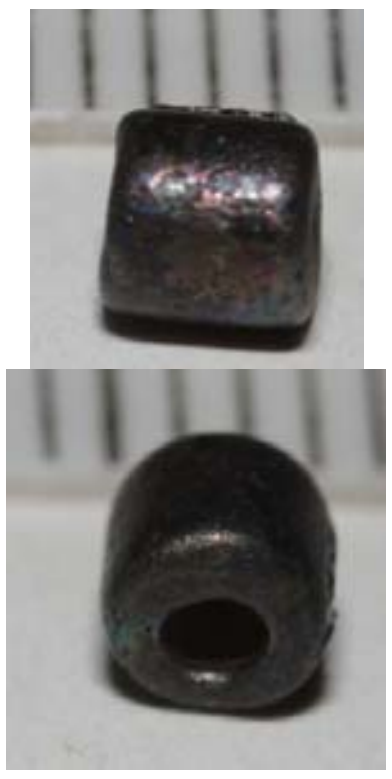 | Accession: **C/762** Box 0041<br>Map. Hill | 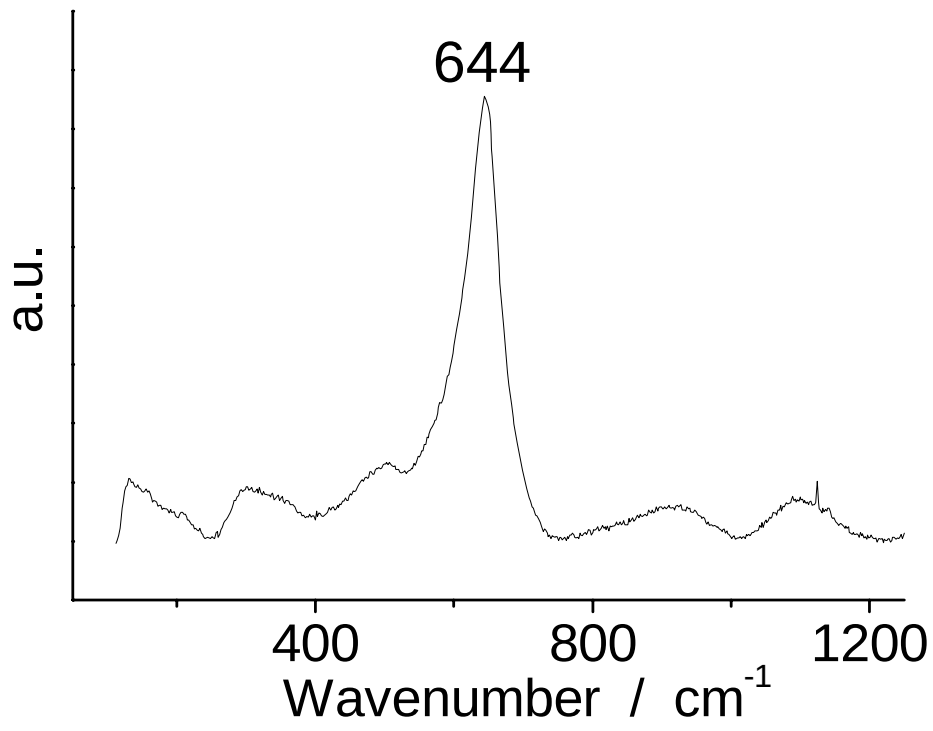 | $Fe_3O_4$ spinel or Co, Cr, Zn |
| 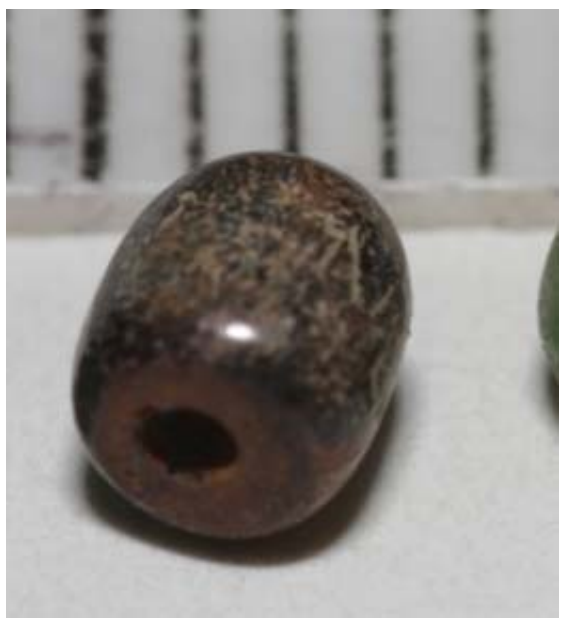 | Accession: **C2378** Box C0054<br>K2 | 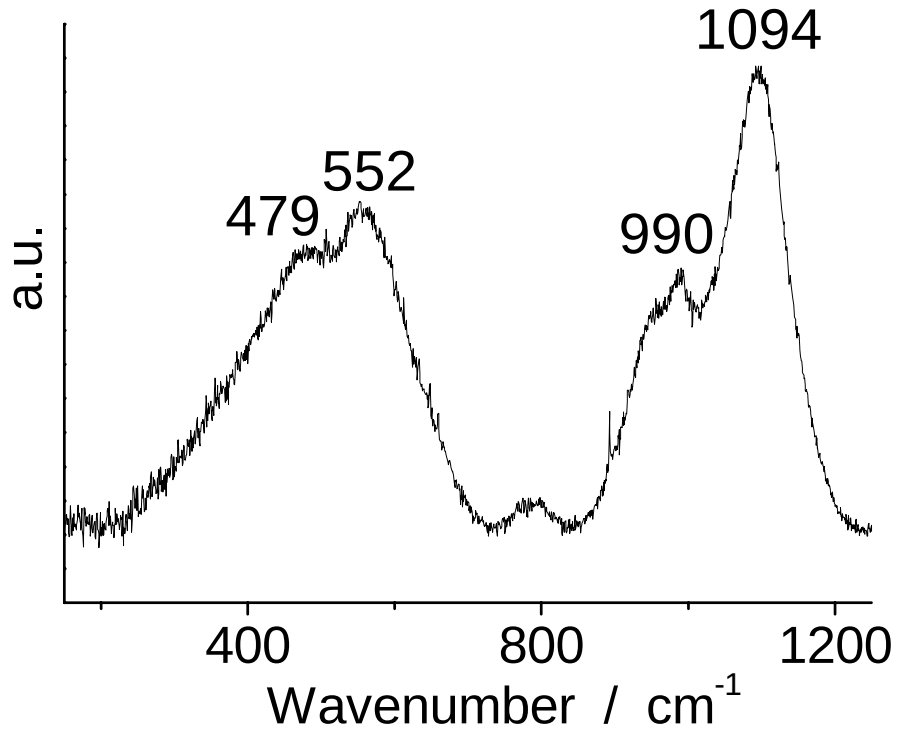 | $Na_2O/CaO$<br>* T64000 |
| 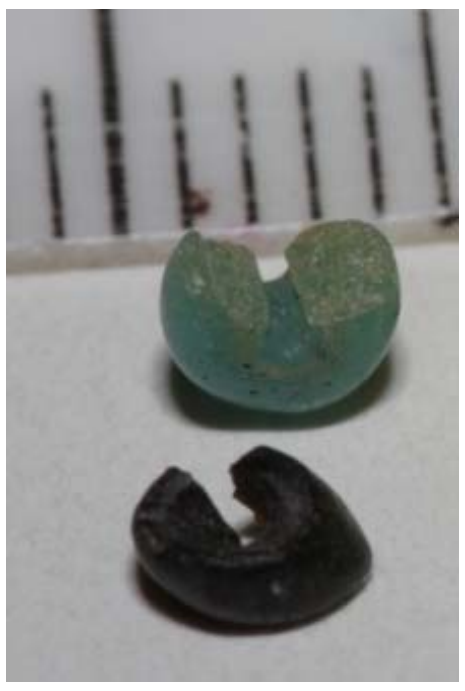 | Accession: **C1515** Box C0054<br>Southern Terrace | 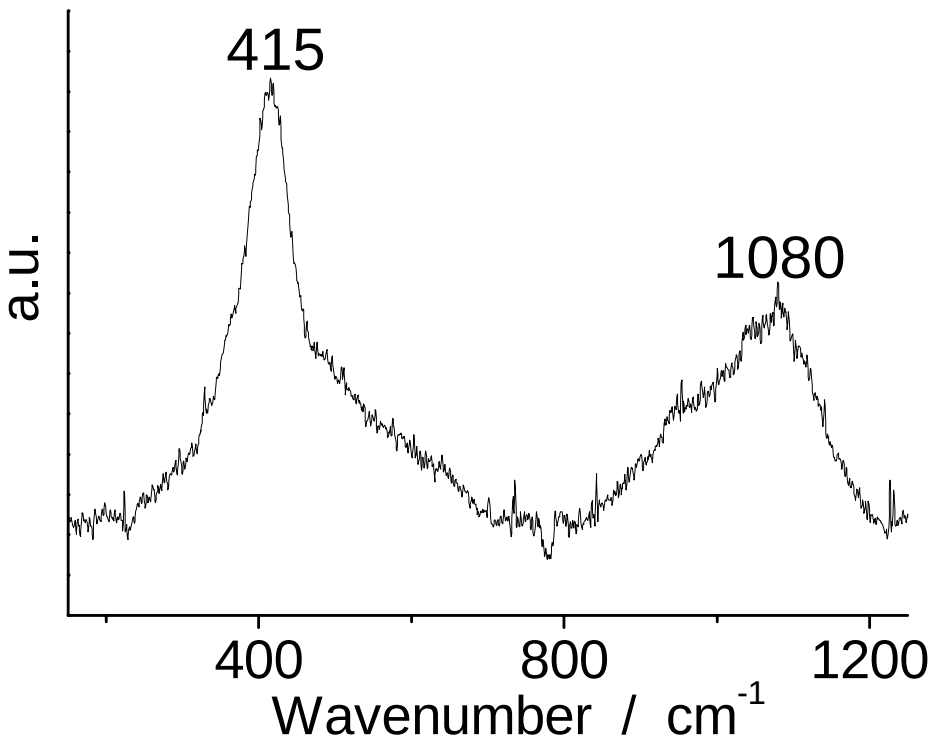 | Fe-S chromophore<br>* T64000 |
| 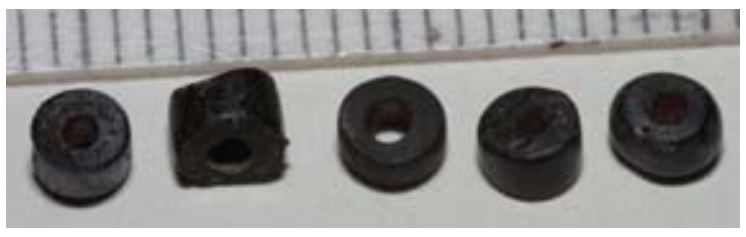 | Accession: **C952** Box C0046<br>Map. Hill | 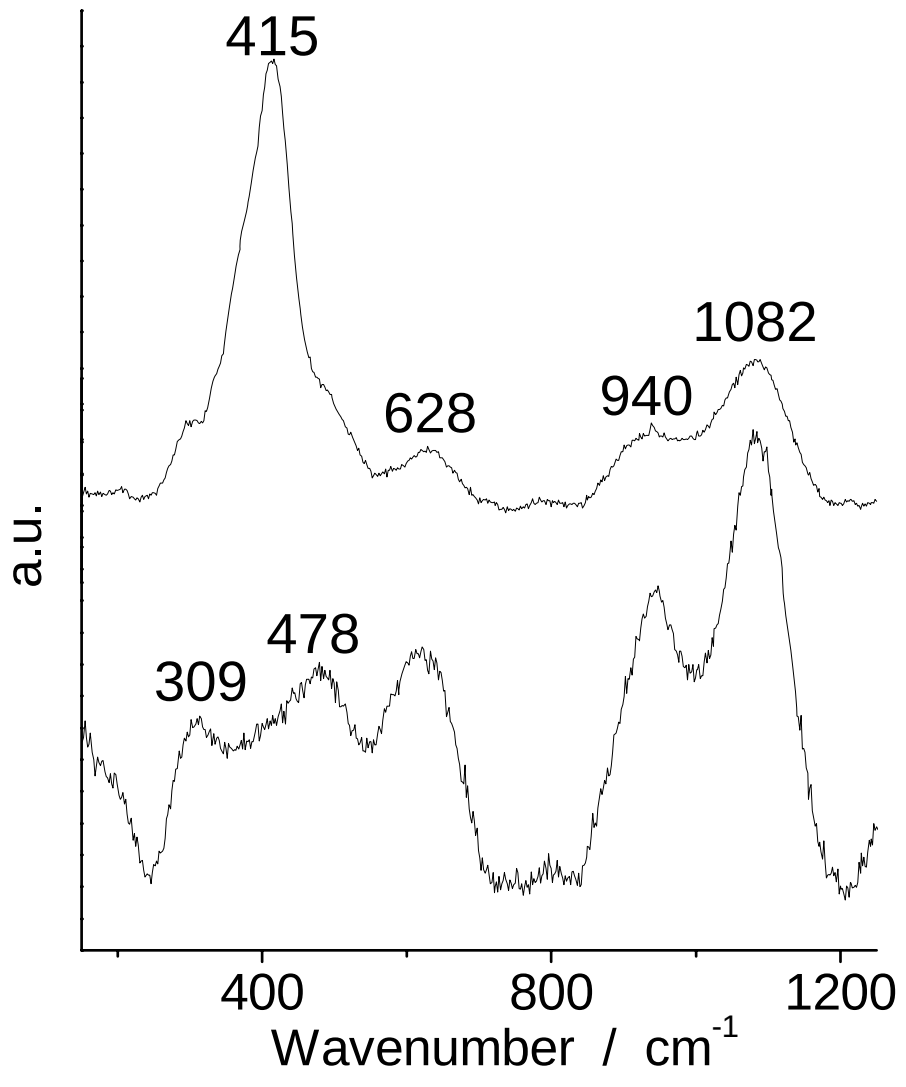 | Fe-S chromophore<br>$Na_2O/CaO$ |

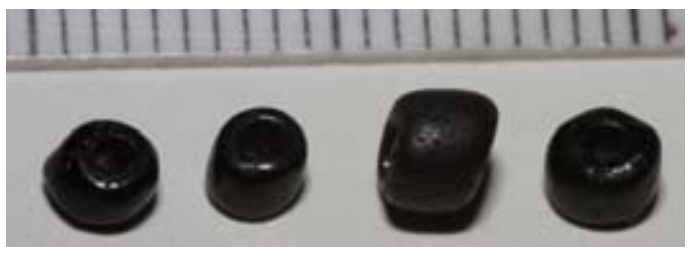

Accession:
**C1873**
Box C0050

K2

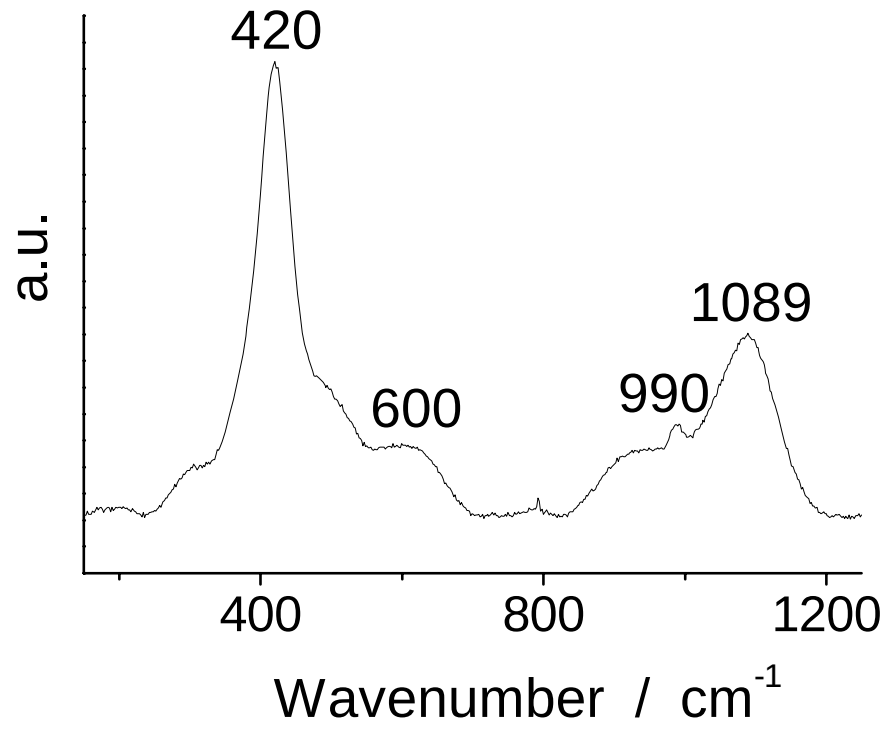

Fe-S chromophore

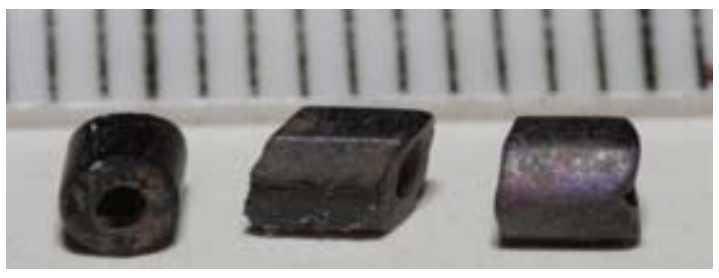

Accession:
**C754**
Box C0038

K2

(~60)

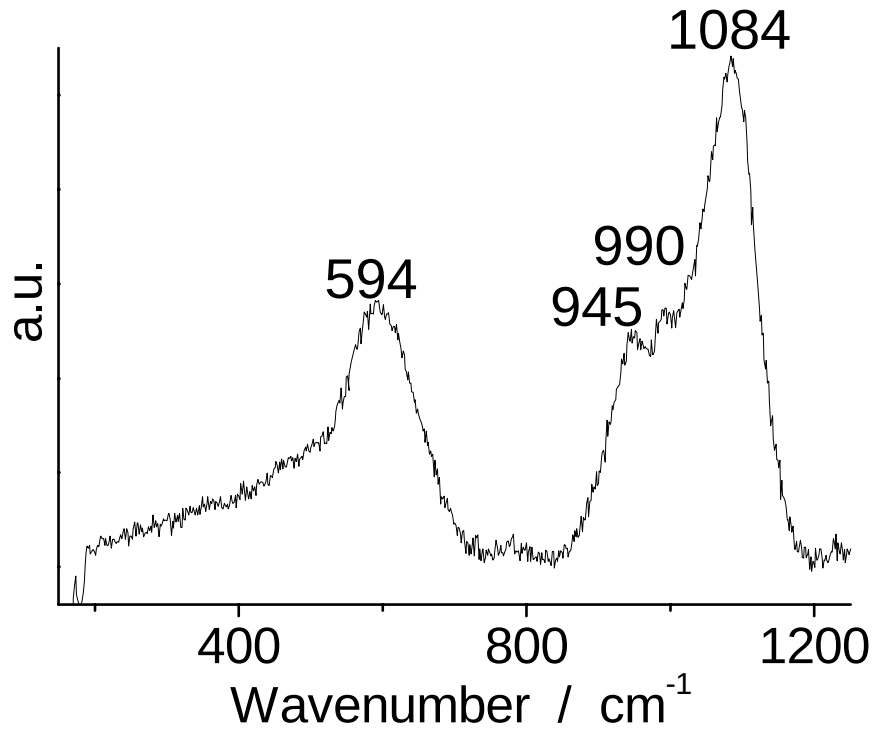

$Na_2O/CaO$

* Infinity

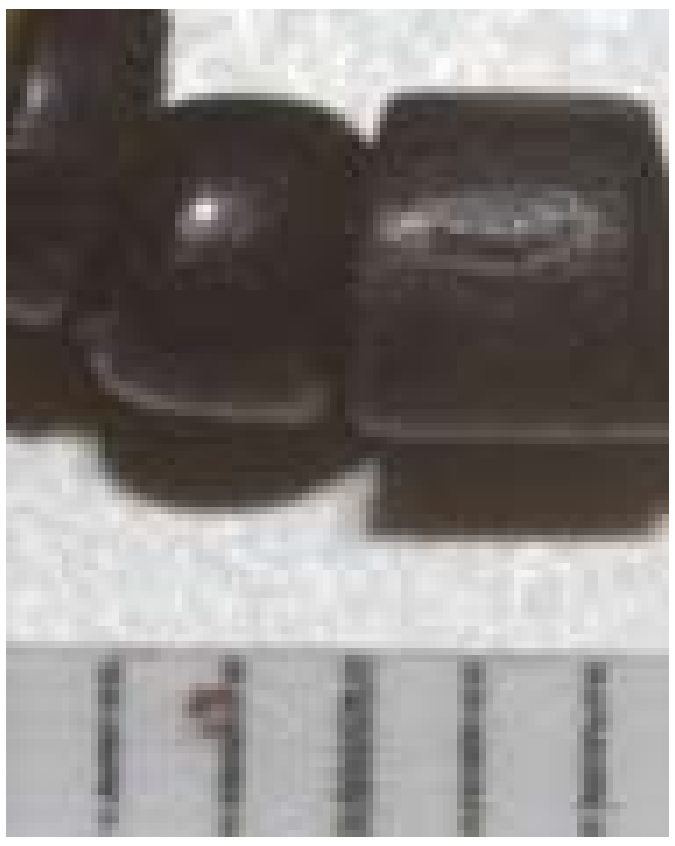

Accession:
C783
Box C0041

Greefswald

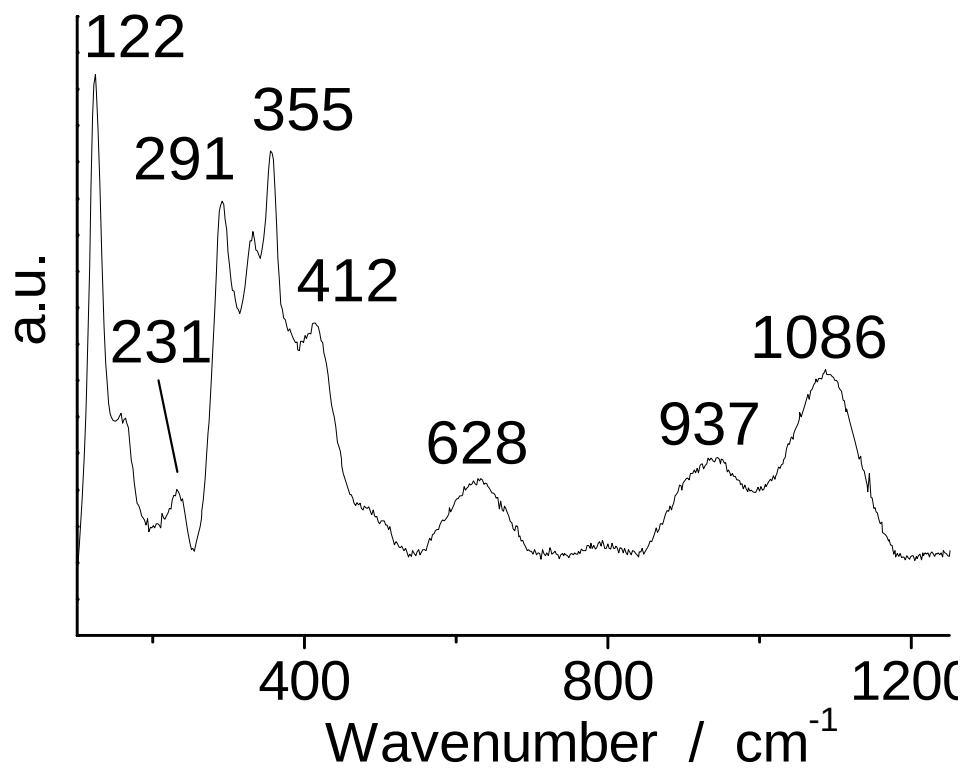

$Na_2O/CaO$

Fe-S chromophore
Mapungubwe oblate

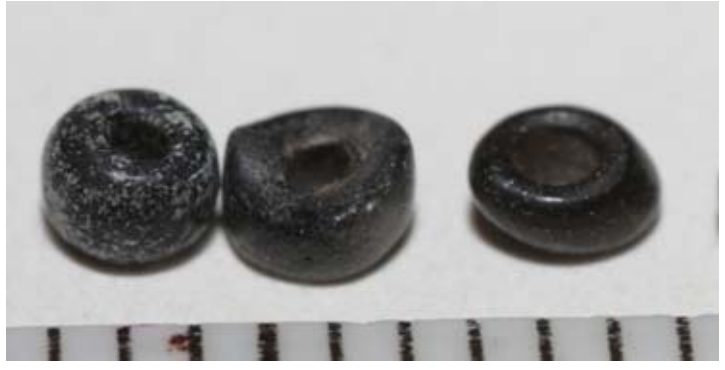

Accession:
C1351
Box C0053

Bambandy-analo (K2)

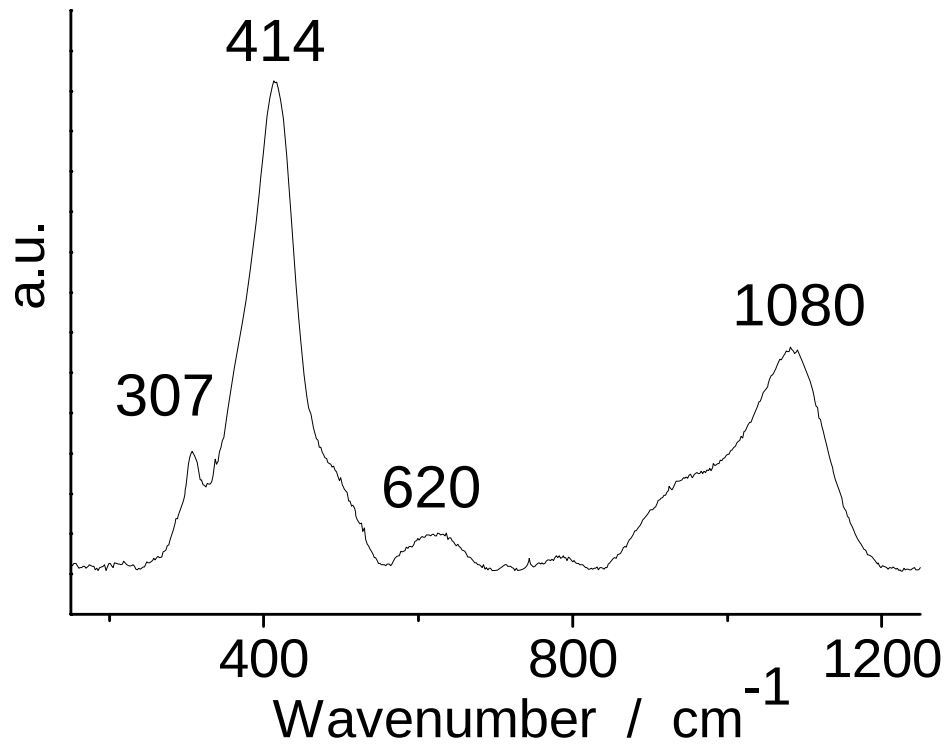

Fe-S chromophore

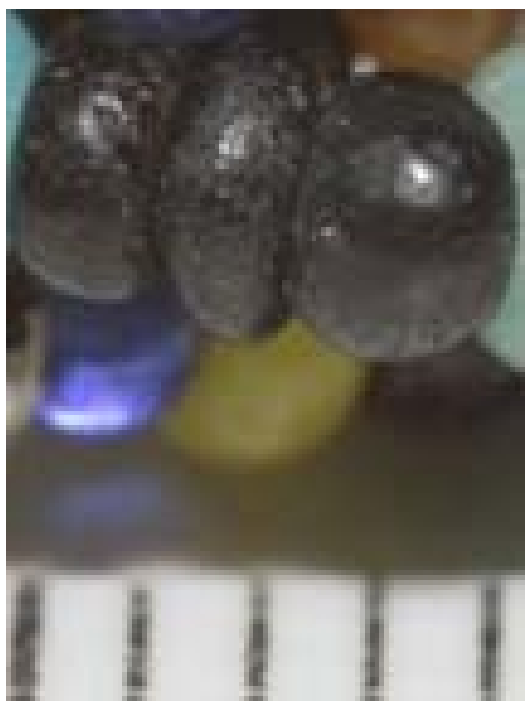

Accession:
C934
Box C0047

K2

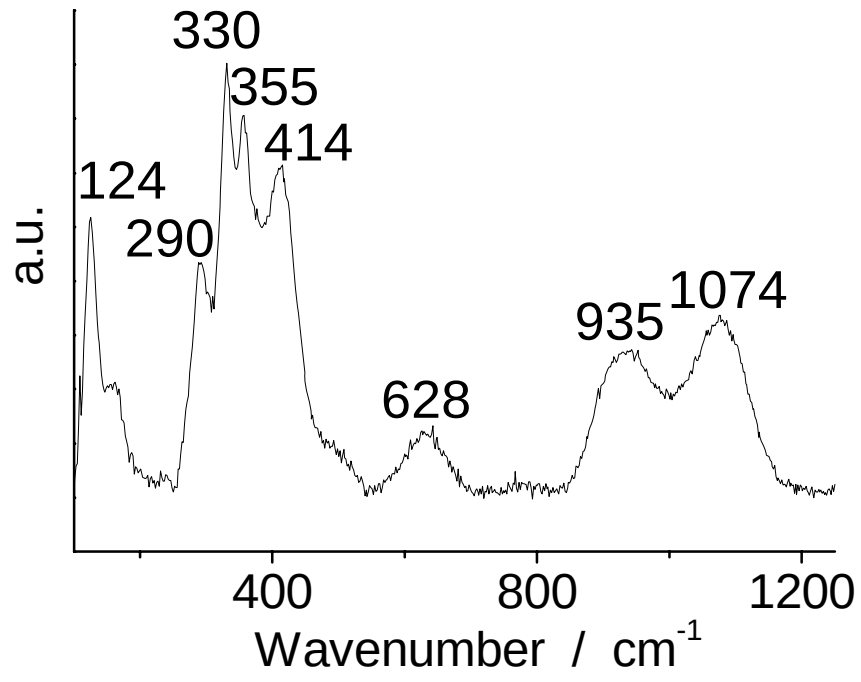

Fe-S
chromophore
Mapungubwe
oblate

**Figure 6:** **WHITE BEADS**

**BEADS**
(scale= 1mm)

**REF. MUSEUM**
(quantity)

**RAMAN SPECTRA**
(after baseline subtraction, a.u. = arbitrary unit)

**GLASS MATRIX PIGMENT**

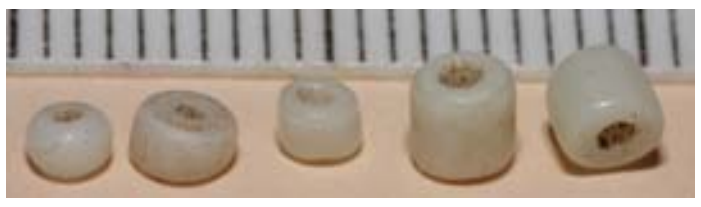

Accession:
**C/952**
Box 0046

Map. Hill

(~400-500)

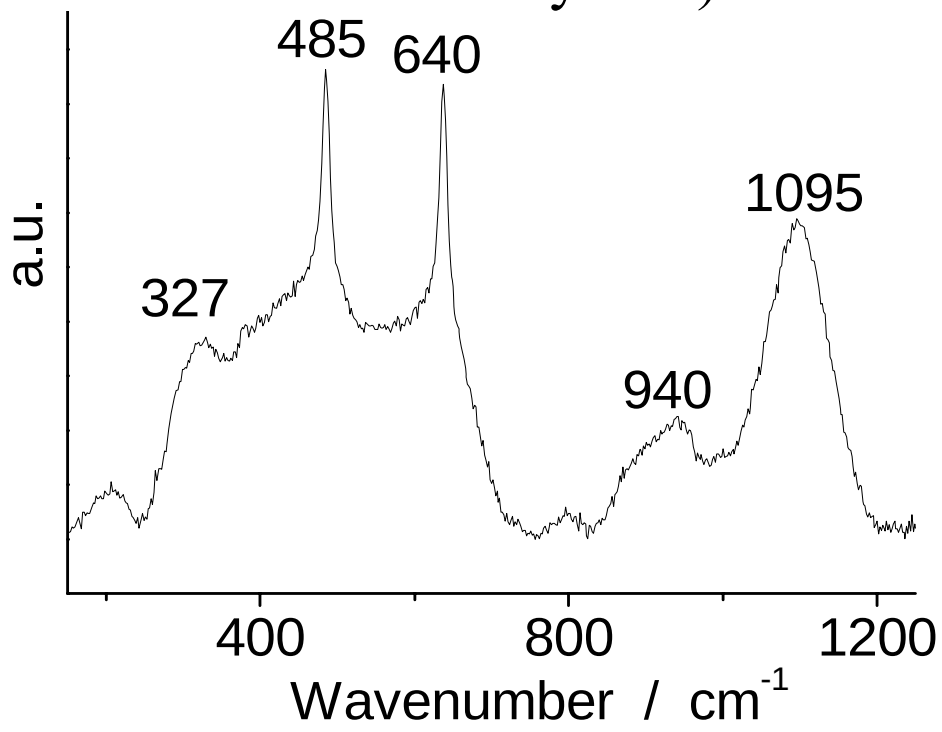

$Ca_2Sb_2O_7$
European
17-18th c.

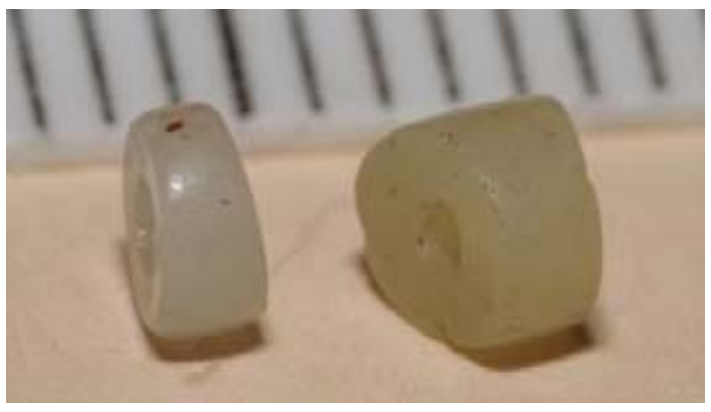

Accession:
**C952**
Box C0046

Map. Hill

(~100)

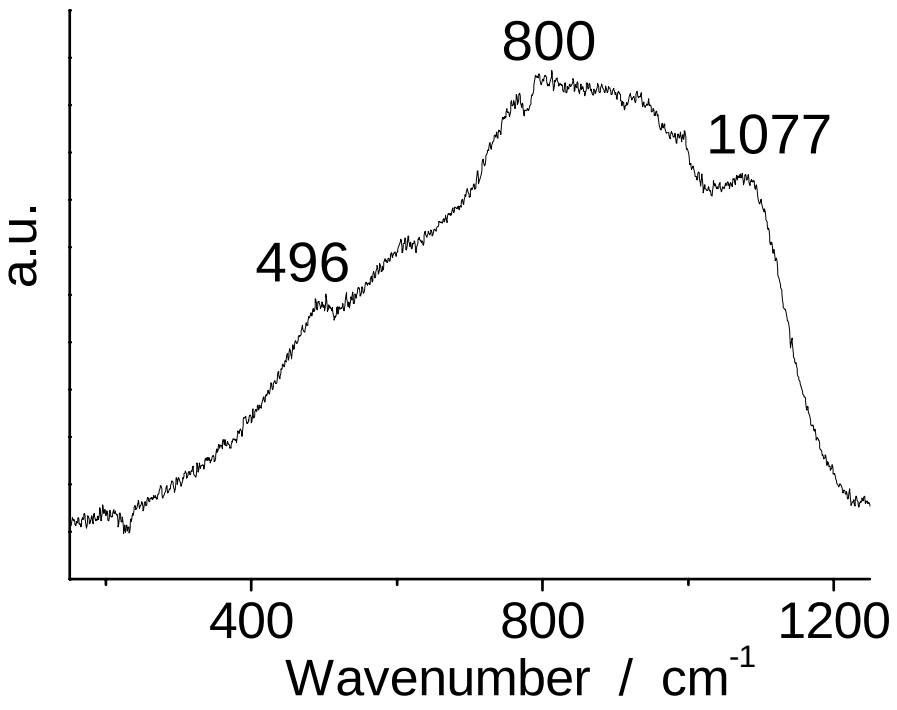

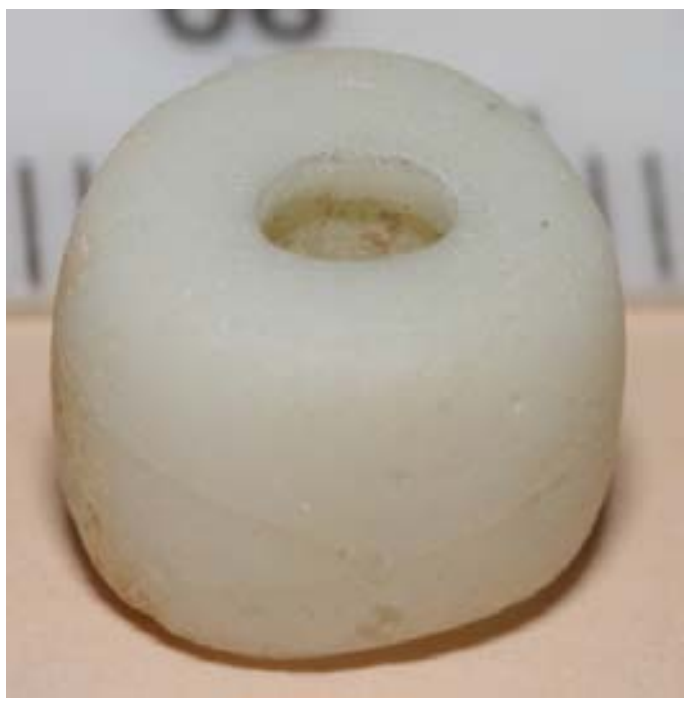

A

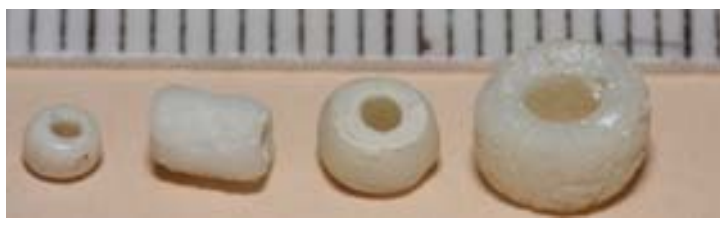

B C D E

Accession:
**C836**
Box C0054

Map. Hill

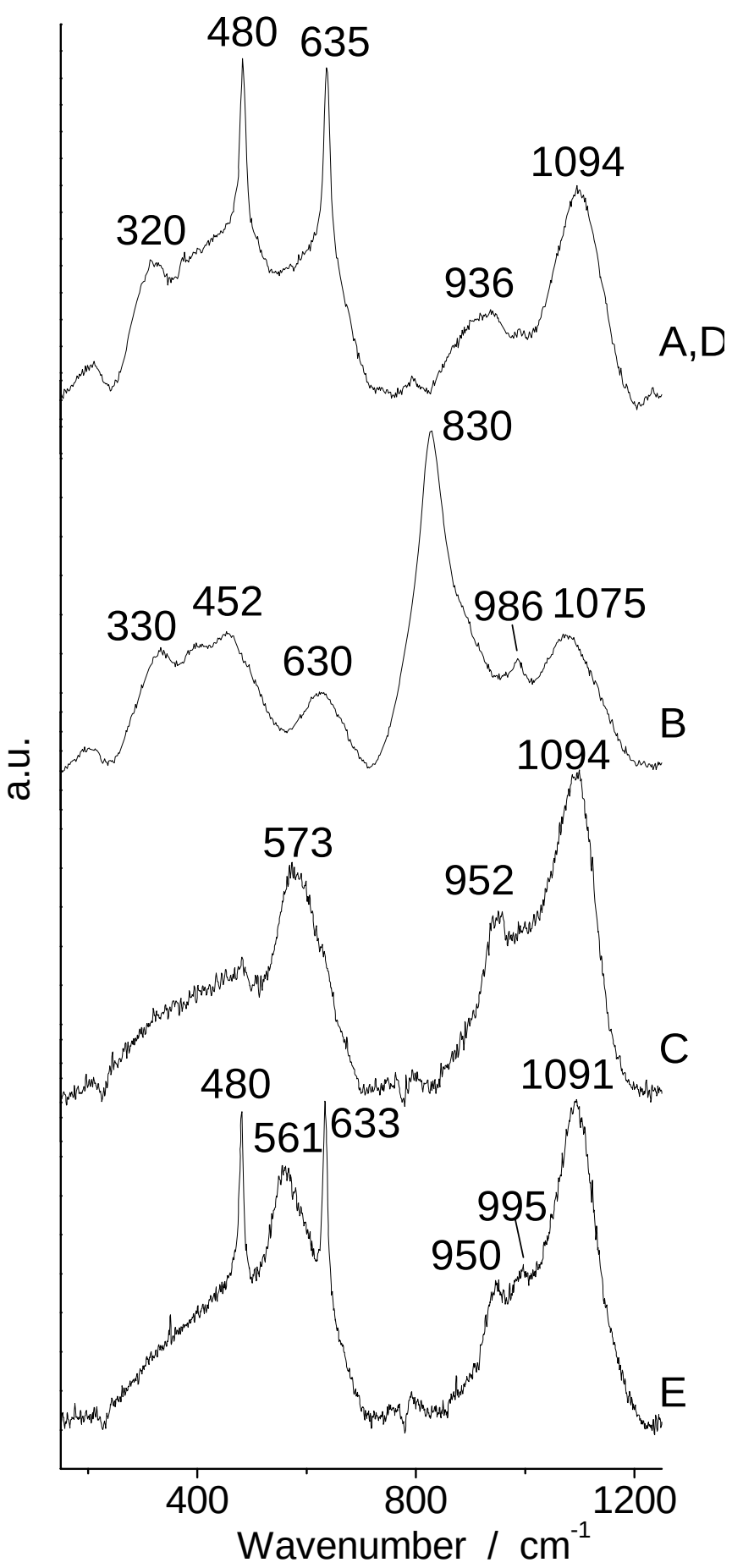

$Ca_2Sb_2O_7$
European
17-18th c.

Ca and Pb
Arsenate
16th c.

$Na_2O/CaO$

*T64000

$Na_2O/CaO$

$Ca_2Sb_2O_7$
European
17-18th c.

* T64000

## Figure 7: *******************OTHER***********************

**BEADS** (scale= 1mm) | **REF. MUSEUM** (quantity) | **RAMAN SPECTRA** (after baseline subtraction, a.u. = arbitrary unit) | **GLASS MATRIX PIGMENT**

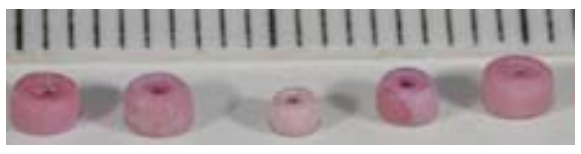

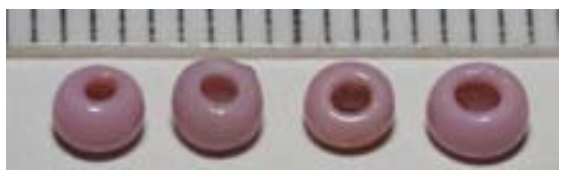

Accession: **C/838** Box C054
Accession: **C1550** Box C0054

K2

(~500-600)

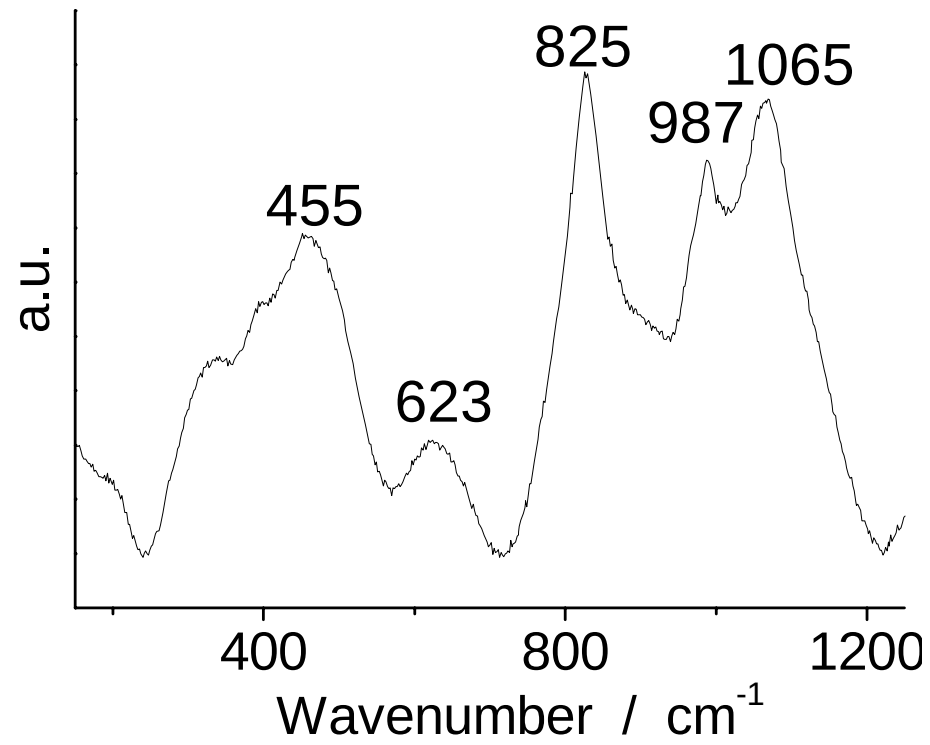

$Na_2O$

Ca and Pb Arsenate 16e c.

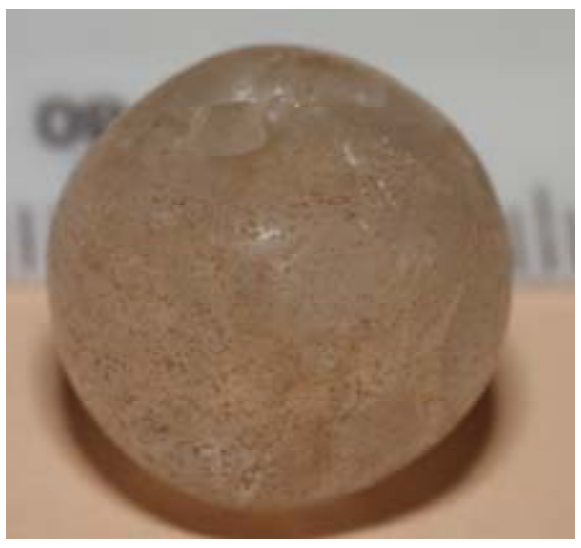

Accession: **C/830** Box 0043

Map. Hill

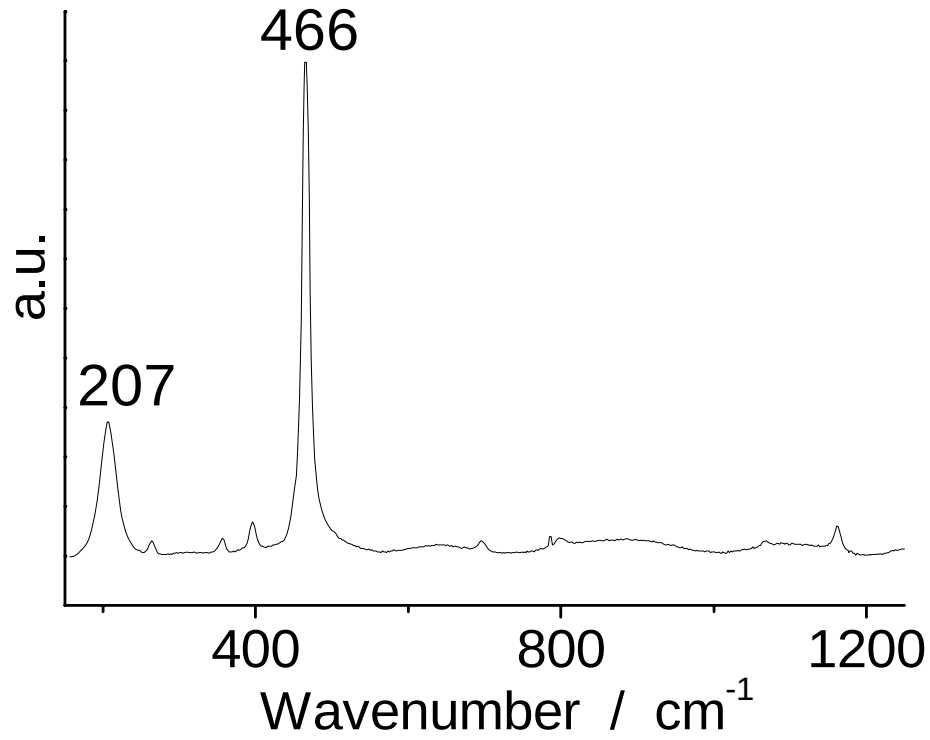

Quartz

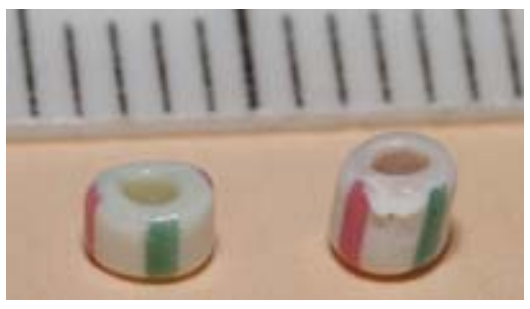

Accession: **C/952** Box 0046

Map. Hill

*Striped*

(~10)

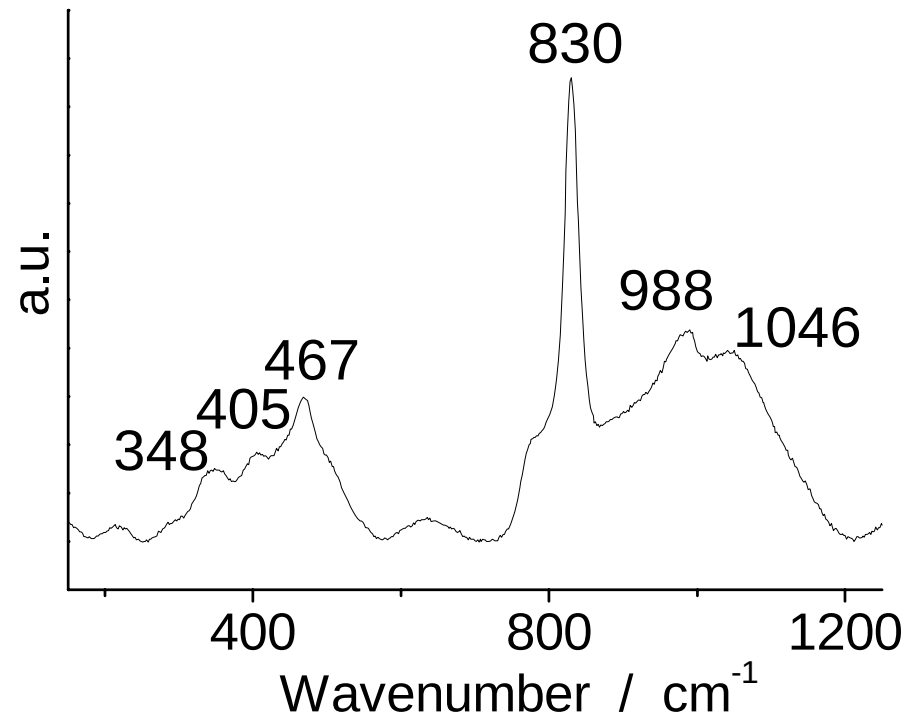

Ca and Pb Arsenate 19th c.

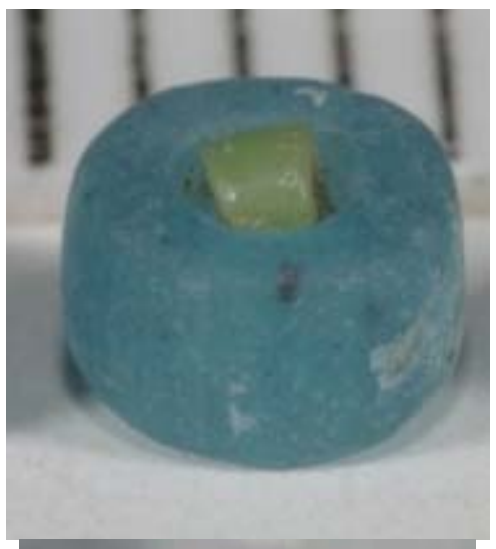

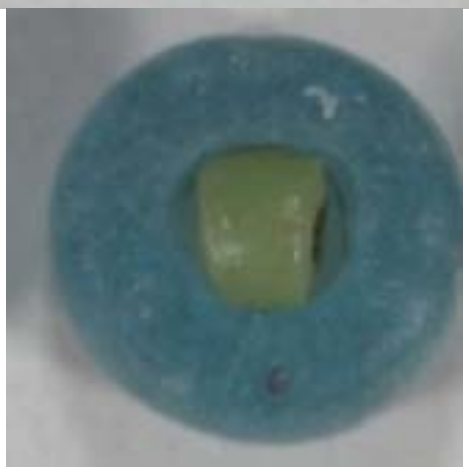

Accession: **C/952** Box C0046

Map. Hill

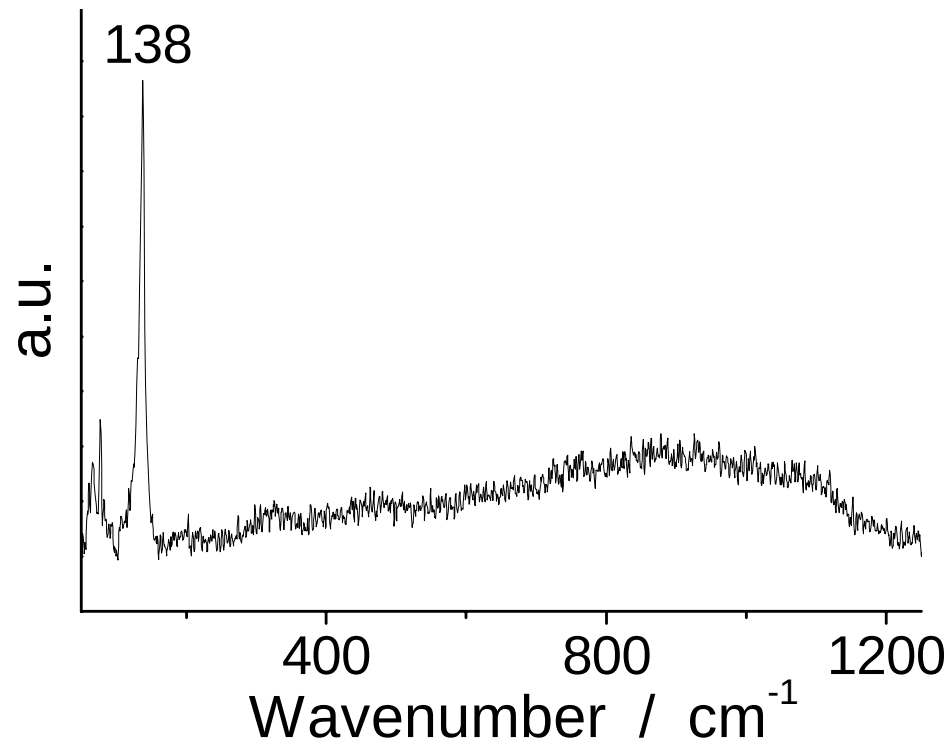

$Pb_2Sb_2O_7$ Naples Yellow Renaissance

* T64000

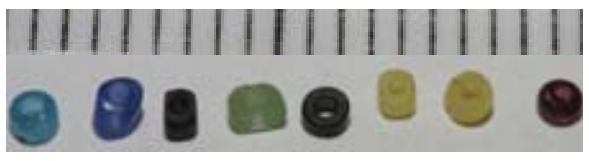

Accession:
**C/952**
Box 0046

Map. Hill

*tiny*

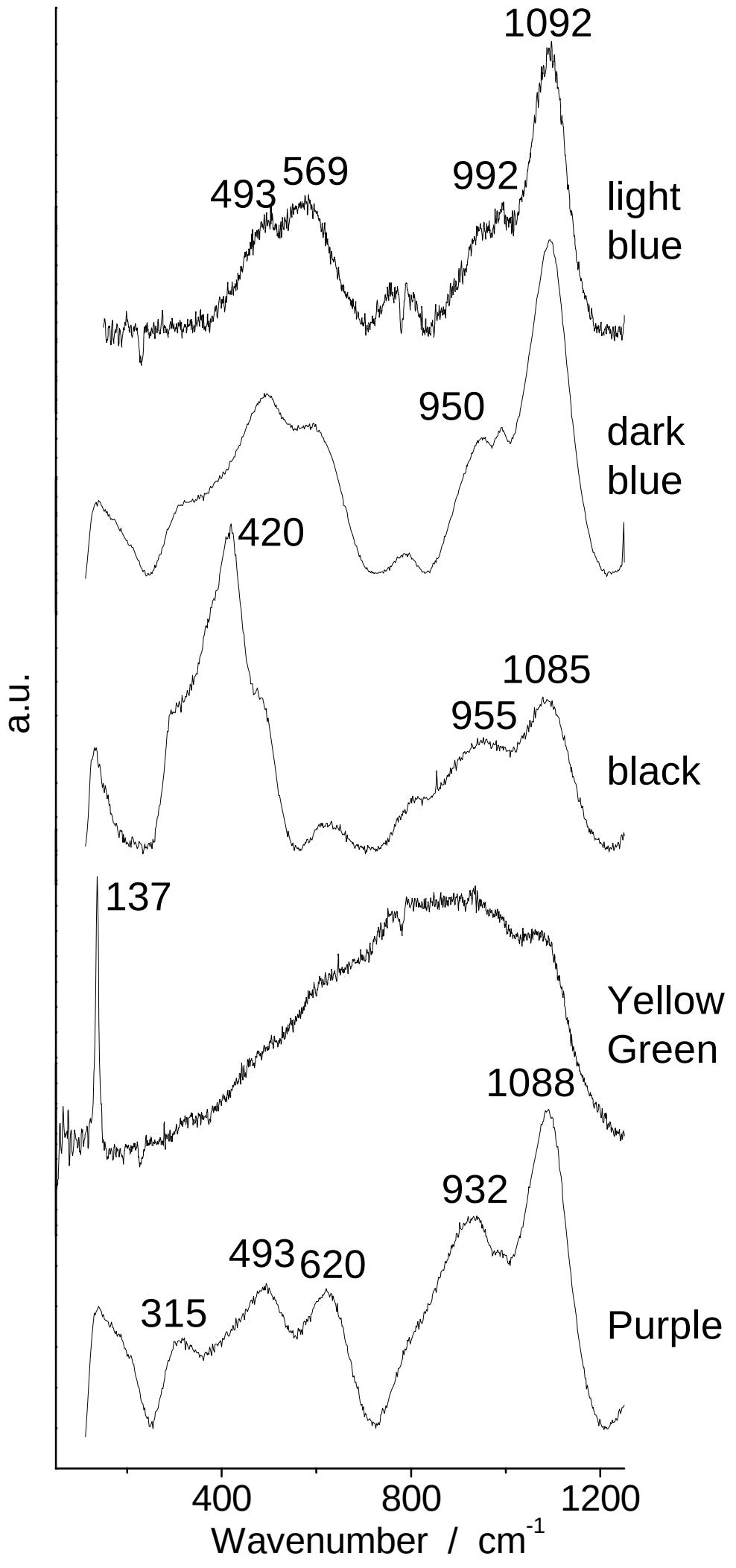

$Na_2O/CaO$

* T64000

$Na_2O$

Fe-S chromophore

$Pb_2Sb_2O_7$
Naples Yellow

* T64000

$Al_2O_3/Na_2O$
$Na_2O$

## 2. MAPUNGUBWE AND K2 ARCHAEOLOGICAL SITES

Mapungubwe is a small flat-topped sandstone plateau situated in the Limpopo valley close to the present-day borders of South Africa, Botswana and Zimbabwe (Figure 8). Excavations on Mapungubwe hill during 1933-1934 exposed three "royal" burials in which gold funerary objects, gold beads and bangles, were found together with imported glass beads (Fouche, 1937). The sheer volume of beads recovered is remarkable; from one burial alone 26 037 glass beads were counted (Saitowitz, 1996). Shortly after the discovery of Mapungubwe another archaeological site, namely Bambandyanalo (later renamed K2), approximately 1 Km to the southwest of Mapungubwe hill, was also discovered. During subsequent excavations at both sites (1934-1940) thousands more glass beads were found (Gardner, 1963).

The excavations on Mapungubwe uncovered several strata of occupation which were clearly separated by layers of ash indicating that the settlements were burnt down between occupations (Gardner, 1963). Pottery styles and bead types varied between the layers and made it possible to classify the beads into three different periods of time namely M1, M2 and M3, with M3 being the top layer (Gardner, 1963). This stratification provided a link to the K2 site as the type of beads that was excavated at K2 is the same as those found in the M1 layer, which is proof that this period of time of occupation on Mapungubwe coincides with the settlement at K2. It also showed that burials at K2 associated with M2 style beads are intrusive from the hill as M2 style beads were not found in the occupational material at K2 (Gardner, 1963). The M3 type of bead excavated in the uppermost layers was regarded as quite modern, some dating from the $18^{th}$-$19^{th}$ centuries (Beck, 1937; Gardner, 1963). Based on this and other information, such as the re-growth of plant species, the last occupation date on the hill was estimated to be the middle of the $19^{th}$ century (Gardner, 1963). Further excavations at the site 1960-1975 used carbon dating to re-date the last occupation date on the hill as 1280 AD (Meyer, 1998; Vogel, 1998; Woodborne *et al.,* 2009).

a)

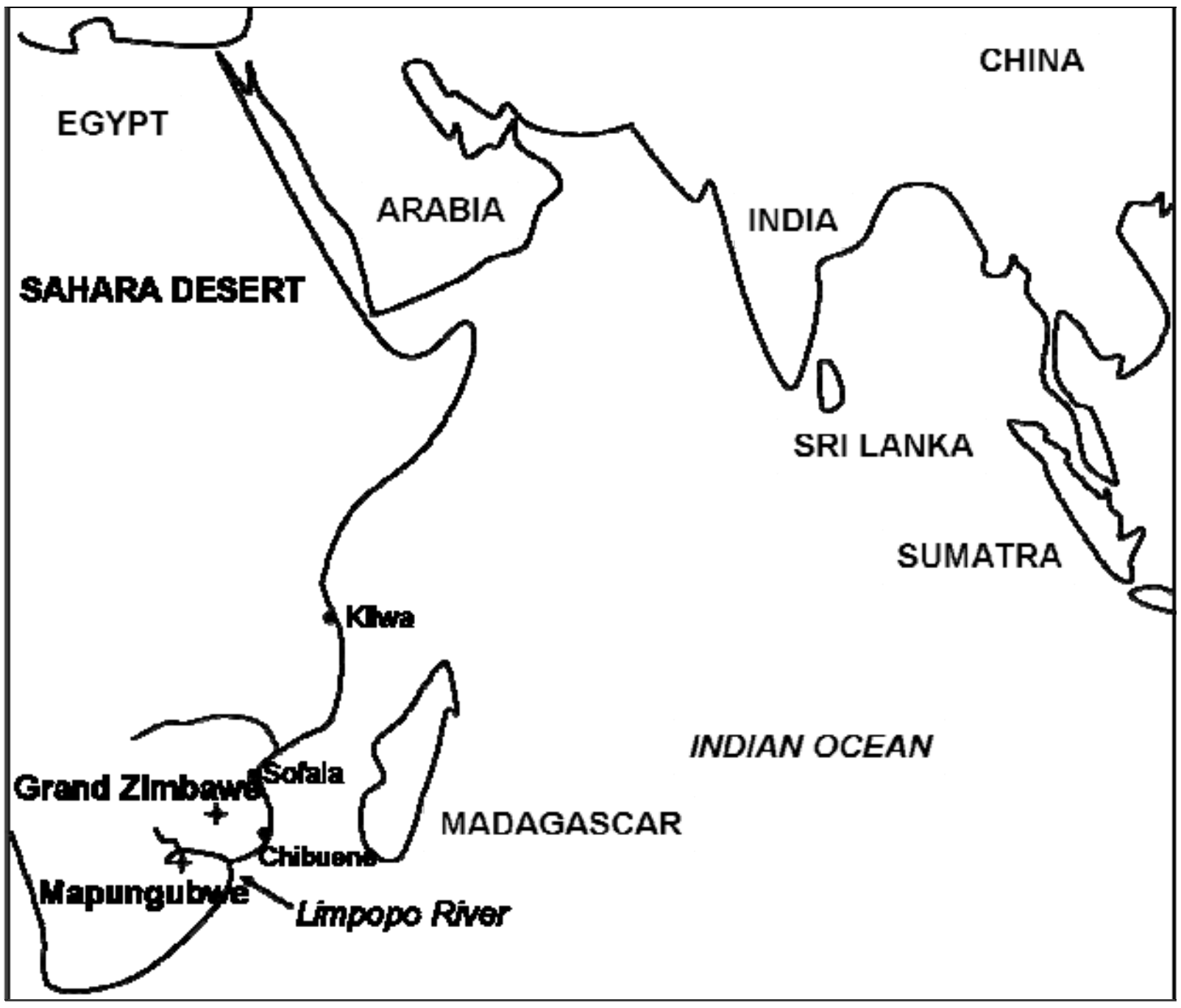

b)

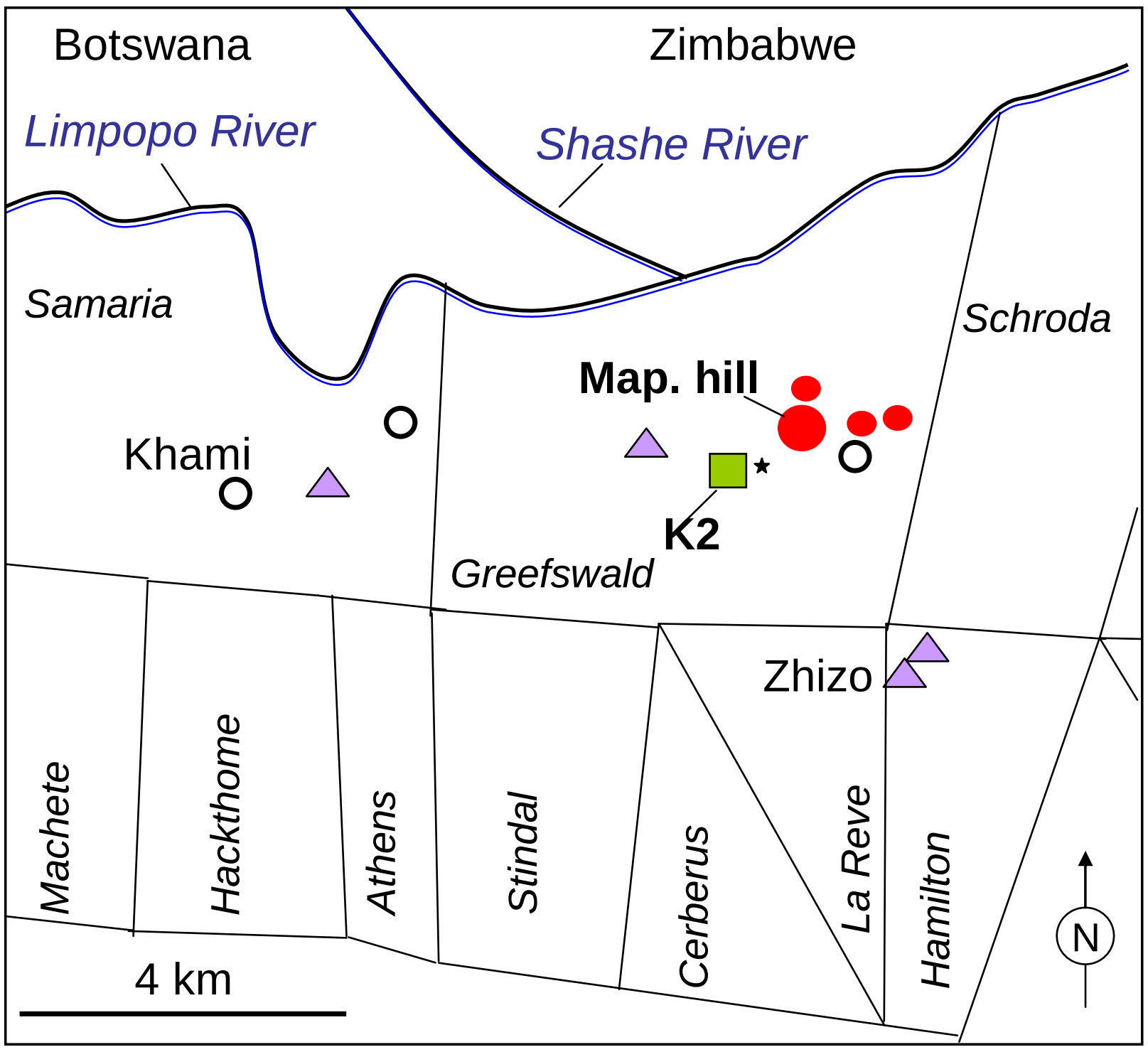

*Figure 8: a) Location of the Mapungubwe sites at the confluence of the Limpopo and Shashe rivers; b) details of the different excavation spots at Mapungubwe.*

Raman and XRD analyses of the glaze of Chinese celadon shards, excavated in 1934 in the M3 layer on Mapungubwe hill, re-dated the shards from its original classification as the Song dynasty (1127-1279 AD) to the Yuan (1279-1368 AD) or early Ming (1368-1644 AD) dynasties, which was an indication that the last occupation date of Mapungubwe may have been one to two centuries later than 1280 AD (Prinsloo *et al.*, 2005). In order to obtain supportive evidence for these results, which will have an impact on the chronology of the whole region, a Raman spectroscopic study of the thousands of glass trade beads, excavated at the same site was initiated. An in-depth study, utilizing Raman spectroscopy and supportive techniques, was conducted on the Mapungubwe oblates, the beads associated with the gold-bearing graves and representing the M2 time period (Prinsloo and Colomban 2008). Although very precise information about the technology used to produce and colour the glass was obtained, the provenance of the beads could not be established.

## 3. ORIGIN OF GLASS BEADS

Exotic imports, when found in an archaeological context, are clear proof of international trade and if the provenance of the object can be determined, light can be shed on trade routes and even roughly date the site. To unravel the early history of sub-Saharan Africa this is of cardinal importance since only a few written records of the pre-Portuguese era exists in western literature and up to the 19$^{th}$ century this is also sparse. Glass beads are the most abundant of the imported trade goods that have been conserved and excavated at archaeological sites all over Africa. The central position of Mapungubwe hill, situated at the confluence of the Limpopo and Shashe rivers, made it from the earliest time accessible through old caravan routes to Central Africa (e.g. Great Zimbabwe), Egypt and the Mediterranean world (Saitowitz, 1996). The Limpopo River also connected Mapungubwe to the African east coast (outposts of the Swahili Kilwa Sultanate: Sofala, Catembe (now Maputo), etc.) and by sea trade as far as the Red sea, Arabic gulf, India, Champa and China (Saitowitz, 1996, Colomban, 2005). Furthermore, trade along the African west coast was accessible via the interior through Botswana and Angola, where Portuguese mariners traded in beads from Europe 150 years before they rounded the Cape of Storms and also dominated the African east coast trade (Saitowitz, 1996). Possible trading partners therefore were numerous and the potential origin of the beads widespread.

Not all the beads were imported. “Garden roller” beads (Figures 1, 9a) shaped like the heavy rollers used to press grass lawns were discovered at Bambandyanalo (K2) in 1934. The blue-green beads were unlike any bead previously excavated at South African Iron Age archaeological sites and the discovery of broken (and one whole) clay moulds at K2, into which the beads fit perfectly proved that the beads were manufactured on site (Gardner, 1963). Garden rollers were also excavated on Mapungubwe (layer M1), but not one clay mould (Figure 9a) was found, indicating that the beads were manufactured at the K2 site and traded to Mapungubwe. It is the most southern site in Africa where evidence of glass-reworking has been discovered and it was shown that the beads were manufactured by heating smaller glass trade beads found in association with the large "Garden Rollers" up to the temperature required for sintering and glass formation (Davison, 1973; Saitowitz, 1996; Wood 2000).

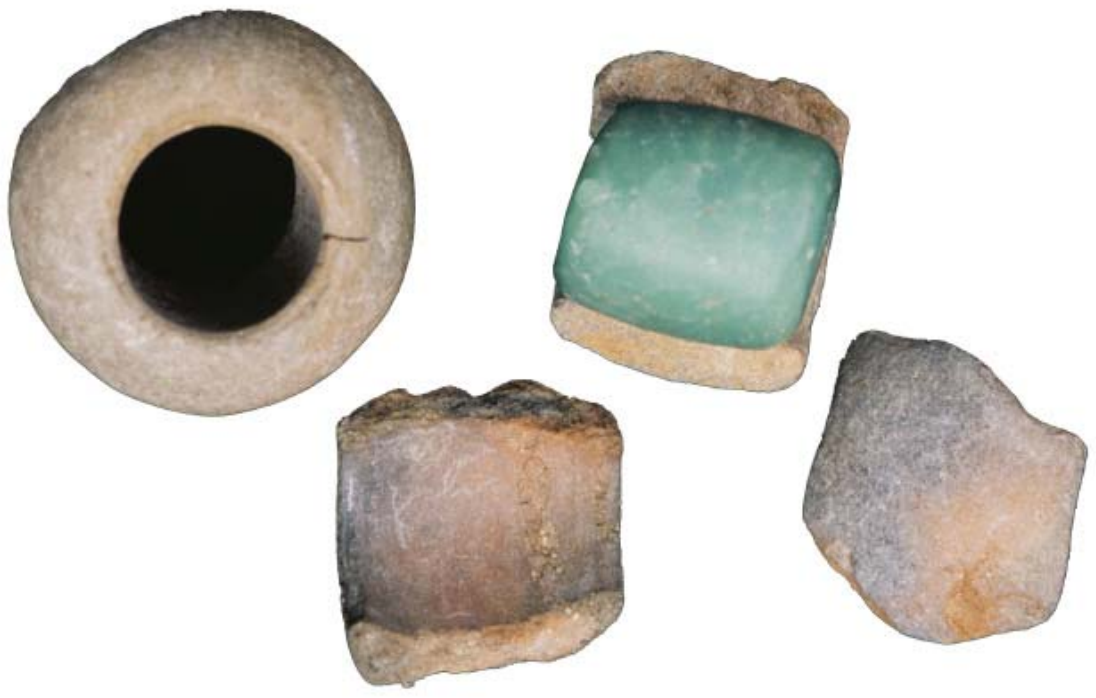

a)

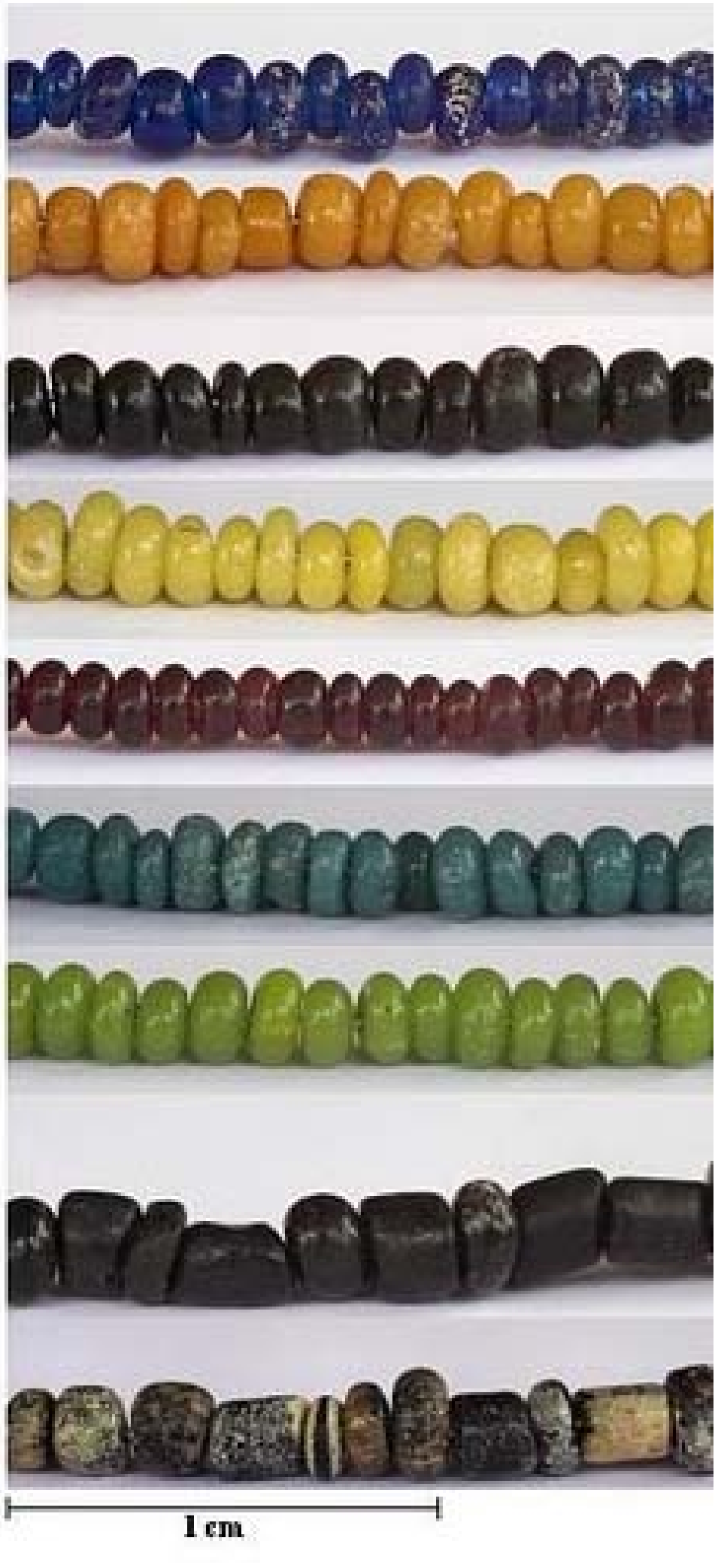

b)

*Figure 9: a) Garden Roller clay moulds and bead. (bead diameter ~ 12-15 mm ; photograph Mapungubwe museum files.), b) Mapungubwe oblates (photograph A. Carr)*

Initial studies on the beads found during the early excavations classified them according to morphological appearances, specific gravity tests and fluorescence measurements (Beck, 1937; Van Riet Lowe, 1955; Van der Sleen 1956; Gardner, 1963).

India, Fustat (Egypt) and Venice were proposed as origin for the older beads, whilst some beads such as the large cobalt blue annular ("Dutch Dogons" made in Germany) and hexagonal ("Russians" made in Czekoslovakia) beads were recognized as originating from Europe. The beads associated with the burials (in particular the three gold-bearing skeletons) comprise 80% of the collection. Davison (1974, 1979) found these Mapungubwe beads to be chemically different to the beads that occur at archaeological sites in southern and eastern Africa and at that time known as "trade wind beads" (after Van der Sleen, 1956). She named the beads the Mapungubwe Oblates and also found that the beads from the K2 site (M1 layer at Mapungubwe) were chemically in a separate group (XRD measurements) and after Gardner named it the M1 group (Davison and Clark 1974; Davison 1979). Extensive studies using XRD, REE and XRF measurements of the burial beads and the beads excavated by Meyer and Eloff (later excavation) were undertaken by Saitowitz who proposed Fustat as the origin of some of the beads based upon REE analysis, but gave inconclusive scientific evidence (Robertshaw *et al.*, 2010; Wood 2000, 2005).

Wood proposed a classification system for the beads excavated in Southern Africa before contact with the west, according to their morphological and technological attributes and dated each series by reference to established radiocarbon chronologies (Wood, 2000, 2005). Table 1 summarizes the bead series, which was also identified by her as occurring at Mapungubwe and K2. She named the earliest beads Zhizo (after excavation site where they were first excavated), K2 (after the K2 site and the same as the M1 beads of Gardner and Davison), Garden rollers (after Gardner), Indo-Pacific (after Francis, 2001 who replaced "trade wind beads" of van der Sleen) and Mapungubwe oblates (after Davison). A few large Islamic beads were found at Mapungubwe, but not analysed in this study. She also described a Zimbabwe and Khami beads series, of which she did not find examples amongst the beads she studied from the Mapungubwe and K2 sites. Not many of the oldest type of beads, namely the Zhizo bead series (Figure 9b), occur at Mapungubwe and they are usually heavily corroded.

The colouring elements used in beads imported into southern African before the 14$^{th}$ century were determined by elemental analysis: iron, lead, tin, copper and cobalt (Robertshaw *et al.,* 2010). The association of Pb with Sn in yellow glass indicated that lead stannate (lead tin yellow type 1 or 2) was the preferred yellow pigment, particularly for plant ash glasses (Tite *et al.,* 1998) Iron dissolved as $Fe^{3+}$ was used to obtain yellow (Perreira *et al.,* 2009) and acts as a reducing agent for copper so that it precipitates as metal or cuprite in brownish-red glasses. (Nassau, 2001; Colomban, 2009; Colomban and Schreiber, 2005). Copper was used to make blue-green and green glasses and cobalt generally for deep blue.

| Bead series | Traded period in southern Africa | Method of manufacture | Size | Colour | Shape |
|---|---|---|---|---|---|
| Zhizo | ca. 8-10 AD | Drawn | 2.5-13 mm diameter up tp 20 mm long | Cobalt, yellow, blue-green, green | Tubes |
| K2 | ca. AD 980-1200 | Drawn | -Small -2-3.5 mm diameter 1.2-4 mm long | Transparent to translucent blue-green to light green | Tubes, cylinders |
| K2 Garden Roller | ca. AD 980-1200 | Reheated K2 series beads in single-use clay mould | -Large -10 to14 mm diameter to 7 to 15 mm long | Transparent to translucent blue-green to light green | Barrel-shaped |
| Indo-Pacific | ca. AD 1000-1250 | Drawn with ends rounded through reheated | Vary but most are 2.5 to 4.5 mm diameter | Black and brownish-red beads are opaque; yellow, soft orange, green and blue-green are translucent | Vary but most are cylindrical |
| Mapungubwe oblate | ca. AD 1240-1300 | Drawn, Heat rounded | Uniform 2 to 3.5 mm diameter | Opaque black the most common Cobalt blue, green, plum, turquoise, bright orange | Uniform Oblate or cylindrical with well-rounded end |
| **Islamic** | Ca. AD 1250-1300 | Not well defined | large | Decorated with patterns made up of glasses of several colours | Not well defined |

*Table 1: Bead series classification based on morphological attributes and elemental analysis according to Wood.*

A comparison of the chemical composition of the main bead types are given in Table 2, the values are given in oxide percentages (Robertshaw *et al*., 2010). The bead series are all soda-lime silica beads. The K2-type, K2 Garden Roller and Indo-Pacific beads are characterized by mineral soda and a high quantity of alumina (Table 2). The K2 series contains the largest quantity of soda (~ 16 %) and the Indo-pacific series is rich in $Fe_2O_3$ (~ 2 %). The Mapungubwe oblate series is distinct from the K2 and Indo-Pacific series because of the high quantity of manganese (~ 6 %) and calcium (~ 7 %).

| Site | *K2* | *K2* | *Indo-Pac* | Map oblate |
|---|---|---|---|---|
| Assignment | **GR** | | | Wood |
| $Na_2O$ | 14.36 | **16.22** | 14.75 | 13.47 |
| MgO | 0.37 | 0.43 | 0.59 | **5.8** |
| $Al_2O_3$ | **16.63** | **11.85** | **13** | 7.67 |
| $SiO_2$ | 61.05 | 64.51 | 63.08 | 61.88 |
| $K_2O$ | 3.39 | 3.34 | 3.46 | 3.47 |
| CaO | 2.85 | 2.34 | 2.85 | **6.66** |
| $Fe_2O_3$ | 1.35 | 1.3 | **2.27** | 104 |
| **Type** | | **2** | | **3** |

*Table 2: Average composition of the main beads series (K2, Garden Roller (K2GR), K2, Indo-pacific and Mapungubwe) excavated at Mapungubwe and K2 according to Wood and Robertshaw et al.*

It has been established that Raman spectroscopy is an excellent and non-invasive method to characterise glasses as both the type of glass matrix and pigments used as colouring agents can be determined from a Raman spectrum (Burgio and Clark, 2001; Colomban, 2003a; Colomban *et al.,* 2003c; Ricciardi *et al.,* 2009a). The Raman spectra and XRF analysis of the Mapungubwe oblates confirmed the glass to be typical soda/lime/potash glass similar to Islamic glass from the $8^{th}$ century (Ommayad) (Colomban *et. al,* 2004a), but with higher levels of aluminium, iron and magnesium. The turquoise, bright green, bright yellow and orange colours were obtained by utilizing a combination of cassiterite ($SnO_2$) and lead tin yellow type II ($PbSn_{1-x}Si_xO_3$). Doping with cobalt and manganese produced dark blue and plum coloured beads. The Fe-S chromophore was detected through its resonance enhanced spectrum in the black beads. Corrosion of the black beads was investigated and an organic phase detected on the beads, which might have influenced the corrosion process. We have shown that Raman spectroscopy can classify the glass matrix and provide unique information about the pigments used that is not possible with other analytical techniques.

## 4. EXPERIMENTAL

Micro-Raman spectroscopy was performed with HR 800, T64000 and Labram Infinity micro-Raman spectrometers from HORIBA Scientific, Jobin Yvon Technology (Villeneuve d’Ascq, France).

All the beads were analysed with the HR 800 with a 514 nm Argon exciting laser (50mW at laser exit), a x50 objective and 600 t/mm grating with recording times ranging between 50-300s. The T64000 and Labram Infinity Raman spectrometers (514 nm laser) was used later to record spectra on some beads where no Raman signature or spectra of very poor quality were obtained with the HR 800 (~10% of the spectra). In many cases where the Raman signal of the beads was low the edge filter of the HR 800 caused interferences in the form of a wavy background and as the T64000 spectrometer uses a triple monochromator system to eliminate contributions from the Rayleigh line, this aspect is eliminated. In the database an asterisk indicates the spectra recorded with the T64000 or Infinity, without asterisk means that we used the HR 800.

The Raman signature of an amorphous silicate is composed of two massifs, one centered around 500 $cm^{-1}$ and the other one around 1000 $cm^{-1}$ (Furukawa *et al.,* 1981; McMillan and Piriou, 1982). The first represents bending vibrations of the $SiO_4$ tetrahedron and the second stretching vibrations. It has been shown that the second peak can be decomposed into five components following the $Q_n$ model, namely $Q_0$, $Q_1$, $Q_2$, $Q_3$ and $Q_4$ which are attributed to stretching vibrations of $SiO_4$ tetrahedra with 0, 1, 2, 3 or 4 bridging oxygens. Because the $SiO_4$ tetrahedron is a very strong chemical and vibrational entity, the Raman signature of silicates depends on their nanostructure and can be directly correlated to the glass composition. The wavenumber maxima of the Si-O bending and stretching massives are well-established tools for the classification of amorphous silicates (Colomban, 2003a; Colomban *et al.,* 2006a; Tournié *et al.,* 2008).

The analyses were made, where it was possible, on a clean part of the bead in order to obtain the Raman signature of the original glass without interference of corrosion products. In order to compare and identify the silicate glasses according to their Raman signatures, the fluorescence background was subtracted and Raman parameters extracted from the spectra such as the maximum position of the bending and stretching massifs respectively ($\delta_{max}$ $SiO_2$ and $\nu_{max}$ $SiO_2$) to use for the classification of the glass matrix.

## 5. GLASS MATRIX CLASSIFICATION

The glass matrix of the beads was classified according to the main parameters extracted from the Raman signature of the $SiO_4$ vibrational unit: the maxima position of the bending ($\delta_{max}$ $SiO_2$) and stretching massifs ($\nu_{max}$ $SiO_2$). This classification tool has proved its efficiency in the study of many glasses and enamels (Colomban *et al.,* 2006a; Colomban and Tournié, 2007, Kirmizi *et al.,* 2010a,b; Ricciardi *et al.,* 2009a,b). The spectra used are free of significant bands originating from pigments, which could disturb the glass matrix. In some cases a decomposition treatment (Colomban, 2003a; Colomban *et al.,* 2006a; Ricciardi *et al.,* 2009a,b) was used that makes it possible to extract the requested Raman parameters even when the pigment signature superimposes on that of the glass matrix.

In Figure 10a the maximum positions of the bending ($\delta_{max}$ $SiO_2$) and stretching massifs ($\nu_{max}$ $SiO_2$) of a very large collection of glass, varying in technology used, production period and origin, are plotted against each other. Specific areas on the plot represents different groups, in some instances compositions are also indicated. In Figure 10b the data for the selection of beads excavated at Mapungubwe hill and K2 are added to the figure. All the beads fit into two main groups, namely soda ($Na_2O$) and soda-lime ($Na_2O/CaO$) glasses, which is in agreement with previous chemical analysis of the beads (Davison and Clark 1974; Davison, 1979; Saitowitz, 1996; Robertshaw *et al.*, 2010). Only one spectrum corresponds to a PbO-rich composition. According to previous studies (Colomban *et al.*, 2006a; Kirmizi *et al.,* 2010a,b; Ricciardi *et al.,* 2009a,b), when the quantity of alumina increases the position of the bending massif shifts towards lower wavenumbers while an increase in the quantity of lead shifts the maximum position of the stretching massif downwards. Lixiviation also induces a wavenumber shift and a decrease in intensity (Tournié *et al.*, 2008). The large scattering of the data shows a variability of composition, which are related to the raw materials used (mineral or sea plant ash for soda glass, continental plant ash for potash-lime glass), the workshop technology or the recycling of glass of different origins. Pigment addition can also shift the position of the bands. The beads excavated on Mapungubwe hill are well distributed into $Na_2O$ and $Na_2O/CaO$ groups while the majority of beads excavated at K2 are in one group $Na_2O$. The diversity of the Mapungubwe hill beads arises from the longer occupation time of the site in comparison to the K2 site and indicates diverse origins of the beads.

a)

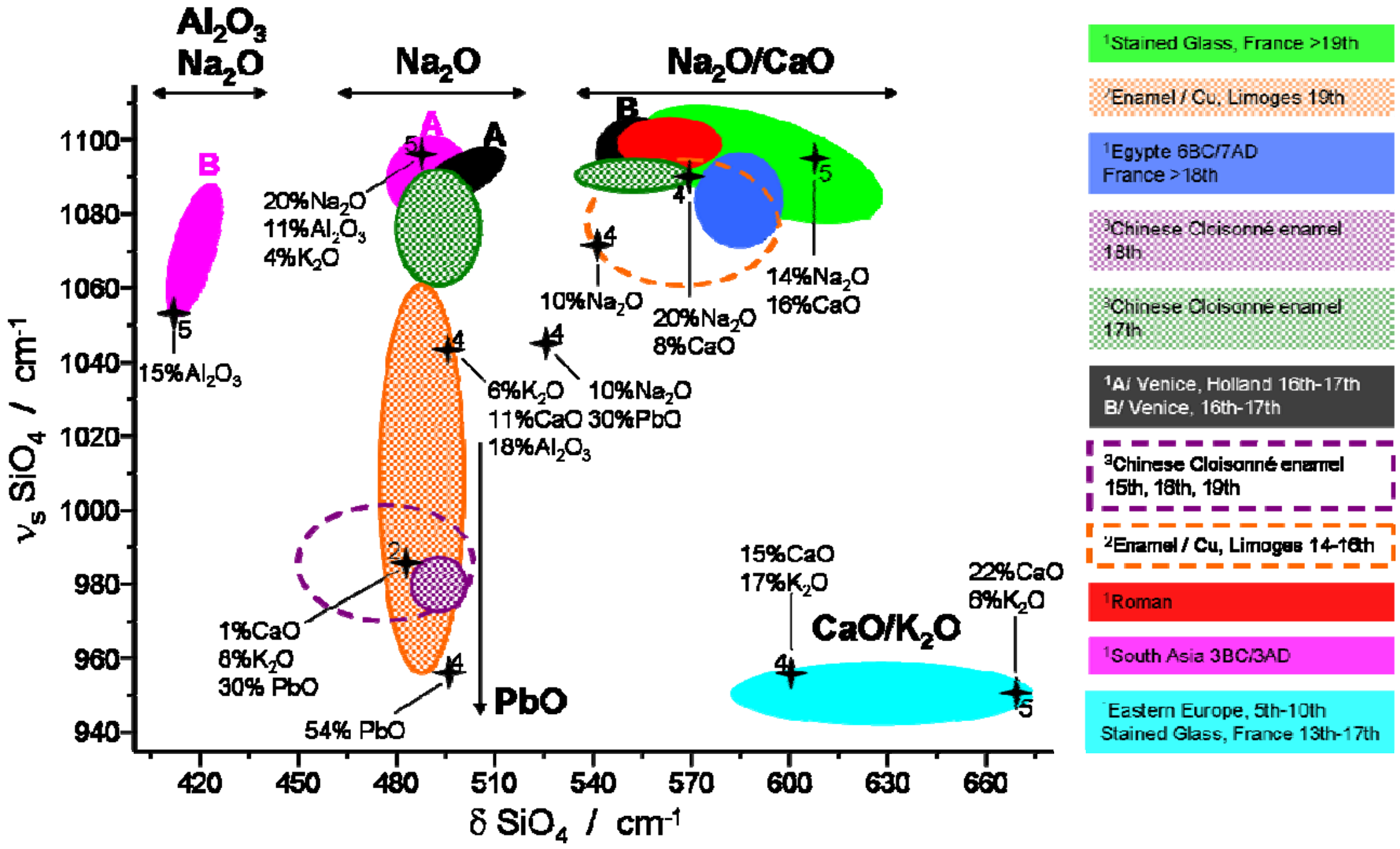

b)

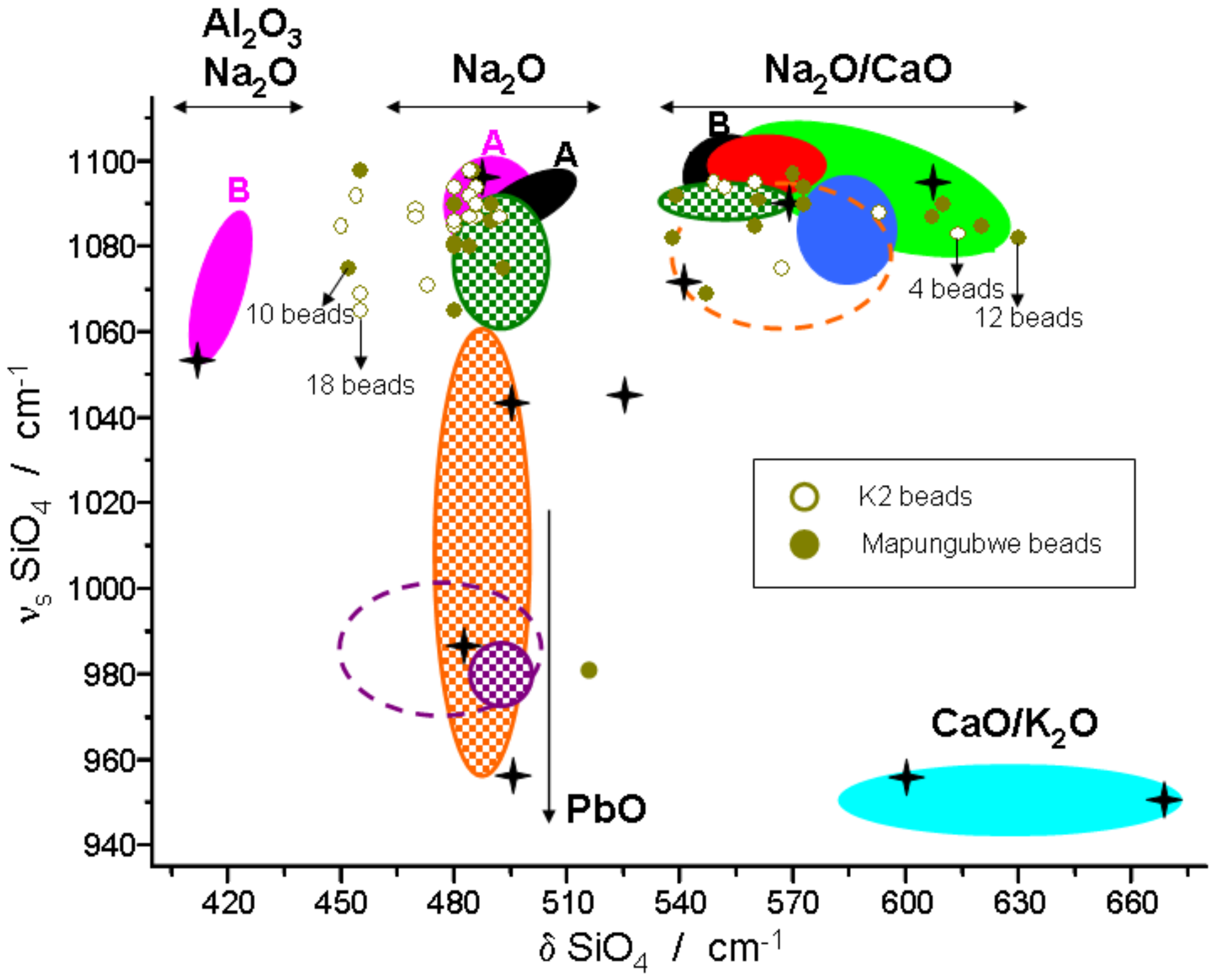

*Figure 10: Classification using the $SiO_4$ bending massive versus stretching massive plot a) of a large glass corpus (1, Ricciardi et al., 2009b; 2, Kirmizi et al., 2010a,; 3, Kirmizi et al., 2010b; 4, Colomban et al., 2006a, ; 5,Tournié, 2009) and b) plus the data of the beads excavated at Mapungubwe hill and K2.*

In Figure 10b it can be seen that the K2 and Mapungubwe beads, with bending massif around 450 cm$^{-1}$, form a new group outside the established groups related to the $Na_2O$ group. The other beads fall into the established groups. The majority of beads from the K2 site (stretching massif around 1090 cm$^{-1}$) are typical of South Asian glass (3BC/3AD) and beads with stretching massif around 1080 cm$^{-1}$ are similar to Chinese *Cloisonné* enamel and glass from Venice and Holland (16$^{th}$-17$^{th}$ c.) with a composition close to 20% of $Na_2O$, 11% of $Al_2O_3$ and 4% of $K_2O$ wt. The beads excavated at Mapungubwe hill, which belong to the $Na_2O/CaO$ group, exhibit bending massifs between 540 and 630 cm$^{-1}$ and are more scattered between groups than the beads from the $Na_2O$ group. The beads with bending massif around 570 cm$^{-1}$, are included in the Roman and Chinese *Cloisonné* enamel (17$^{th}$ c.) areas with a composition close to 20% of $Na_2O$ and 8% of CaO wt. Mapungubwe beads with spectra with bending massif around 615 cm$^{-1}$ fall into the area of the stained glass windows from 19$^{th}$ c. with a composition close to 14% $Na_2O$ and 16% CaO. The Zhizo beads (Figure 9b) are corroded to a great extent, but it was possible to record spectra on both a corroded part of the glass and an original piece. The spectrum of the corroded part has a stretching massif at 1075 cm$^{-1}$ and falls into the group for corroded glass and the spectrum of the original glass (568, 1094 cm$^{-1}$) falls into the sodium group.

A more detailed classification of the glass types can be made by examining the shape of the vibrational signature that is directly related to the glass nanostructure and hence not only to the composition but also the production technology. In Figure 11, five representative Raman signatures of the glass matrix of all the collected spectra (details available in Tournié *et al.,* 2010) are shown. *Type 1* is characterized by three bands more or less equal at 320, 497 and 595 cm$^{-1}$ for the bending massif and three bands (2 small and 1 big) respectively at 952, 991 and 1092 cm$^{-1}$ for the stretching massif. *Type 2* is also characterized by 3 bands for the bending massif, but only two bands for the stretching massive at 943 and 1084 cm$^{-1}$. *Type 3* is a typical Roman glass (Colomban *et al.,* 2006a).

*Type 1* can be considered as an intermediate between *Type 2* and *Type 3* and *Type 4* is due to corrosion by acidic water. Previous studies (Colomban *et al.,* 2006b; Tournié *et al.,* 2008) have demonstrated that $K^+/H^+$ substitution (lixiviation) induces an increase in intensity of the ~480 cm$^{-1}$ component in the bending multiplet and a downshift of the stretching maximum. Comparing the spectra of *types 3* and *4* it is clear that the intensity of the 492 cm$^{-1}$ component has increased in *Type 4*, which indicates a modification of the O-Si-O angle in the $SiO_4$ polymeric framework due to proton insertion (Scanu *et al.,* 1994; Sharma *et al.,* 1983). Also, a downshift from 1090 to 1075 cm$^{-1}$ is observed for the stretching component and corresponds to a lengthening of the Si-O bond due to the interaction of the oxygen atoms with the inserted protonic species. Finally *Type 5* is a lead-rich glass easily identified by a strong stretching mode at 972 $cm^{-1}$. A narrow component at ~985 $cm^{-1}$ could correspond to sulphate traces (Konijnendijk and Buster, 1977; Lenoir *et al.*, 2009).

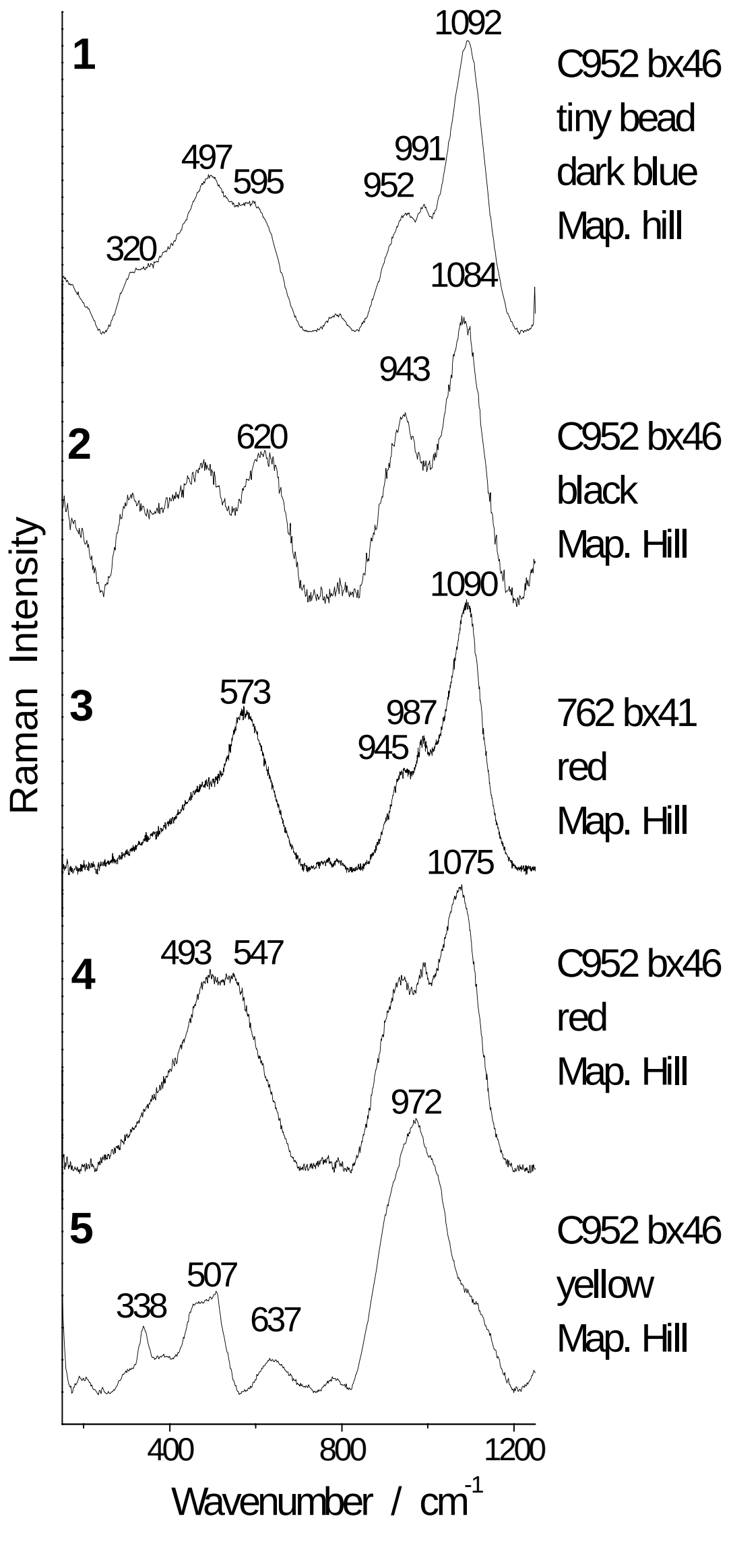

*Figure 11: Representative Raman signatures of the five types glass matrix observed among the corpus of glass beads. The type 2 is a soda glass. The type 3 is soda-lime glass, the type 4 is a corroded glass and the type 5 corresponds to the lead. The type 1 seems to be an intermediate between the type 2 and 3.*

Figure 12 focus only on the data from beads excavated at the Mapungubwe hill and K2 sites. When it is possible the type of glass series according to the Wood and Gardner classifications are indicated. The glass matrix of the majority of K2 beads are *Type 2* and the data scattering is weak; however two spectra belong to *Type 1* and three to *Type 4*. In contrast, the beads from the Mapungubwe hill excavation are distributed among all the types of glass with a small majority of *Type 2*. These observations confirm the previous results obtained from Figure 10 namely that the data of the assemblage of beads from Mapungubwe hill is scattered broadly and may be related to many different origins and periods of production.

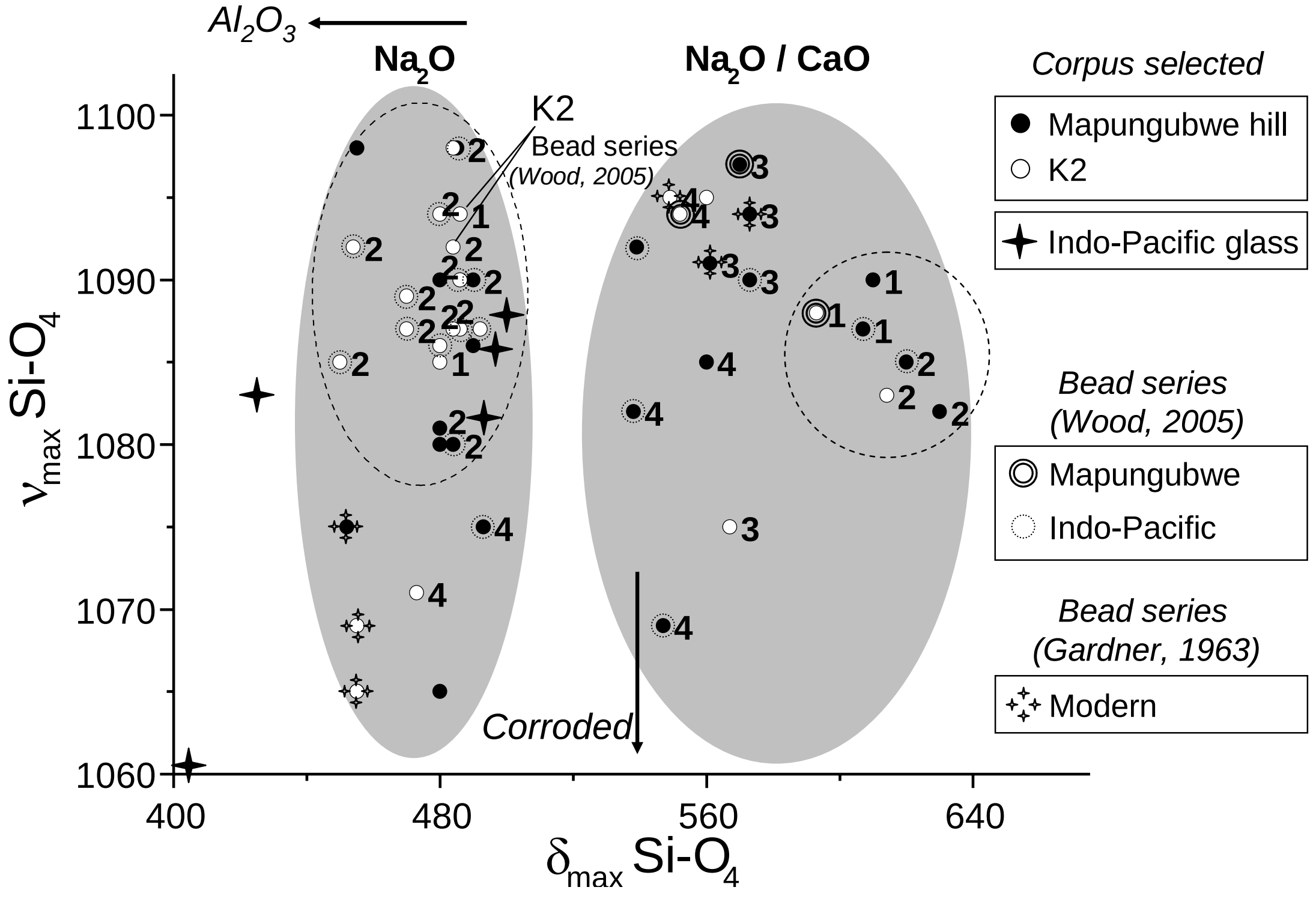

*Figure 12: Classification of the glass beads from Mapungubwe hill and K2 sites according to the bending and stretching massive of $SiO_4$. The type of glass matrix ($Na_2O$ and $Na_2O$/CaO), the tendency of glass matrix ($Al_2O_3$, PbO or corroded) and the bead series (K2, Mapungubwe, Indo-Pacific and European) are also indicated.*

If we compare the bead series from the Wood classification, we notice that among the Mapungubwe excavated beads there is Indo-Pacific beads (mostly, *Type 2*) and modern European beads (*Type 3* and corroded *Type 4*). Down shift is due to lead addition and/or corrosion. *Types 1* and *2*, both from Mapungubwe hill and K2 sites, fit into both glass matrixes, $Na_2O$ and $Na_2O$/CaO, while *Type 3* is concentrated in one group $Na_2O$/CaO. Note previous elemental analysis does not consider the corrosion and this induces a systematic error in the comparison of the compositions. The Raman parameters of these glasses are in the same area of *Type 1* and *2* beads classified by Wood as Indo-Pacific.

## 6. PIGMENTS

Raman spectrometry is a very efficient method to determine the type of pigment or chromophore used to colour glass and we have illustrated how Raman spectroscopy combined with XRF measurements can characterise the precise technology that was used to colour a group of beads (Prinsloo and Colomban, 2008). Table 3 summarises the pigments and other crystalline phases identified in this study and incorporates the results of the study of the Mapungubwe Oblates. Bead colour, main Raman peaks, first production or origin and some references are also indicated.

| **Pigment** | **Colour bead** | **Phases** | **Raman peaks / $cm^{-1}$** | **1st production origin** | References |
|---|---|---|---|---|---|
| Lazurite | **turquoise** | **$S_3^-$ and $S_2^-$ in $Na_8(Al_6Si_6O_{24})S_n$** | **540** | **Synthetic from 1828** | Burgio and Clark, 2001 |
| Calcium antimoniate | **Blue, green, white** | **$CaSb_2O_7$** | **480, 634** | **Antiquity** | Gedzeviciûtë *et al.*, 2009 Mass *et al.*, 1998 |
| Ca / Pb Arsenate | **Blue, white, pink, striped** | **$CaAs_2O_6$** | **830** | **16th c. in Italy** | Verità, 2000 |
| Chromate | **blue** | **$CrO_4$ ions** | **844** | **18th c.** | Colomban *et al.*, 2004 Edwards *et al.*, 2004 |
| Copper | **Blue, red** | **Cu** | **490** | **Antiquity** | |
| Cobalt | **blue** | **Co** | **605** | **Antiquity** | |
| Yellow Naples | **Yellow, green** | **$Pb_2Sb_2O_7$ pyrochlore** | **144** | **1st synthesis 1570-1293 BC in ancient Egypt** | Kirmizi *et al.*, 2009 |
| Stannate | **Yellow, green** | **$PbSnO_4$ pyrochlore** | **129, 196 457** | | Bell *et al.*, 1997 Burgio and Clark2001 |
| Clinopyroxene diopside | **green** | **?** | | | |
| Pb-Sn solid solutions | **orange** | **pyrochlore** | **112, 136, 280, 427, 528** | | Prinsloo, Colomban, 2008 |
| Amber | **Black** | **Fe-S chromophore** | **415** | **From India, Sri Lank, Mapungubwe** | Prinsloo and Colomban, 2008 |
| Amber | **Black** | **Fe-S chromophore Mapungubwe Oblate** | **290, 332, 355, 410** | **Mapungubwe** | Prinsloo and Colomban, 2008 |
| **Spinel** | Black | $Fe_{3-x}$ $[Co, Cr, Zn]_xO_4$ | 640 | | |

*Table 3: Pigments and their main Raman signature observed in the glass beads.*

### *6.1 Blue*

Eight different Raman signatures were observed for the blue beads (Figure 1). The lazurite signature, due to the sulphide ion chromophore, typical of natural lapis lazuli (Colomban, 2003b) or of synthetic ultramarine (Picquenard *et al.,* 1993, Burgio and Clark, 2001) is observed for several glass beads (Figure 1, Table 3). Differentiation between lapis lazuli and ultramarine is not possible from their Raman spectra. Lapis lazuli was identified in some Iranian ceramic glazes made during the 13$^{th}$/14t$^{h}$ c. (Colomban, 2003b) and is very common in Mameluk glasses from 13$^{th}$-15$^{th}$ c., (Colomban *et al.,* to be published, Ward, 1998; Brill, 2000). The use of lapis lazuli in these beads as pigment led us to assign them to an Islamic (Fustat or Iran) production.

The Raman spectrum of the blue bead with an intense peak at 844 cm$^{-1}$ is typical of chromate ($CrO_4$) ions (Figure 1, Table 3) as found in Iznik chromites (Colomban *et al.,* 2004a) and in some 18$^{th}$ century porcelain pigments (Edwards *et al.,* 2004). The peak at 995 cm$^{-1}$ is assigned to a Ca/Al silicate (Colomban, 2004). Actually some Iranian and Asian cobalt sources consist of Co-Cr (Fe) or Co-Mn (Fe) ores (Colomban, 2003b; Colomban *et al.,* 2004b; Ricciardi *et al.,* 2009b; Gratuze *et al.,* 1992).

In some the spectra of Figure 1 a pair of peaks at 454 and 624 cm$^{-1}$ appears, which is typical of calcium antimonate (Gedzeviciûtë *et al.,* 2009; Ricciardi *et al.,* 2009a). This pigment was used from antiquity (Mass *et al.,* 1998; Ricciardi *et al.,* 2009a) as opacifier or white pigment in Egypt. In Europe a transition was made from tin-rich to antimony-rich opacifiers during the 17$^{th}$ century, which was again replaced by arsenic in the 18$^{th}$ and 19$^{th}$ centuries (Sempowski *et al*., 2000).

A Raman signature of arsenate was observed (Figure 1, Table 3) for milky blue beads. The use of arsenate was developed in Italy (Venice) during the 16$^{th}$ c. (Verità, 2000) to make *lattimo* glass but the use only spread out during the 19$^{th}$ c. we assign this production to a 19$^{th}$ c. import. The transition from Sn to Sb to As is then clearly illustrated in the beads found on Mapungubwe hill (Sempowski *et al.,* 2000) and indicates a long trading period.

The Raman spectrum of a large cobalt blue hexagonal bead shows a narrow peak at 960 cm$^{-1}$, characteristic of calcium phosphate as observed in Islamic glazes and glazed Medici porcelain (Colomban *et al.,* 2004a). These beads were produced in Czekoslovakia (Bohemia) and used during the slave trade. Examples occur at many archaeological sites in Africa and are popularly known as “Russians”.

The Garden Roller (Figures 1, 9) bead analysed in this study are of the light green variety and some Raman spectrum in Figure 1 shows a sulphate signature ($SO_4^-$ stretching peak at ~988 $cm^{-1}$, Lenoir *et al.,* 2009; Konijnendijk and Buster, 1977) characteristic of a glass melt that have not been sufficiently annealed. This Raman signature is very close to a Raman signature of glass containing a high quantity of alumina. It has been shown that the Al content of the Garden Rollers are more that the small beads that they were made of and it was attributed to contamination from the clay mould in which the beads were fired (Wood, 2000). This spectrum is also very similar to that of a corroded glass (*Type 3*) because of the peaks at 480 and 988 $cm^{-1}$ previously attributed to corroded glass (Tournié *et al*., 2008). This spectrum is not representative of the Garden Roller beads as there is a very large variation in colour and degree of vitrification between the beads and some spectra closely resembles a *Type 3* glass (Roman) (Prinsloo *et al.,* 2007).

In some spectra of Figure 1 there is no Raman signature of a pigment but only that of the glass matrix. Spectrum a in Figure 1, characterized by an intense bending band at 548 $cm^{-1}$, is observed both for the blue (light, transparent, oblate, Figure 3 c6,9, when Cu ions are dissolved in the glass, the Raman signature of the glass matrix is preserved) and red beads (tube and cylinder, Figure 2). The blue beads belong to the K2 series and the red beads to the Indo-Pacific series according to the Wood classification.

Spectra of the deep cobalt blue beads show an intense bending band at 605 $cm^{-1}$ (Figure 1). The absence of a pigment signature indicates that the colour was obtained by dissolution of the Co transition metal ions (Colomban *et al.,* 2001).

The turquoise colour of Mapungubwe oblate beads was obtained by adding CuO to a small presentation of lead tin yellow type II and cassiterite as opacifier (Prinsloo and Colomban, 2008).

### *6.1 Yellow and green*

The bright yellow, for the Mapungubwe oblates, was obtained with lead tin yellow type II, as identified through Raman spectroscopy as well as XRF measurements. The Raman spectrum has bands at 138 (vs), 324 (m, br), which is very close to that of Naples Yellow (lead antimonite) (both pigments have the same pyrochlore structure) with bands at 140 (vs), 329 (m, br) and 448 (w, br) (Bell *et al.*, 1997), so it is not possible to distinguish unambiguously between the two pigments through their Raman spectra. Naples yellow as pigment was developed during the renaissance but it has been claimed that it was already present on tiles in Babylon from 16$^{th}$ c. BC (Mass *et al.*, 2002; Burgio and Clark, 2001).

Raman signatures of yellow beads are shown in Figure 3 and green beads in Figure 4. The position of the strongest peak in the spectra varies between 136-144 $cm^{-1}$ (Figures 3, 4), which shows that the pigments are all Pb–Sn-based pyrochlore solid solutions (Krimizi *et al.*, 2010a). Lead (II) stannate, $Pb_2SnO_4$ (lead tin yellow Type 1) with two very strong peaks at 129 and 196 $cm^{-1}$ was not observed.

Another type of Pb-Sn solid solution was identified (Figure 3) for the bright orange beads that forms part of the Mapungubwe oblates (Prinsloo and Colomban 2008). Raman bands originating from lead tin yellow type II can be distinguished, but in addition a very strong signal at 113 $cm^{-1}$ and a peak at 529 $cm^{-1}$ are observed (Figure 3). The orange colour was obtained by adding lead to the mix and the extra peaks occur at wavenumbers very near to that of red lead oxide (Prinsloo and Colomban, 2008). A similar spectrum has previously been reported in Islamic ceramics from Dougga in Ifriqiya, one sample dating from the 11$^{th}$-12$^{th}$ century (Zirides period), the other from the 17$^{th}$-18$^{th}$ century (Ottoman period). (Colomban *et al.,* 2001a)

The bright green of the small opaque Mapungubwe oblates was obtained by a combination of lead tin yellow type II and copper ions (Prinsloo and Colomban 2008). Two other Raman signatures were observed for green beads (Figure 4). One is assigned to calcium antimonate, which confers to the bead a milky aspect, and the colouring element is certainly copper because copper ions do not form specific phases that can be recognised in a Raman spectrum.

The second one corresponds to a large wound bead (Figure 4) found on Mapungubwe hill and the sharp peaks at 326, 363, 391, 669 and 1014 $cm^{-1}$ is characteristic of the clinopyroxene diopside. The Raman signature of diopside has previously also been detected in a glass paste vase from Egypt (6th–5th century BC) (Ricciardi *et al.,* 2009b).

### *6.3 Black and white*

The Raman signatures of black beads are shown in Figure 5. The black pigment of the thousands (>100 000) of black beads associated with the middle period of Mapungubwe was identified as a Fe-S chromophore (Prinsloo and Colomban, 2008). The majority of the black beads are Mapungubwe Oblates ,which vary in shape on different positions on the bead. For larger cylindrical and round beads the peak at 415 $cm^{-1}$ (Table 3) is assigned to the resonance enhanced Fe-S chromophore (Prinsloo and Colomban, 2008), a chromophore used to produce amber glass. The chemical composition of the round and cylindrical beads differs from the oblates and therefore have different origins. A similar spectrum of the Fe-S chromophore of the large cylindrical and round beads is also obtained for black South Indian and Sri Lanka glasses (Prinsloo and Colomban, 2008). In this study we identified another pigment for black beads corresponding to the spinel $Fe_3O_4$ or Co, Cr, Zn-substituted homologues (Colomban *et al*., 2008) and this is an indication that the black beads had even a fourth production centre.

Two pigments were detected for the white beads; calcium antimonite and Ca/Pb arsenate (Figure 6, Table 3). The arsenate were also detected on the pink (Figure 7) and striped beads (Figure 7) which are typical European beads and in agreement with Gardner's classification

## 7. GLASS MATRIX, PIGMENTS AND PLACE/PERIOD OF PRODUCTION

The glass matrixes of the lazurite coloured beads (glass *Types 2* and *3*, Figure 11) and beads containing calcium antimonite are distributed into both the $Na_2O$ and $Na_2O/CaO$ (Figure 13) groups. This implies that these blue beads have two different origins or periods of production. The glass matrixes of the arsenate-based blue beads and copper blue beads fit into the $Na_2O$ groups with a majority displaying the *Type 1* Raman signature. The cobalt blue beads fit in the $Na_2O/CaO$ groups (Figure 13) with a majority of *Types 1* and *2* and a few of *Type 3* Raman signatures.

The red beads belong in both groups $Na_2O$ and $Na_2O/CaO$ and the majority of the Raman spectra is *Type 4* (corroded glass, Figure 13). A red coloured glass was obtained from Neolithic times (Colomban, 2009) by distributing copper nanoparticules in a glass matrix and very large quantities of red glass was produced during Roman times (Ricciardi *et al.,* 2009a) and continued in Islamic productions. Copper corrosion is easy and explains the beads degradation.

The green and yellow beads with Ca antimonate and Naples yellow / lead tin yellow type 2 belong to the $Na_2O$ group except one green bead (Figure 13). For other beads also coloured with Naples yellow / lead tin yellow type 2, the Raman signature of the green bead is type 2 and the yellow one is type 4 (Roman like glass corroded).

The white beads coloured with arsenate fit into to the $Na_2O$ group while Ca antimonate white beads belong to the $Na_2O/CaO$ group (*Type* 3). Finally, the black beads belong to the $Na_2O/CaO$ groups.

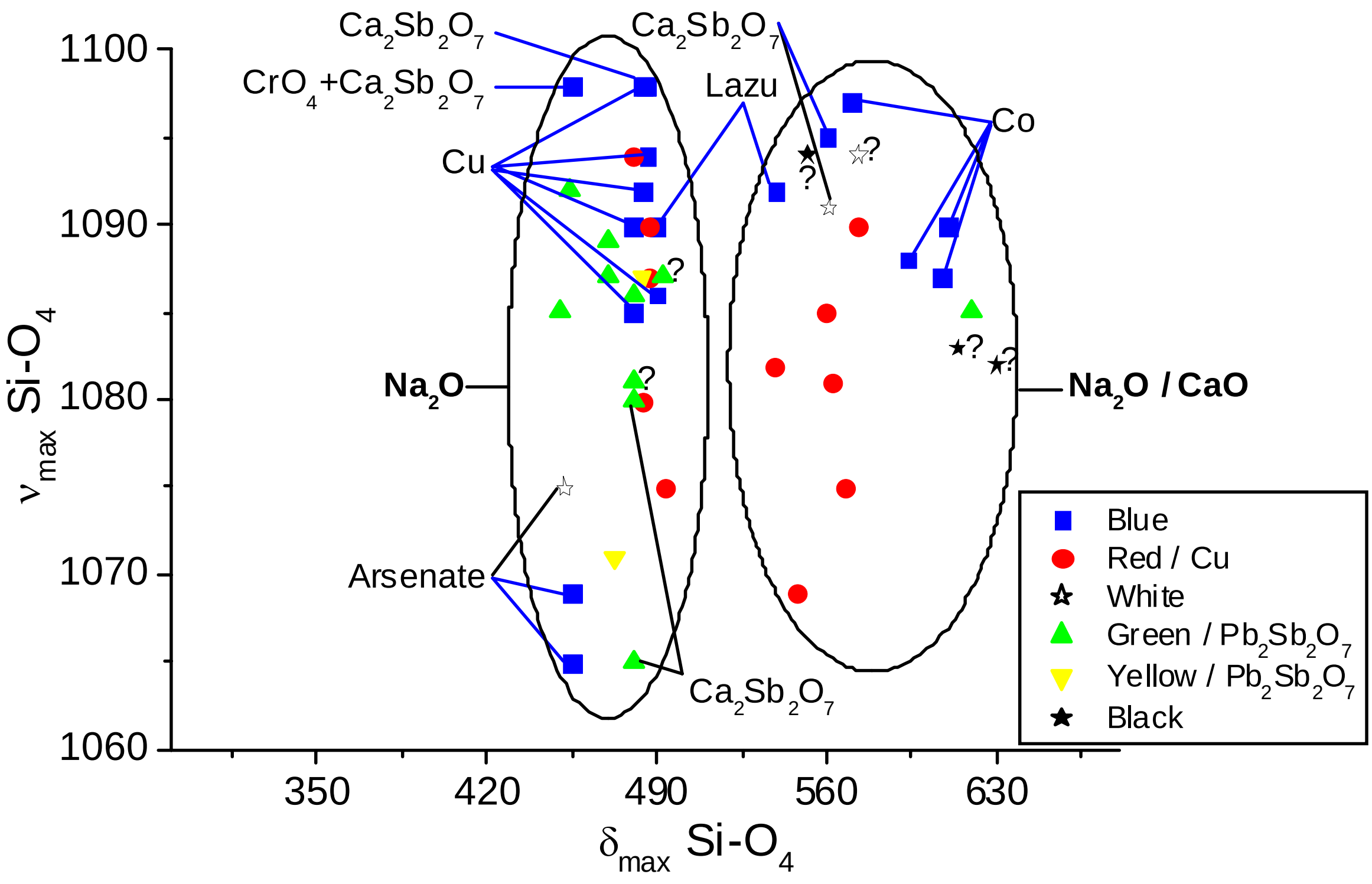

*Figure 12: Classification of the glass beads from Mapungubwe hill and K2 sites. The colours and pigments of the beads are indicated.*

Table 4 summarizes the results of this study according to colouring agent (pigment/chromophore), the glass matrix, the Raman signature type and when possible the origin and period of production with the assignment of Wood classification (Wood 2000, 2005). The quantity of beads found in the Mapungubwe Museum collection for the more modern beads (M3 Gardner) is also reported. Using this Table it is possible to speculate about the number of possible origin and/or period of production for each colour of Mapungubwe hill and K2 beads. Concerning the blue beads, 6 pigments were used and among the pigment some have two different glass matrixes, for example lazurite and blue antimonate. So, in total there might be approximatly 8 possible origins or periods of production for blue beads spread between Europe, Indo-Pacific and Middle East. Regarding the white beads, 2 pigments were observed with Raman spectrometry. Everyone is agreed to say that these bead are modern and come from Europe. We can guess as well that pink and striped beads made from arsenate originate from the same location as the white beads containing arsenate.

| Pigment/Chromophore | Quantity | Glass matrix | | | **Origin and/or traded period** | |
|---|---|---|---|---|---|---|
| | | **$Na_2O$** | **$Na_2O/CaO$** | **Type** | **Assignment this work** | Previous classification* |
| Blue/Lazurite | **400** | **x** | **x** | **3, 2** | **Fustat, Iran** | Indo-Pacific |
| Blue/Arsenate | **400** | **x** | | | **Italy, 19th** | European, 19th |
| White/Arsenate | **50** | **x** | | | **Italy, 19th** | European, 19th |
| Striped/Arsenate | **10** | | | | **Italy, 19th** | European, 19th |
| Pink/Arsenate | **550** | | | | **Italy, 19th** | European, 19th |
| Blue/copper | | **x** | | 1, **2** | **Indo-Pacific** | Indo-Pacific |
| Red/copper | | **x** | **x** | **2,3,**4 | **Indo-Pacific** | Indo-Pacific |
| Blue/Cobalt | **930** | | **x** | 1, 2, **3** | | |
| Blue/chromate | **50** | | | | **Middle East, Iran** | |
| Blue/Ca antimonate | **450** | **x** | **x** | **2** | | European |
| Green/Ca antimonate | **10** | **x** | | | | European |
| White/Ca antimonate | **500** | | **x** | **3** | | European |
| Green/Lead Tin yellow Type II | | **x** | **only 1** | 2 | | Mapungubwe / Indo-Pacific |
| Yellow/Lead tin yellow Type II | **100** | **x** | | **4** | | Mapungubwe / Indo-Pacific |
| Black/Spinel | | | **x** | | | Mapungubwe |
| **Black/Fe-S chromophore** | | | x | | | Mapungubwe |

*Table 4: Summary of beads analysed in this study regarding colour, glass matrix, Raman signature type and quantity of beads that do not fit into the Wood classification Wood, 2000, 2005, Robertshaw et al., 2010.*

The beads that we analysed in this work that we found that did not fit into the Beads Series defined by Wood Table 4 amount to almost 2000 beads. This is a significant quantity and indicates a longer occupation time on the hill which lasted well after the 14$^{th}$ c. and at least until the 19$^{th}$ c.

## 8. CONCLUSION

It was possible to classify the beads into two groups using the parameters extracted from their Raman signatures, namely soda and soda/lime glass. The classification is in accordance with previous work where other analytical techniques were used to determine the type of glass matrix. It provides an easy non-destructive method for archaeologist to discriminate non-destructively between bead types, when it is not possible to differentiate between beads on morphological factors alone.

The strength of Raman spectroscopy was illustrated in identifying the pigments used to colour the beads. Many of them was first used after the 13th century some even dating from the 19th century. This supports our previous results, which found that the celadon shards excavated on the hill dates from a later period than the original classification and indicates a later occupation date for the hill. This calls for more research to find a way to reconcile the carbon dating of the hill with the physical evidence of the modern beads excavated on the hill. A few pigments were identified that is typical of Islamic productions and is supportive of previous authors who claimed Fustat to be a possible production site.

The information obtained from identifying both pigments and type of glass matrix of the beads indicates a large number of production sites and therefore a considerable quantity of trading partners over a long period of time. Future, more in-depth studies of the beads, where other parameters extracted from their Raman signatures such as polymerisation index are also used might even reveal more production origins.

To complement this first study of beads excavated in Africa, which focussed on the K2 and Mapungubwe sites, we plan to analyse beads from the same period from other sites in Africa such as Great Zimbabwe. After the Portuguese rounded the Cape of Good Hope and the scramble for Africa began - millions of beads were brought into Africa as trade goods and ended up in artefacts made by the various tribes of Africa. Such as initiation dolls, aprons and every imaginable utensil. We plan to analyse these artefacts, which are part of museum collections all over the world. In this we hope to trace the beads to the country of origin and so make a contribution in reconstructing the early history of Africa.

## 9. ACKNOWLEDGEMENTS

The authors thank Ms Sian Tiley and Ms Isabelle Barrier, Curator and Assitant Curator of the Mapungubwe Museum for their help and advice. Dr Leskey Cele and the Tshwane University of Technology are kindly acknowledged for the permission to use their Raman facilities.

We also thank Dr Alan Carr for the photographs of the bead samples. Dr B. Gratuze and L. Dussubieux are kindly acknowledged for the glass references.